\documentclass[smallextended]{svjour3}       
\smartqed  
\usepackage{graphicx}
\usepackage{graphicx}
\usepackage{subfigure}
\usepackage{amssymb,amsmath,amsfonts}
\usepackage{epsf}
\usepackage{wrapfig}
\usepackage[normalem]{ulem}
\usepackage{eurosym}
\usepackage{epsfig}
\usepackage{float}
\usepackage{framed}
\usepackage{lineno}
\usepackage[hidelinks, pdfborder={0 0 0}]{hyperref}
\usepackage{doi} 
\usepackage{booktabs} 
\usepackage{cancel}
\usepackage{pdflscape}
\usepackage{blindtext}
\usepackage{graphicx}
\usepackage{everypage}
\usepackage{environ}
\usepackage[figuresright]{rotating}

\usepackage[usenames,dvipsnames]{xcolor}  

\newcommand{\mbu}{\mathbf{u}}
\newcommand{\mby}{\mathbf{y}}

\newcommand{\xgh}{x_{\text{g/h}}}
\newcommand{\xhref}{x_0}
\newcommand{\xh}{x_\text{h}}
\newcommand{\xg}{x_\text{g}}

\newcommand{\MCC}{N_\text{MCC}}
\newcommand{\CBItwo}{N_\text{CBI-2}}
\newcommand{\CBIthree}{N_\text{CBI-3}}
\newcommand{\CBIfour}{N_\text{CBI-4}}
\newcommand{\Bsixfour}{N_\text{B64}}
\newcommand{\Beight}{N_\text{B8}}
\newcommand{\Bthreeeight}{N_\text{B38}}
\newcommand{\Bfour}{N_\text{B4/B5}}
\newcommand{\Bthreeone}{N_\text{B31/B32}}
\newcommand{\Bsix}{N_\text{B6/B9/B3}}
\newcommand{\Btwenty}{N_\text{B20}}
\newcommand{\Bseven}{N_\text{B7}}
\newcommand{\Bforty}{N_\text{B40/B30}}

\newcommand{\MechLips}{[\text{lips}_\text{mech}]}
\newcommand{\ChemLips}{[\text{lips}_\text{chem}]}
\newcommand{\MechGrasper}{[\text{grasper}_\text{mech}]}
\newcommand{\unbroken}{[\text{unbroken}]}

\newcommand{\xvec}{\begin{bmatrix}
\xg\\
\xh
\end{bmatrix}}

\renewcommand{\land}{\,\,}
\newcommand{\lneg}{\,!\,}
\renewcommand{\lor}{\parallel}

\newcommand{\red}[1]{\textcolor{black}{#1}}

\newenvironment{amssidewaysfigure}
  {\begin{sidewaysfigure}\vspace*{.6\textwidth}\begin{minipage}{\textheight}\centering}
  {\end{minipage}\end{sidewaysfigure}}

\begin{document}

\title{Control for Multifunctionality 
}
\subtitle{Bioinspired Control Based on Feeding in \textit{Aplysia californica}}

\titlerunning{Control for Multifunctionality}        

\author{Victoria~A.~Webster-Wood \and
        Jeffrey~P.~Gill \and
        Peter~J.~Thomas \and
        Hillel~J.~Chiel
}


\institute{
    V.A.~Webster-Wood \at
    Department of Mechanical Engineering; Department of Biomedical Engineering; McGowan Institute for Regenerative Medicine\\
    Carnegie Mellon University, 5000 Forbes Ave., Pittsburgh, PA, 15213 \\
    \email{vwebster@andrew.cmu.edu}           
        \and
    J.P.~Gill \at
    Department of Biology \\
    Case Western Reserve University, 2080 Adelbert Road, Cleveland, OH 44106-7080
        \and
    P.J.~Thomas \at
    Department of Mathematics, Applied Mathematics and Statistics;  Department of Biology\\ Department of Cognitive Science; Department of Electrical, Computer and Systems Engineering\\
    Case Western Reserve University,
    10900 Euclid Avenue,
    Cleveland, Ohio 44106-4901
        \and
    H.J.~Chiel \at
    Department of Biology; Department of Neurosciences; Department of Biomedical Engineering\\
    Case Western Reserve University,
    2080 Adelbert Road,
    Cleveland, OH 44106-7080
}

\date{Received: date / Accepted: date}

\maketitle

\begin{abstract}
Animals exhibit remarkable feats of behavioral flexibility and multifunctional control that remain challenging for robotic systems. The neural and morphological basis of multifunctionality in animals can provide a source of bio-inspiration for robotic controllers. However, many existing approaches to modeling biological neural networks rely on computationally expensive models and tend to focus solely on the nervous system, often neglecting the biomechanics of the periphery. As a consequence, while these models are excellent tools for neuroscience, they fail to predict functional behavior in real time, which is a critical capability for robotic control. To meet the need for real-time multifunctional control, we have developed a hybrid Boolean model framework capable of modeling neural bursting activity and simple biomechanics at speeds faster than real time. Using this approach, we present a multifunctional model of \textit{Aplysia californica} feeding that qualitatively reproduces three key feeding behaviors (biting, swallowing, and rejection), demonstrates behavioral switching in response to external sensory cues, and incorporates both known neural connectivity and a simple bioinspired mechanical model of the feeding apparatus. We demonstrate that the model can be used for formulating testable hypotheses and discuss the implications of this approach for robotic control and neuroscience.

\end{abstract}

\section{Introduction}
\label{intro}

Multifunctionality, a basis for behavioral flexibility, is critical for navigating and adapting to a complex changing environment. In animals as well as humans, multifunctionality is observed across a wide range of behaviors. Living systems must smoothly shift from one behavior to another while varying specific behaviors to handle changing environmental conditions. Even relatively simple organisms demonstrate multifunctional control. For example, grass-cutter ants use their mandibles to cut stalks of grass, carry them to the nest and manipulate them once in their nests \cite{Roschard2003}; frogs exhibit swimming, walking, and hopping gaits \cite{Stehouwer1992,Stamhuis2005}. The tremendous range and adaptability of control is observed even more strongly in human manipulation: humans can use their hands to lift barbells substantially heavier than their own body weight, but also use the very same hands to play complex piano concertos.

Despite the obvious importance of multifunctionality for animal systems, truly multifunctional control remains a challenge for robotics \cite{Royakkers2015}. To develop robotic controllers for multifunctional behavior, one possible approach would be to develop a methodology that can map multifunctional biological systems onto simulated devices or robots. Such a methodology would enable researchers to develop control architectures through rapid prototyping and simulation. The controllers could then be effectively improved by comparison to the original biological system, and by assessing their effectiveness as a simulated controller for an artificial device. Including known neurons, connections, and biomechanics underlying multifunctional behavior allows the models to immediately suggest testable experimental hypotheses, clarifying the biological mechanisms of multifunctionality. At the same time, to make the simulation useful for artificial or robotic devices, the modeling framework should run faster than real-time. A computationally efficient, biologically relevant framework could then lead to direct real-time control of the original biological system and of an artificial robotic system, and thus provide a bridge from neuroscience to robotics.

What mediates multifunctional behavior in biological nervous systems, and what can we learn from them for robotic control? Three alternative neural architectures have been proposed for multifunctional control: dedicated control circuits, population-based control circuits, and re-organizing control circuits \cite{MORTON1994413}. The first alternative dedicates a control circuit to each behavioral function. For example, an escape circuit might suppress and override a swimming circuit \cite{Huang1990}. Functionally decomposing behavior, and assigning dedicated control to each function, has historically been the traditional approach to robotic control, such as in traditional finite state machines \cite{McCulloch1943,S.C.Kleene1951,Mulgaonkar2016,Ravn1995}. The drawback is that controlling a wide repertoire of behaviors can lead to a combinatorial explosion, making this approach impractical in general, and it is clearly not used for most animal behaviors.  A second alternative is encoding solutions through the activity of a neuronal population. For example, the direction of a motor response may be encoded by a broadly tuned population of neurons \cite{Georgopoulos1692,Georgopoulos1988,Schwartz1988}. Population encoding is the basis of many machine learning approaches to robotic control \cite{Sewak2019}. A drawback of this solution is that it is difficult to isolate subnetworks with specific functionalities, so it can be difficult to understand how the system works. A third possibility, which appears to be a more common solution in biological systems \cite{MORTON1994413}, is that of reorganizing circuits: by varying the timing and phasing of activity and incorporating feedback from the periphery (body), single circuits can be reconfigured to produce several multifunctional behaviors. For example, the same multifunctional circuit in crustacea can be reconfigured to generate qualitatively different ingestive behaviors \cite{selverston1992dynamic,SELVERSTON1976215}. Despite increasing evidence that this third alternative may be the most common for biological control, relatively few robotic control architectures are based on this solution.

How can one effectively implement any of the three neural architectures for multifunctionality? One possibility is to use machine learning to allow the controller to ``learn'' a multi-functional network architecture. Machine learning has led to many applications and predictive modeling approaches by relying on large training datasets and intense computational power \cite{Hausknecht2015,Jaques2017,Xie2018,Hosman2019,Glaser2017,Wang2018,Sussillo2012,Aggarwal2019}. Since the relationship between network form and function is often very complex, it has not been easy to understand how the resulting networks actually function, or to use them to direct experimental analyses of an actual biological system. Thus, another approach has been to develop detailed models of individual neurons and networks based on actual experimental measurements; the detailed dynamics of individual neurons can be approximated using multiconductance, multicompartment biophysical models \cite{Hodgkin1952,Ekeberg1991,koch1998methods}. The drawback of this approach is that large numbers of parameters must be set experimentally, and given the variability within nervous systems, the resulting network may not capture the original dynamics of the system \cite{Marder2011,Golowasch2002,Beer1999,prinz2004similar}. A potential third way has been to use more phenomenological neural models to capture aspects of neural architecture and dynamics with a greatly reduced set of parameters, and these have been successfully used for biological modeling and control \cite{Prescott2016} including those inspired by insects \cite{Szczecinski2017a,Szczecinski2017,Buschges2005,BEER1990169,Beer1992b}, lobsters \cite{Ayers1995,Ayers2020,Ayers1977}, \textit{Pleurobranchaea} \cite{Brown2018}, lampreys \cite{Prescott2016,Buschges2008,KamaliSarvestani2013} and fish \cite{Ekeberg1993}, salamanders \cite{Bicanski2013,Harischandra2010},  and other tetrapods \cite{Hunt2015,Hunt2017}. 

Possible nominal model approaches to capture neural circuit dynamics include the use of integrate-and-fire neurons \cite{IZHIKEVICH2000,Mihalas2009a,Tal1997a}, rate models \cite{Wilson1973}, discrete asynchronous event-based models \cite{Bazenkov2020}, and at the simplest level, Boolean models or finite automata \cite{Ayers1995,Rosin2013,S.C.Kleene1951}. Integrate-and-fire nodes have been successfully implemented in synthetic nervous systems using neuron pool circuit models for robotic control \cite{Szczecinski2017a,Hunt2015,Hunt2017,Szczecinski2017}. However, the complexity of the animal models used for bio-inspiration precludes the possibility of capturing full circuit connectivity or individual identifiable neurons. Population firing rate models are often used to represent neural activity for therapeutic brain-machine interface technologies development \cite{Nicolelis2009,LU2012137,Latash1999}. Firing-rate models of neural networks go back at least to the Wilson-Cowan equations \cite{Wilson1973,Wilson1972,Destexhe2009}, and have helped understand neural behaviors as diverse as  spontaneous pattern formation \cite{Ermentrout2010}, processing of sensory input signals \cite{Beck2007,Brosch2014,Wood2017}, and motor control \cite{Shea-Brown2006,Heuer1995,Shoham2005,Sussillo2012,Beer1999a,Beer1999}.
Boolean network models, being closely related to finite state machines, originated with the seminal study by McCulloch and Pitts \cite{McCulloch1943} and have found application in robotics and as reduced models of biological systems such as gene regulatory and signal transduction networks  \cite{Oishi2014,DeJong2002,Payne2013,Saadatpour2010,Giacomantonio2010,Edwards2001,kauffman1993origins,Packard1985}. Of these nominal models, Boolean network models likely provide the lowest computational cost, while still capturing the overall on/off behavior of neuronal bursting. Such models have been used to describe neural activity recorded during multifunctional behaviors \cite{Ayers1995}, and are the foundation of finite automata \cite{McCulloch1943}. Boolean network models can be used to capture a wide range of biological phenomena \cite{Oishi2014,Payne2013,Dallidis2014,Harris2002,Siegle2018} and can even be extended to capture stochastic processes \cite{RiveraTorres2018}. 

In many neural models, the focus is on the network controller, without accounting for the dynamics of the periphery, or body. For applications in bioinspired robotic control, a computational modeling approach is needed that captures both the dynamics of the neural circuitry and the critical interactions between the brain, the body, and the environment \cite{Chiel1997,Chiel2009}. As with neural components, a variety of models have been developed to capture the nonlinear properties of individual muscles, and their organization into muscular structures  \cite{Zajac1989}. The complexity of such muscle models can vary from capturing muscle biochemical kinetics using a cross-bridge model \cite{yu1997nonisometric,Eisenberg1980,Haselgrove1973,Piazzesi1995,Zahalak1990} to spring-damper representations such as used in the linear Hill muscle model \cite{A.V.Hill1938,Shadmehr1970}. Such models can be fit to match muscle physiology observed in animal systems \cite{Yu1999}, and used to model complex musculature such as muscular hydrostats \cite{Chiel1992}. \red{When modeling or fitting muscle models to experimental data, individuality and variability of model parameters is important, and one should not use average values
\cite{BluemelEtAlBueschges2012a-Hill,BluemelEtAlBueschges2012b-Using,BluemelEtAlBueschges2012c-Determining}. However, fitting to individual values creates additional computational overhead, and it is possible that the model may not capture the behavior of a range of individuals.} Fundamentally, the role of these muscle models is to capture the integration of muscle activation dynamics into a resulting tension. Once again, although these muscle models are important for modeling mechanical systems, complex structures involving both the kinematics and kinetics of multiple muscles, in general, will have high computational overhead \cite{Sutton2004a,Sutton2004b,Novakovic2006}. Thus, \red{if a simulation is to run faster than real time and be able to be rapidly fit to a given individual, one must use simplified models. For the work presented here, we were interested in generating a model that can run faster than real time, and have done this using a very simplified muscle model for the biomechanics.}

To meet the need for computationally efficient, explainable, multifunctional controllers, we have developed a hybrid Boolean network model framework, i.e. primarily using Boolean network elements but using continuous mechanical models. This framework combines discrete Boolean logic calculations of neural activity with simplified semi-continuous second-order muscle dynamics and peripheral mechanics. To our knowledge, mixed Boolean (neural) / continuous (biomechanical) models have not previously been used for motor control. The use of Boolean logic for capturing neural activity results in a computationally efficient control algorithm that can run faster than real-time. The use of a simplified model of the peripheral mechanics provides sufficient sensory feedback for the controller to adjust to changing environmental conditions, and allows key characteristics of each of the multifunctional behaviors to be observed. To demonstrate mapping from a known multifunctional biological system, an animal model is needed with a relatively small neural network controlling a well-understood musculature. Therefore, we demonstrate this model framework for multifunctional control using the experimentally tractable model system of feeding in the marine mollusk \textit{Aplysia californica}. The resulting model controls a simplified mechanical model of the feeding apparatus and successfully demonstrates ingestive behaviors, including biting and swallowing, as well as rejection of inedible materials. In this paper, we will first describe prior models of the \textit{Aplysia} feeding neural circuitry and periphery, then describe our novel Boolean model framework. 
We will demonstrate how the hybrid Boolean controller is developed based on observations from behavioral, biomechanical, and electrophysiological experiments and the existing literature. 
Finally, we will use the resulting model to show  multifunctional control, and illustrate how it can be used to make testable experimental predictions.

\section{Prior Models of \textit{Aplysia} Feeding and Neural Circuitry}
\label{sec:prior-models}

Feeding behavior in \textit{Aplysia} is multifunctional and has been well characterized. Three key feeding behaviors are observed in the intact animal: biting, swallowing, and rejection. Animals flexibly switch between behaviors as sensory inputs vary, e.g., switching from biting to swallowing once food (seaweed) is successfully grasped. Moreover, as the animal encounters seaweeds that impose varying mechanical loads, the animal may robustly adjust the magnitude and duration of force it exerts to ingest the seaweed \cite{Lyttle2017,GillChiel2020}. These multifunctional behaviors provide a model system for intelligent robotic grasper control. Previous models of the neural circuitry and peripheral mechanics have been reported that form the foundation for the hybrid Boolean model presented here, and we will briefly review the relevant aspects of these previous models.

\subsection{Prior Neural Circuit Analyses and Models}
\label{sec:prior-models-neural}

There is a wealth of information on the circuitry controlling feeding in the experimentally tractable \textit{Aplysia} nervous system. 
The tractability is a result of several factors: there are relatively few neurons responsible for feeding behavior (on the order of 200 motor neurons and dozens of key interneurons \cite{Cash1989,SUSSWEIN2012304}); neurons in \textit{Aplysia} are large, pigmented, and have similar synaptic inputs, outputs, morphology and biophysical properties from one animal to the next, and can thus be identified as unique individuals \cite{kandel1976cellular}; the somata, which are the largest parts of the neurons, are electrically excitable and electrically compact, so that stimulating or inhibiting the neuron at the soma controls its outputs \cite{Connor1986}. 
Together, these features make it possible to determine detailed neural circuitry that applies across all animals. In particular, the neural circuitry involved in \textit{Aplysia} feeding has been extensively studied \cite{Cropper2019}. Two ganglia contain the primary neurons responsible for generating the relevant multifunctional behavior: the cerebral and buccal ganglia. The buccal ganglion contains many of the primary motor neurons that innervate the musculature of the feeding apparatus, as well as interneurons and sensory neurons involved in feeding \cite{Cropper2019}. The cerebral ganglion is the primary locus for many key interneurons responsible for guiding behavioral switching \cite{Cropper2019}. Many of the neurons of the feeding circuitry can be consistently identified between animals due to their location, size, and electrical characteristics \cite{Weiss1986,Hurwitz1996,Chiel1988,Susswein1988,Kabotyanski1998,TEYKE1991307,Chiel1986}. In particular, many of the motor neurons innervating key muscles have been previously identified, including B3/B6/B9 innervation of the I3 retractor muscle \cite{Church1993,Church1994}, B31/B32/B61/B62 innervation of the I2 protractor muscle \cite{Hurwitz1994}, B7 innervation of the hinge retractor \cite{Ye2006aSwallows}, B8a/b innervation of the grasper \cite{Morton1993b,Church1994,Evans1998}, and B38 activation of the anterior region of the I3 retractor muscle \cite{Church1993}. The coordination of these motor outputs is mediated via many known interneurons both in the buccal and cerebral ganglia \cite{Cropper2019}.

\begin{figure}
\centering
\includegraphics[width=\textwidth]{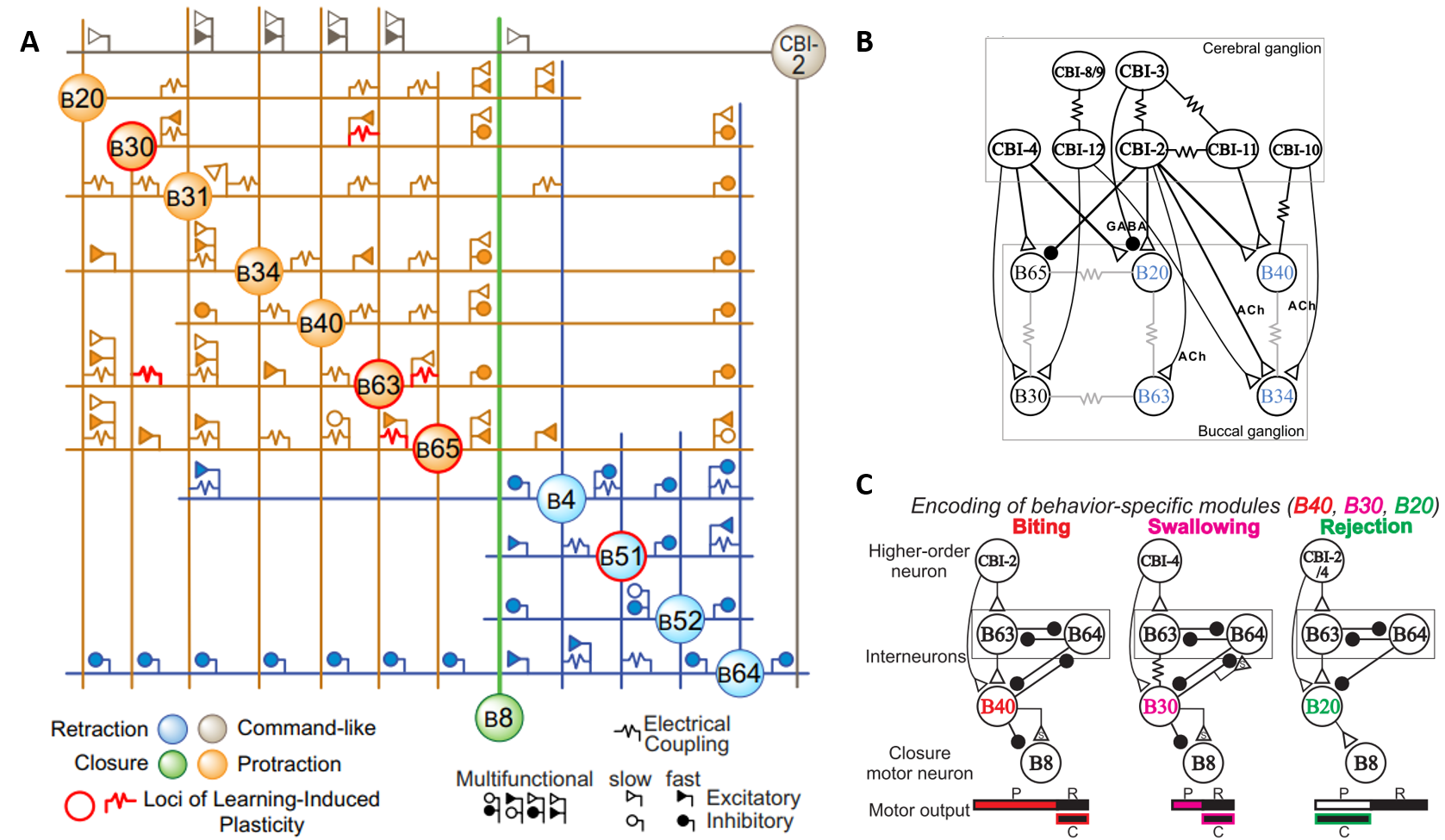}
\caption{\red{\textbf{A} The SNNAP model of key \textit{Aplysia} feeding circuitry recently presented by Costa et al.~\cite{Costa2020} (reproduced with permission). This model includes many key motor neurons and interneurons as well as CBI-2. \textbf{B} A network diagram reproduced with permission from Jing et al.~showing key cerebral-buccal interneurons responsible for behavioral switching in feeding \cite{Jing2017}. \textbf{C} Behavior-specific modules proposed by Jing et al. (Copyright 2004 Society for Neuroscience), reproduced from \cite{Jing2004} per The Journal of Neuroscience reuse permission guidelines.}}

\label{fig:priormodels}
\end{figure}

The neural circuitry controlling feeding in isolation from the musculature that mediates feeding has been modeled in detail. 
Cataldo et al.~\cite{Cataldo2006} developed a network model with Hodgkin-Huxley-type neurons incorporating known data on conductances and the roles of important second messengers in individual identified neurons mediating feeding behavior (using the SNNAP modeling platform \cite{ziv_simulator_1994}) which has been recently updated by Costa et al.~\cite{Costa2020}.  
\red{Figure \ref{fig:priormodels}.A shows details of the neural circuitry included in this model}. 
This model includes key motor neurons and interneurons in the buccal ganglia. Using this approach, they were able to generate ingestion-like and rejection-like neural activity. 
However, the model did not take into account the role of many relevant cerebral-buccal interneurons (CBIs) in switching between the behaviors, and thus could not differentiate between bite-like and swallow-like patterns, nor did it provide control of a simulated periphery, and thus could not incorporate sensory feedback during feeding, which we have argued may play a critical role in generating robust feeding behavior \cite{Lyttle2017,Shaw2015}. \red{Alternative conceptual models have included a wider assortment of CBIs (Figure \ref{fig:priormodels}.B), such as the circuits proposed by Jing et al.~\cite{Jing2004,Jing2017,Jing2001,Jing2002,Jing2005}, in which the CBIs involved in switching from ingestive to egestive behavior, as well as from biting to swallowing, are included (Figure \ref{fig:priormodels}.C). 
These models highlight the importance of the CBIs in behavioral multifunctionality, but do not include a model of the periphery to capture the role of biomechanics in behavior. 
This prior work provides a foundation for the development of real-time or faster than real-time controllers based on this animal model by identifying key neurons involved in the multifunctional behavior.}

\subsection{Prior Mechanical Models}
\label{sec:prior-models-mech}

While many studies have investigated the neural circuitry underlying \textit{Aplysia} feeding behavior, and some have developed detailed models of that circuitry, fewer have considered the critical role of the peripheral biomechanics on the control architecture and behavior. However, the parallel evolution of the peripheral musculature and control circuitry result in a tightly coupled system \cite{Chiel1997}. To understand and create multifunctional controllers, we must understand the interactions of the complete system. 

Kinematic and kinetic models of the \textit{Aplysia} feeding apparatus have previously been reported in the literature. Based on MRI images during feeding, kinematic models have been developed to capture the mechanics of feeding \cite{Neustadter2002a,Drushel1998}. These models highlight the morphological computation inherent in the feeding apparatus. In particular, the kinematic changes observed during feeding reveal how shape changes in the grasper can change the mechanical advantage of key muscles \cite{Novakovic2006,Sutton2004a}. In addition to kinematic models, basic kinetic models have been proposed which capture the dynamics of key structures throughout feeding \cite{Sutton2004a}. Such models can be extended through the inclusion of kinematic reconfiguration observed through MRI imaging \cite{Novakovic2006}. However, the existing mechanical models do not yet include the neural circuitry needed for controller development.

\subsection{Prior Neuromechanical Models}
\label{sec:prior-models-neuromech}

Abstract neuromechanical models, which combine neural control and biomechanics into a unified model of \textit{Aplysia} feeding, have also be developed, such as a stable heteroclinic channel model \cite{Shaw2015,Lyttle2017}. This model captures the CPG-like activity of the feeding circuitry using three mutually inhibitory nodes representing pools of motor neurons. Though the nodes do not map precisely to known neural connectivity, their dynamics can be simulated rapidly, connected to basic kinematic models of the periphery, and respond to changes in sensory inputs, such as the load on the seaweed. Furthermore, stable heteroclinic channel controllers have been successfully translated to robotic applications \cite{horchler2015designing,Webster2013}. However, such models do not provide insight into the detailed neural mechanisms underlying multifunctional control.

\section{Models and Methods}
\label{sec:methods}

Our approach to modeling begins with experimental observations from intact animals of both their feeding behavior and recordings of the major motor neuronal activity controlling feeding. These observations of the functional outputs of the system motivated an outside-to-inside modeling approach: first, a minimal set of peripheral structures and muscles are represented; second, the direct controllers of those muscles (motor neurons) are added; finally, layers of local and ultimately global control mediated by interneurons are added.

\subsection{Experimental Methods and Data Analysis}
\label{sec:experimental-methods}

The activity patterns of identified neurons during distinct feeding behaviors were obtained experimentally from intact animals via chronically implanted electrodes.
Materials and procedures are described in detail by \cite{GillChiel2020} and are summarized here.

Adult \textit{Aplysia californica} (200--450 g) were anesthetized via injection of isotonic magnesium chloride solution (333 mM) and immersion in chilled artificial sea water (1-5$^{\circ}$ C) for at least 10 minutes.
A small incision in the body wall near the head was made which permitted access to the feeding apparatus (buccal mass).
Differential electrodes, comprised of twisted pairs of fine (25-$\mu$m diameter), insulated stainless steel wires (see \cite{CullinsChielJoVE2010} for fabrication details), were implanted on the protractor muscle I2, the radular nerve (RN), and buccal nerves 2 (BN2) and 3 (BN3).
Together these recording sites permitted extracellular monitoring of nearly all of the major motor neurons of the circuitry controlling feeding (I2: B31/B32/B61/B62; RN: B8a/b; BN2: B38, B6/B9, B3; BN3: B7), as well as an important pair of multiaction interneurons (BN3: B4/B5) \cite{LuJoVE2013}.
The incision was closed with a suture.
Animals recovered after 1-3 days.

Instrumented animals were presented with different food stimuli to elicit different feeding responses.
To elicit bites, which are failed attempts to grasp food \cite{Kupfermann1974}, dried nori (Deluxe Sushi-Nori, nagai roasted seaweed, Nagai Nori, USA INC, Torrance, Ca) was touched to the rhinophores, anterior tentacles, and perioral zone until protractions of the feeding grasper were visible.
To elicit swallows, animals were permitted to grasp and ingest the food.
To elicit rejections, animals were first enticed to partially swallow a polyethylene tube by simultaneously touching nori to the perioral zone;
after several centimeters of tubing were swallowed, the nori was removed, and eventually the animal rejected the tube by pushing it out of the mouth using its grasper.

For some swallows, an unbreakable food stimulus (double-sided tape between two uniform strips of dried nori) was anchored to a force transducer and suspended vertically over the animal.
The animal attempted to swallow the strip, but because it was anchored and unbreakable it could only make progress until tension began to develop in the anchored strip.
After this, the animal continued to attempt to swallow despite the increase in load for up to several minutes.

An electromyogram from the protractor muscle I2 and extracellular nerve signals from RN, BN2, and BN3 were digitally recorded, along with swallowing force measured by the force transducer.
Video was captured simultaneously so that behaviors could be reviewed during analysis.

Analysis of experimental data was aided by the Python package \emph{neurotic} (NEURoscience Tool for Interactive Characterization) \cite{gill_neurotic_2020}, and analysis procedures were similar to those described by \cite{GillChiel2020}.
Briefly, spikes were detected using window discriminators.
Units corresponding to identified neurons can be identified from nerve recordings because axonal nerve projections and the relative amplitude and timing of spikes is consistent from animal to animal \cite{LuJoVE2013}.
Amplitude thresholds were determined manually.
Spikes were grouped into bursts using firing frequency criteria (see \cite{GillChiel2020} for details;
for B7, the burst initiation and termination frequencies were 20 Hz and 10 Hz, respectively, based on observations by \cite{Ye2006aSwallows,Ye2006bRejections}).
Video was used to determine the timing of inward movement of food during swallowing and outward movement of tubing during rejection.

\subsection{Simplified Model Framework for Multifunctional Control}
\label{sec:framework}

To develop a simplified model of multifunctional control, we employed a demand-driven complexity approach: rather than modeling the complex dynamics of all possible units in the neural network, and the detailed biomechanics, we identified key neuronal elements based on functional outputs during behavior, modeled minimal associated peripheral mechanics, and refined both models to reproduce multifunctional behaviors. The result is a hybrid Boolean model consisting of the peripheral biomechanics and neural circuitry. Neural activity is represented using discrete Boolean units, whereas the biomechanics are calculated continuously in space using a semi-implicit integration scheme (see Appendix \ref{integration}). In the following sections, we present the proposed hybrid Boolean model framework applied to the multifunctional feeding behavior of \textit{Aplysia}.

\subsection{System Identification Based on Key Biomechanical and Neural Elements} 
\label{sec:system-id}

\textit{Aplysia's} feeding behavior is multifunctional. As an animal attempts to ingest food, it \textit{bites} (a failed grasp); once it succeeds in grasping food, it pulls it into the buccal cavity (i.e., it \textit{swallows}). If it encounters inedible material, it pushes it out of the buccal cavity (i.e., it \textit{rejects} food). The animal must continuously shift flexibly among these different behaviors as it encounters the changing properties of food (e.g., mechanical load, toughness and texture). Based on the known neural circuitry in the buccal ganglia and our recordings of each of the three feeding behaviors in intact behaving animals, we identified critical motor neurons and musculature necessary to reproduce multifunctional feeding in simulation. 

\begin{amssidewaysfigure}
\includegraphics[width=7in]{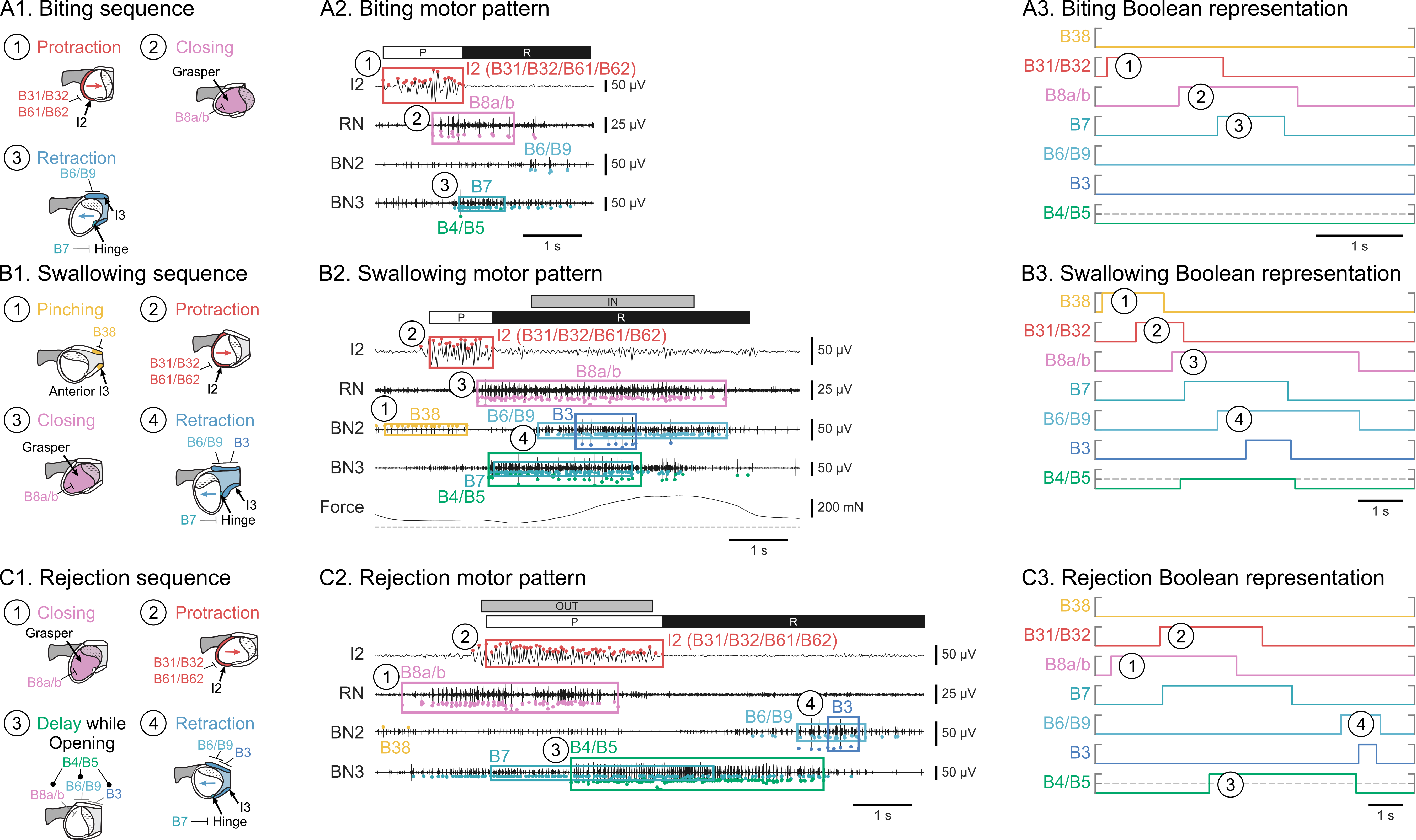}
\caption{See next page for caption.}
\label{fig:translation}
\end{amssidewaysfigure}

\addtocounter{figure}{-1}
\begin{figure} [t!]
  \caption{(See Figure on previous page.) Biting, swallowing, and rejection have distinct functional, kinematic, and neural control characteristics.
\textbf{A1} Biting is illustrated schematically in a sequence of cross-sections of the feeding apparatus. Biting begins with strong protraction of the open grasper towards the jaws (to the right), mediated by the protractor muscle I2 and the motor neurons B31/B32 and B61/B62; the grasper closes (indicated by a shape change from circular to elliptic) near the peak of protraction through the action of closure motor neurons B8a/b; having failed to grasp food, the closed grasper retracts weakly towards the esophagus (to the left), mediated by activation of the hinge muscle via B7 and the jaw muscle I3 via B6/B9.
\textbf{A2} Outputs of the motor control system (muscle and nerve activity) were recorded during biting, allowing timing of identified motor neuron activity to be determined. An understanding of the biomechanics (A1) and the neural control permits mapping the motor pattern to the kinematic sequence (circled numbers). Colored boxes around spikes indicate bursts of activity sufficiently intense to elicit functional movements. Bars indicate the protraction (P) and retraction (R) phases.
\textbf{A3} A simplified, discrete representation of the bursts of motor neuronal activity in A2. In this column, the I2 motor pool is abbreviated to ``B31/B32'' for brevity.
\textbf{B1} Swallowing begins with pinching the anterior jaws, mediated by the motor neuron B38, to prevent loss of food while the open grasper protracts; protraction is weaker than in biting; the grasper closes; the closed grasper retracts strongly to deliver food to the esophagus through recruitment of the jaw motor neuron B3, as well as intensified activation of B6/B9 and B7.
\textbf{B2} The motor pattern contains indications of each kinematic difference between biting and swallowing. Swallowing force and time of inward movement of food are also indicated. The multi-action interneurons B4/B5 are also active at a moderate level during swallowing and may act to delay the jaw motor neurons.
\textbf{B3} A discrete representation of the motor neuronal activity in B2, with B4/B5 active at a moderate level (dashed line).
\textbf{C1} Rejection begins with closing of the grasper; the closed grasper strongly protracts, expelling inedible material; the jaws are delayed from closing, giving the grasper enough time to open (indicated by a shape change from elliptic to circular) so that food will not be pulled back in during retraction; the open grasper retracts.
\textbf{C2} An essential difference between rejection and the ingestive behaviors is the timing of grasper closure, seen in the motor pattern as B8a/b activity during protraction. B4/B5 is very intensely activated during rejections and is responsible for the delay in jaw closure. Outward movement of the inedible material is also indicated.
\textbf{C3} In the discrete representation of C2, B4/B5 intensity is elevated relative to swallowing.
Note that motor patterns (A2, B2, C2) are plotted on identical time scales to emphasize differences in duration; discrete representations (A3, B3, C3) are rescaled for direct comparison of burst phasing. B1 and B2 are modified from \cite{GillChiel2020} with permission.}
\end{figure}

\paragraph{Biting: } In \textit{Aplysia}, \textit{biting} is characterized by strong protraction of the grasper, which closes prior to peak protraction as it attempts to grasp food, followed by weak retraction when food is not grasped (Figure \ref{fig:translation}.A1\red{; circled numbers indicate specific phases of the kinematics and are also used for the corresponding bursts of neural activity in parts A and B of the figure, respectively}). \red{Since biting is an attempt to grasp food, the power stroke is the protraction phase. See Figure \ref{fig:SchematicIntro} and \ref{fig:Allschematics}.A}. As grasping attempts are unsuccessful in this behavior, no force is applied to the seaweed. Key muscles and motor neurons involved in this behavior include the protractor muscle I2 and its associated motor neurons B31/B32 and B61/B62 \cite{Hurwitz1994}, the grasper closer muscle I4 and its motor neurons B8a/b \cite{Morton1993b,Church1994,Evans1998}, and to a lesser extent the jaw closer muscle I3 and its motor neurons B6/B9  \cite{Church1994}. In the experimental data shown in Figure \ref{fig:translation}.A2, the very limited B6/B9 activity is probably insufficient to mediate the level of retraction observed in previously reported magnetic resonance imaging data \cite{neustadter2007kinematics}. It is therefore likely that additional muscle units are required for the onset of retraction. Indeed, previous biomechanical models suggest that the hinge muscle, which is activated by neuron B7  \cite{Ye2006aSwallows}, plays a critical role in retraction during biting \cite{Sutton2004a}. As a result of this analysis, the demand-driven model should incorporate four muscle groups (I3, I2, grasper closure, and hinge) and four neural groups (B6/B9, B31/B32, B8a/b, and B7) to produce biting.

\paragraph{Swallowing: } If seaweed is successfully grasped at peak protraction during a bite, \textit{swallowing} is initiated (Figure \ref{fig:translation}.B1). \red{Since during swallowing, animals ingest food, the power stroke is the retraction phase with the grasper closed on seaweed. See Figure \ref{fig:SchematicIntro} and \ref{fig:Allschematics}.B}. To re-position the grasper to pull more seaweed inwards, it is then weakly protracted while open. If it were protracted too strongly, it might push seaweed out. Thus, during the retraction phase, the animal exerts strong forces on seaweed, whereas during protraction, it exerts minimal or even slightly negative forces. Similar muscle groups are activated in swallowing as are in biting, but with changes in duration and intensity. In addition, to prevent seaweed from slipping out, the anterior region of the I3 jaw muscle is pinched closed by activating the B38 motor neuron \cite{McManus2014}. The changes in motor neuronal timing (Figure \ref{fig:translation}.B2) can be understood from the biomechanics: First, to ensure that seaweed is not pushed out during protraction, the activation of the grasper closure motor neurons B8a/b occurs near the end of protraction (rather than overlapping the end of protraction, as is observed during biting). Second, to ensure that protraction is weaker, the protractor muscle I2 is less strongly activated than in biting. Third, to ensure that the grasper releases near the end of retraction, the grasper motor neurons B8a/b and the jaw motor neurons B6/B9 cease activity at about the same time. Fourth, the major jaw motor neuron B3 is recruited to generate greater retraction force. Finally, to maintain a hold on seaweed after the grasper opens, the B38 motor neuron is activated during the protraction phase to pinch the anterior of the jaw muscle onto seaweed.

\paragraph{Rejection: } If an inedible object is detected as a result of the combined sensory cues in the esophagus (e.g., a noxious mechanical stimulus), grasper (a lack of chemical stimulus), and at the lips (a lack of chemical stimulus), the inedible material will be rejected. This is a critical behavior for the animal, as it must be able to free the buccal cavity of inedible material in order to locate edible food. \red{ In \textit{rejection}, the power stroke is characterized by strong protraction with the grasper closed}, followed by retraction with the grasper open (Figure \ref{fig:translation}.C1). \red{See Figure \ref{fig:SchematicIntro} and \ref{fig:Allschematics}.C.} Similar to swallowing and biting, the I3 muscle, the I2 muscle, and the I4 muscle are all activated. However, the timing changes (Figure \ref{fig:translation}.C2): First, the grasper closer motor neurons B8a/b are activated during the protraction phase (i.e., during activation of the I2 protractor muscle and the B31/B32/B61/B62 motor neurons), rather than during the retraction phase, ensuring that the grasper closes and pushes out inedible material \cite{Morton1993,Morton1993b}. Second, since the inedible material is not retained during the protraction phase, the B38 motor neuron is not activated, and no pinch is observed. Finally, since it is critical to retract the grasper with its halves open (so as not to pull inedible material back in), the jaw motor neurons (B6/B9/B3) are initially inhibited at the onset of retraction by the B4/B5 multiaction neurons (since closure of the jaw muscles would push the grasper halves shut); instead, initial retraction is mediated by the hinge muscle (activated by motor neuron B7) \cite{Ye2006bRejections}.

This analysis of the animal data allows us to identify the key muscles and motor neurons necessary to produce the multifunctional behaviors of interest (see Table \ref{tab:motor}) and allows us to develop a simplified biomechanical model of the periphery to integrate into our controller model.

\begin{table}
\caption{Key muscle and motor neurons included in the hybrid Boolean network model of the \textit{Aplysia} feeding apparatus.}
\label{tab:motor}
\begin{tabular}{llll}
\textbf{Muscle} & \textbf{Role}        & \textbf{Motor neurons} & \textbf{References}\\ \hline
I2              & protraction          & B31/B32/B61/B62    &    \cite{Hurwitz1994}          \\
I3              & retraction; pinch & B3/B6/B9; B38         &  \cite{Church1993,Church1994,McManus2014} \\
I4              & grasper closure      & B8a/b              &   \cite{Morton1993b,Church1994,Evans1998}       \\
hinge           & retraction           & B7                 &    \cite{Ye2006aSwallows}   
\end{tabular}
\end{table}

\begin{table}
\caption{Key interneurons in both the cerebral and buccal ganglia included in the hybrid Boolean network model of the \textit{Aplysia} feeding apparatus.}
\label{tab:interneurons}
\begin{tabular}{lll}
\textbf{Neuron} & \textbf{Primary behaviors}    & \textbf{References}\\ \hline
CBI-2           & biting and rejection          &    \cite{Jing2004}\\
CBI-3           & biting and swallowing          &   \cite{Jing2001,Morgan2002,Jing2017}\\
CBI-4           & swallowing and rejection      &   \cite{Jing2004}\\
B64             & protraction-to-retraction transition &   \cite{Hurwitz1996} \\
B4/B5           & rejection                     &   \cite{Jing2001}\\
B20             & rejection                     &   \cite{Jing2001,Jing2002}\\
B40             & biting           &   \cite{Jing2002,Jing2004}\\  
B30             & swallowing &   \cite{Jing2004}
\end{tabular}
\end{table}

\subsection{Biomechanical Model}
\label{sec:biomech}

\begin{figure*}
\centering
  \includegraphics[width=0.8\textwidth]{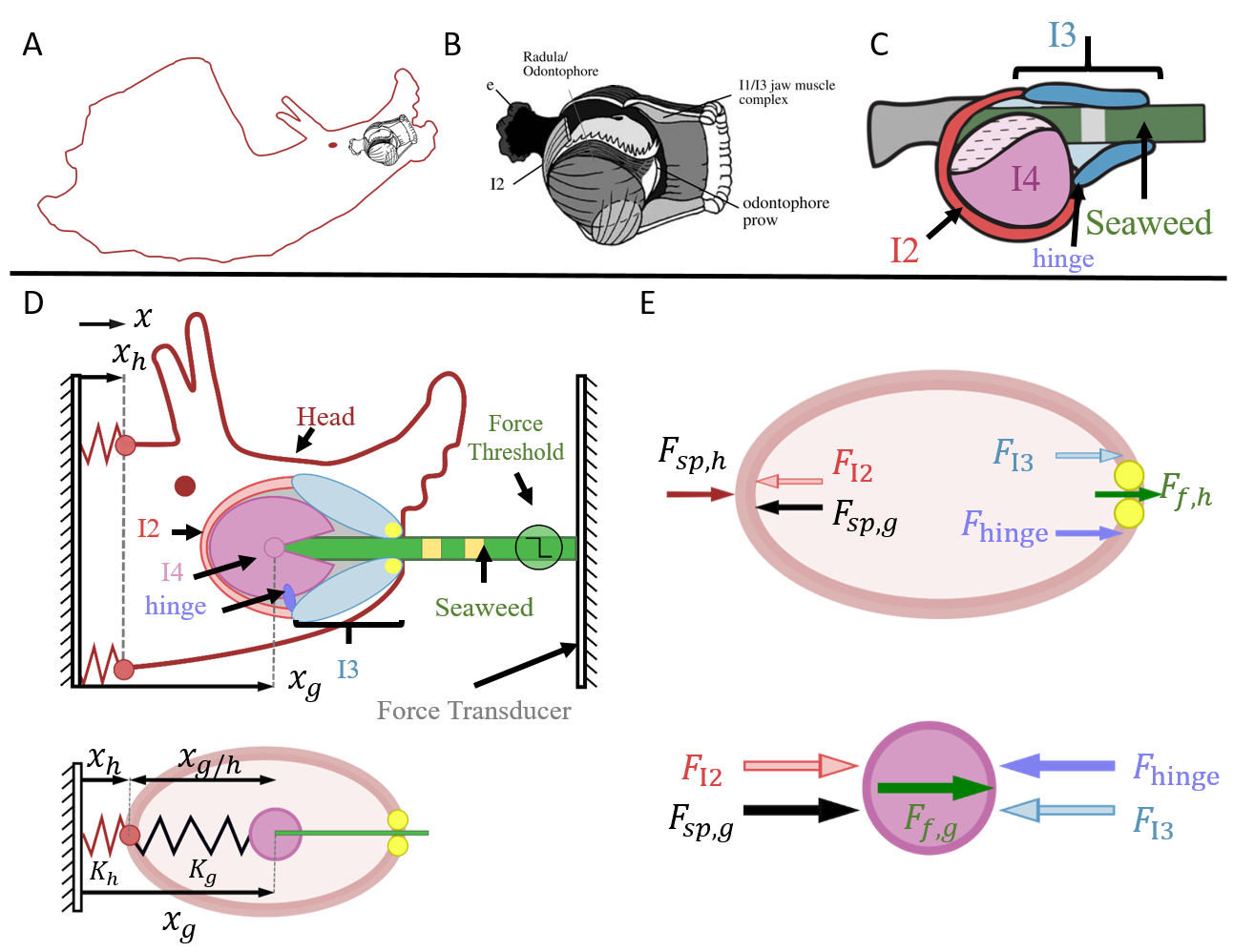}
\caption{A novel biomechanical model of the feeding system. \textbf{A} A schematic showing the position of the feeding apparatus (black) within the body of \textit{Aplysia} (red). \textbf{B} A sketch of the buccal mass, drawn by Dr. Richard F. Drushel, with a cut-away to reveal internal musculature, reproduced with permission from \cite{Sutton2004b}. \textbf{C} A planar schematic of the buccal mass showing the I2, I3, I4, and hinge muscles, and the position of seaweed during ingestion. \textbf{D} A 2D schematic representation (top) of the 1D biomechanical model (bottom) implemented in the hybrid Boolean network framework. This model does not account for the masses or shapes of any of the components. All positions and forces are constrained to the x-direction. The effect of the body and neck on the head is represented by the spring constant $K_h$ where the reference position of the spring $\xhref$ corresponds to the ground plane (i.e. $\xhref=0$). The presence and fixation of seaweed to a force transducer varies based on the behavior being simulated. During swallowing, a force threshold determines whether the seaweed is fixed to the force transducer, or breaks away. This allows the mechanical strength of the seaweed to be varied. \red{The details of the stimuli are presented in Figure 3 and the associated text.}  \textbf{E} All possible forces on the head (top) and grasper (bottom). To model interactions of the seaweed with the jaws (yellow circles, top) and with the grasper (magenta circle, bottom), the friction forces between the seaweed and the relevant jaws ($F_{f,h}$) and grasper ($F_{f,g}$) are calculated, based on the pressure at the location and user-specified coefficients of friction (see Appendix \ref{biomechanics}). }
\label{fig:SchematicIntro}       
\end{figure*}

Understanding the biomechanics of the periphery is important for developing effective multifunctional control. In \textit{Aplysia}, there are several key muscle groups that contribute to feeding behavior. Protraction of the grasper is primarily mediated by the I2 muscle, innervated by neurons B31/B32, which have both interneuronal and motor neuronal functions, and by motor neurons B61/B62 \cite{Hurwitz1994}. The motor neurons B8a/b activate the I4 muscle which results in closing of the grasper and pressure on the seaweed \cite{Morton1993b}; in strong swallows, in which the grasper is more protracted, grasper closure also induces a retraction force at the onset of grasping \cite{Ye2006aSwallows}. Retraction is primarily mediated by the activity of the I3 muscle; in addition, when the grasper is very strongly protracted, the hinge contributes to retraction during biting and rejection \cite{Sutton2004a}. Additionally, during swallowing, the anterior region of the I3 muscle tightens down on seaweed to prevent its release and expulsion during the protraction phase when the grasper is open \cite{McManus2014}; we will refer to this action as a pinch. These key muscle groups provide the foundation for the biomechanical model (Figure \ref{fig:SchematicIntro}.A-C). 

\begin{figure*}
\centering

  \includegraphics[width=\textwidth]{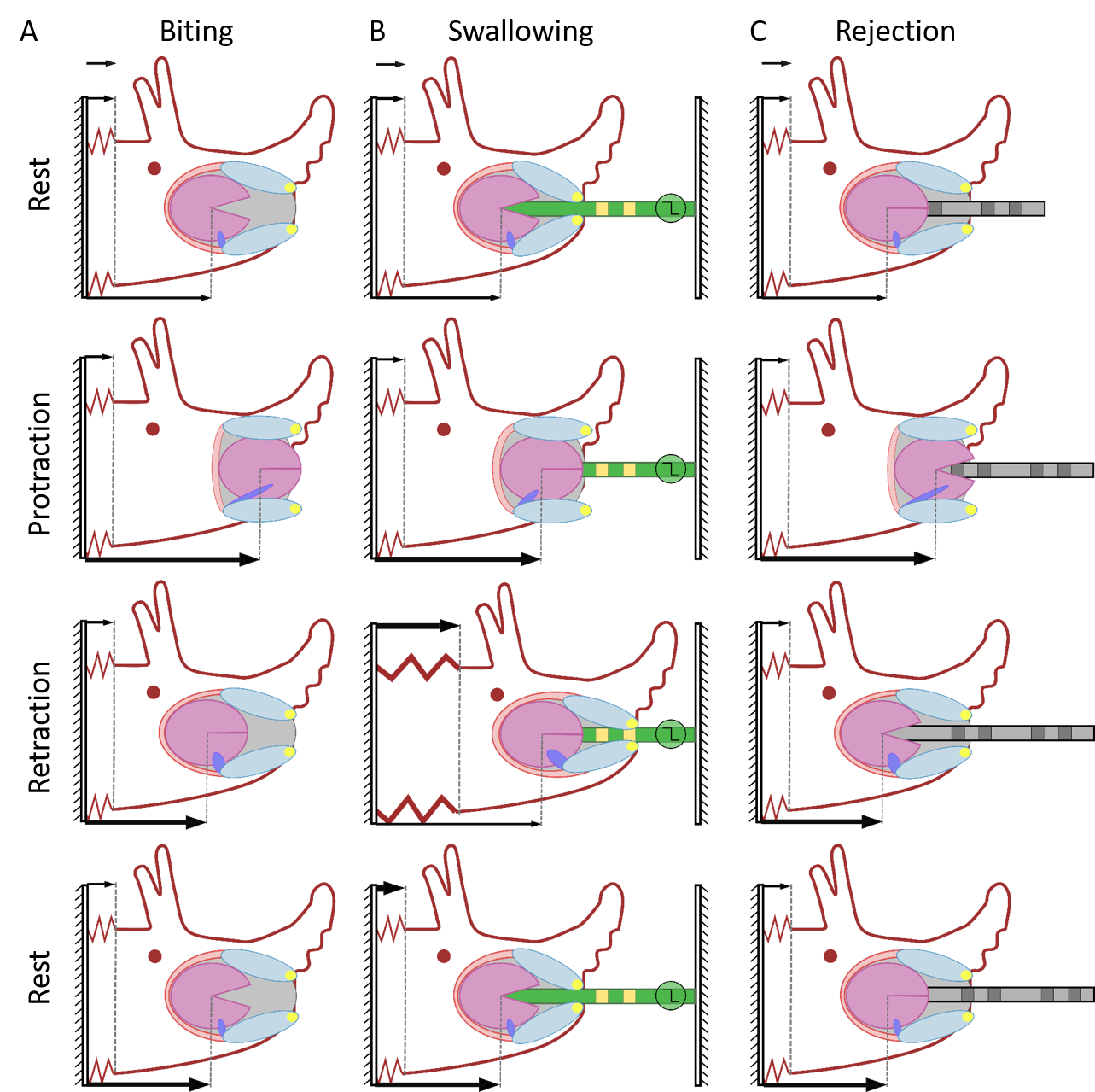}

\caption{Demonstration of one cycle for each of the multifunctional behaviors. Bold arrows at the bottom of each schematic indicate changes in grasper position from one behavioral phase to the next, whereas bold arrows at the top indicate changes in head position. \textbf{A} A schematic representation of the simplified model during biting. No seaweed or tube is present and only the grasper moves throughout the cycle.  \textbf{B} A schematic representation of the simplified model during swallowing. Seaweed is present and fixed to a force transducer. Of particular note is the motion of the head during swallowing. Because the seaweed is fixed to the rigid force transducer, the activation of the I3 muscle when the seaweed is being firmly grasped results in the head being pulled forwards along the seaweed (B, Retraction panel), so long as the force on the seaweed does not exceed the force threshold. \textbf{C} A schematic representation of the simplified model during rejection. Rather than seaweed, a tube is simulated to provide mechanical stimulation without any chemical cues. The tube is not fixed to an external object and is therefore free to be pushed forward during the rejection. Note the outward movement of the marks on the tube after the rejection cycle concludes (C, Rest panel). For both biting and rejection there is a slight forward motion of the grasper from the fully retracted position to the rest configuration shown here.}
\label{fig:Allschematics}       
\end{figure*}

Using these muscle groups, we derived a simplified model of the feeding apparatus that captures the basic mechanics of the head, grasper, and seaweed along a one-dimensional axis. In our model, mechanically tough seaweed is firmly affixed to a force transducer as described in Section \ref{sec:experimental-methods} (Figure \ref{fig:SchematicIntro}). In this preparation, the mechanically tough seaweed is unable to move relative to the force transducer during swallowing so long as the seaweed is unbroken. As a consequence, rather than the animal being stationary and pulling the seaweed into the esophagus, the animal grasps the seaweed and pulls its head forward along the seaweed, so that seaweed moves into the head during the retraction phase but may then move out again as the animal releases the seaweed under tension (Figure \ref{fig:Allschematics}.B). Thus, in this simplified biomechanical model, the forces and motion of the grasper vary with the type of behavior (Figure \ref{fig:Allschematics} A-C) and depend on the friction exerted by the grasper on the seaweed, the friction exerted by the jaws (anterior portion of the I3 muscle) on the seaweed, and the mechanical strength of the seaweed. For this simplified model the masses of the bodies are neglected due to the quasi-static nature of \textit{Aplysia} feeding muscle movements wherein the inertial forces are low relative to the viscous and elastic forces.

Muscle forces are determined by a first-order relationship between normalized motor neuron activity, $N$, and normalized muscle activation, $A$, and a first-order relationship between activation and normalized tension, $T$; numerically, this is implemented using a first-order accurate semi-implicit integration scheme based on operator splitting (see Appendix \ref{integration}) which makes it relatively easy to add new components to the control network:

\begin{equation}
\frac{dA}{dt}=\frac{N(t)-A(t)}{\tau_m}\quad
\longrightarrow
\quad 
    A(t+h) = \frac{\tau_m A(t)+h N(t)}{\tau_m+h} 
\end{equation}

\begin{equation}
\frac{dT}{dt}=\frac{A(t)-T(t)}{\tau_m}\quad
\longrightarrow
\quad 
    T(t+h) = \frac{\tau_m T(t)+h A(t)}{\tau_m+h} 
\end{equation}

\noindent
where $A$ is the muscle activation, $\tau_{m}$ is the activation time constant of the given muscle, $N$ is the activity of the innervating motor neuron \red{(see Section \ref{sec:boolean-network} and Appendix \ref{logic})}, $T$ is muscle tension, and $h$ is the time step. Similar equations can be used to express the grasper and pinch pressures. Muscle tensions are combined with equations for normalized mechanical advantage and a maximum force parameter in units of force to calculate applied force on the grasper, head, and food objects. Activation-Tension, Activation-Pressure, Tension-Force, and Pressure-Force relationships for each muscle as appropriate can be found in Appendices \ref{MuscleForces} and \ref{biomechanics}.

The subsequent motion of the grasper and head are calculated based on the contributions of individual muscles and the friction applied to the seaweed by the grasper and jaws. In the absence of external forces, the motions of the head fall between $\xhref =0$ (i.e., rest)  and 1 (i.e., full extension of the head, Figure \ref{fig:SchematicIntro}.D). Similarly, the grasper motion falls between 0 (i.e., full retraction), and 1 (i.e., full protraction). 

The motions of the head and grasper are calculated based on quasi-static equations of motion. In a system with inertial, viscous, elastic, and \red{applied} forces, the equations of motion can be expressed in the form:

\begin{equation}
    F_{\red{\text{app}}} -k x - c \dot x = m \ddot x 
\end{equation}

\noindent
In \textit{Aplysia} feeding, accelerations and masses are small, so inertial forces are negligible \cite{Sutton2004b}. Therefore the equations of motion simplify to have the form:

\begin{equation}
    F_{\red{\text{app}}} - k x = c \dot x
\end{equation}

\noindent
which can be written as:

\begin{equation}
    \dot x = \frac{\sum F}{c}
\end{equation}

\noindent
\red{where $\sum F$ includes both the elastic forces, $-kx$, as well as any applied external forces}. Therefore, the motions of the head and grasper are calculated as:

\renewcommand{\arraystretch}{1.5}
\begin{equation}
    \frac{d}{dt} \xvec = {\begin{bmatrix}
\frac{F_g}{c_g}\\
\frac{F_h}{c_h}
\end{bmatrix}}
\end{equation}

\noindent
where $\xg$ and $\xh$ are the positions, $F_g$ and $F_h$ are the net forces, and $c_g$ and $c_h$ are viscous damping coefficients for the grasper and head, respectively. For convenience we set $c_g=c_h=1$.

For this model, the total force on the grasper ($F_g$) includes the forces due to contraction of the I2 muscle (${F}_\text{I2}$), I3 muscle (${F}_\text{I3}$), and hinge (${F}_\text{hinge}$), a spring connecting the grasper to the head ($F_{sp,g}$) representing the surrounding musculature and connective tissue, and friction due to the interaction of the grasper with an object, e.g. seaweed or tube, (${F}_{f,g}$):

\begin{equation}
    F_g = {F}_\text{I2} + F_{sp,g} - {F}_\text{I3} - {F}_\text{hinge} + {F}_{f,g}
\end{equation}

The total force on the head ($F_h$) includes those listed above as well as friction between the jaws and the object ($F_{f,h}$) and a spring representing the musculature and connective tissue connecting the head to the rest of the body ($F_{sp,h}$). In Appendix \ref{biomechanics}, we show how this simplifies to:

\begin{equation}
    F_h = F_{sp,h} + F_{f,g} + F_{f,h}
\end{equation}

Details for calculating each of these component forces can be found in Appendix \ref{biomechanics}, and a table of symbols can be found in Appendix \ref{symbols}. This model represents a substantial simplification of the continuum mechanics of the soft bodied structures that make up the \textit{Aplysia} feeding apparatus. However, the model captures the key muscle groups identified in the animal experiments, as well as some of the configuration-dependent mechanical advantages of these muscles. Additional muscle groups and kinematic effects can be easily added into the hybrid Boolean model expressions.

\subsection{A Boolean Network Model of Neural Circuitry}
\label{sec:boolean-network}

Rather than using a complex neural model, each neuron in the hybrid controller presented here is represented by a Boolean logic statement. In contrast to highly detailed models of neural activity which are relatively computationally expensive, such as leaky-integrate-and-fire models \cite{IZHIKEVICH2000,Mihalas2009a,Tal1997a}, simple spiking models \cite{Izhikevich2003}, or Hodgkin-Huxley neurons \cite{Hodgkin1952}, the Boolean representation of neurons approximates neural activity based on bursts of activity observed in the animal data (Figure \ref{fig:translation}).

\begin{figure}

  \includegraphics[width=0.35\textwidth]{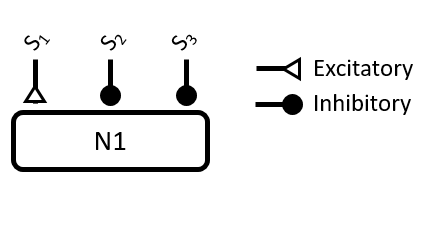}

\caption{An example node in the Boolean network model. This example neuron is innervated by one excitatory input, $S_1$, and two inhibitory inputs, $S_2$ and $S_3$.
}
\label{fig:example}       
\end{figure}

As a first approximation, such bursts can be represented as \emph{on} during firing activity and \emph{off} when quiescent. In the Boolean representation, the activity of the neuron (whether it is active or inactive) is determined by the combined logic of the inputs. For example, a simplified neural unit, $N_1$, with three inputs is shown in Figure \ref{fig:example}. If these inputs are equally weighted, then node $N_1$ can only be active if $S_1$ is active and if inputs $S_2$ and $S_3$ are not active. Therefore, the activity of the node $N_1$ at the next discrete step, $(j+1)$, can be calculated using Boolean logic based on the state of the synaptic inputs at the current discrete step, $(j)$, as follows:

\begin{equation}
    N_1(j+1) = S_1(j) \land (\lneg S_2(j)) \land (\lneg S_3(j))
\end{equation}

\noindent
where the inputs $S_i$ and output $N_1$ have numeric values 0 (off) or 1 (on), $\lneg$ represents Boolean negation, and the AND operator is implemented using multiplication. To account for neurons with variable bursting intensities, we extend the Boolean framework to include three-state model neurons such that quiescence is represented as 0, weak firing as 1, and strong firing as 2. If a normal logical AND were used, the difference between states 1 and 2 would be lost, but this difference remains when variables are multiplied. \red{In our model, all neurons are standard Boolean elements, i.e.~either on or off, except for the B4/B5 interneuron, which has been implemented using the ternary representation based on observations in our animal experiments. For this ternary unit, the Boolean negation, $\lneg N_{B4/B5}$, is undefined. Instead, our implementation tests whether the neuron is off ($N_{B4/B5} < 1$), firing weakly ($N_{B4/B5} == 1$), or firing strongly ($N_{B4/B5} \geq 2$). Each of these statements can then be negated using standard Boolean negation ($\lneg$). Details of the implementation of this ternary neuron can be found in Appendix \ref{logic}.} 

To account for known variations in the strength of inputs to neurons, additional logical calculations can be added to refine the activation of a given model neuron. For example, if $N_1$ is active if $S_1$ is activated or if $S_2$ is not activated but is still inhibited by any activation of $S_3$, the logic calculation could be modified as follows:

\begin{equation}
    N_1(j+1) = (S_1(j)\lor (\lneg S_2(j)))\land(\lneg S_3(j))
\end{equation}

\noindent
where $\lor$ represents the OR operator. When modeling three-state neural inputs that can fire strongly, ORs can be implemented using addition to preserve the magnitude of firing.

\subsubsection{Motor Control Layer}
\label{sec:boolean-network-motor}

\begin{figure*}
\centering

  \includegraphics[width=\textwidth]{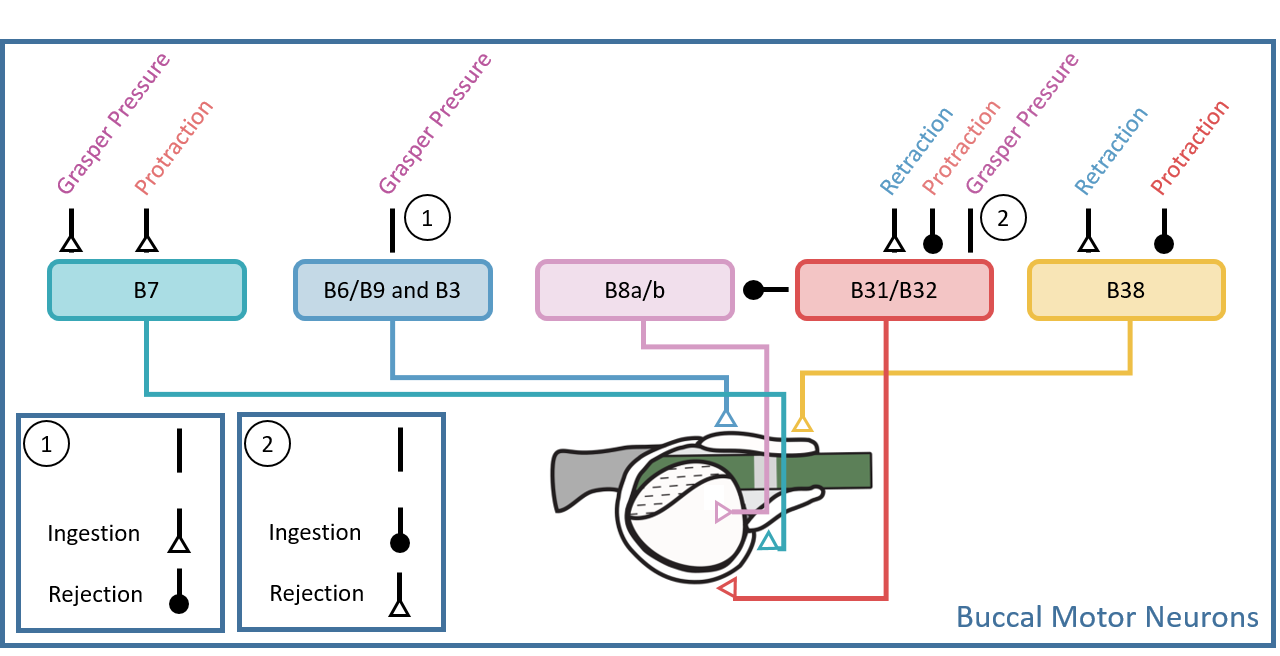}

\caption{Schematic of the motor control layer based on the key motor neurons and musculature identified previously (see Figure \ref{fig:translation}). See Appendix \ref{logic} for logic formulations. Sensory feedback pathways (inputs to motor neurons) are hypothesized based on the relative timing of motor neuronal activity. These sensory pathways are implemented in the model as acting directly on the motor neurons, but are likely to act via sensory neurons and interneurons in the actual neural circuitry. In this diagram, the Grasper Pressure inputs to B6/B9 and B3 and to B31/B32 are shown without end caps. These connections vary depending on the behavioral state as indicated by the insets in the bottom left corner of the image. \red{Behavioral state is controlled in the Boolean logic equations based on the activity of Cerebral-Buccal Interneuron 3 (CBI-3) and the sensory state of the mechanical stimuli in the grasper. See Appendix \ref{logic} for detailed circuit specifications and Figure \ref{fig:3layer} for the full network diagram.}}
\label{fig:1layer}       
\end{figure*}

In our simplified modeling framework, once the key musculature has been identified, a motor control layer is implemented. This layer consists of motor neurons known to innervate the key musculature (Table \ref{tab:motor}) and is built using Boolean model neurons (Figure \ref{fig:1layer}): B31/B32/B61/B62 for activating the I2 protractor muscle; B8a/b for closing the grasper; B6/B9/B3 for activating the I3 retractor muscle; B38 for pinching the anterior jaws; and B7 for activating the hinge muscle.

Though much is known about the \textit{Aplysia} feeding circuitry, there are still open areas of investigation, including exact sensory feedback pathways. Therefore, in the proposed model, we have implemented proprioceptive sensory feedback based on the kinematics of the grasper. In some cases, these sensory pathways directly innervate motor neurons in the current framework. However, it is likely that these are mediated through sensory neurons and interneurons in the animal. Such additional units could be easily added to the framework as they are identified. These sensory feedback inputs are implemented based on tunable thresholds of the grasper position and pressure of closing. Such thresholds may allow the model to be fit to individual animals or enable modulation of the network by tuning the values of thresholds. It should also be highlighted that in this model the larger motor pool B31/B32/B61/B62 is sometimes abbreviated as B31/B32; the rationale for doing so is that B31/B32 have both motor neuronal and interneuronal properties \cite{Hurwitz1996}.

\subsubsection{Local Coordination}
\label{sec:boolean-network-local}

\begin{figure*}
\centering

  \includegraphics[width=\textwidth]{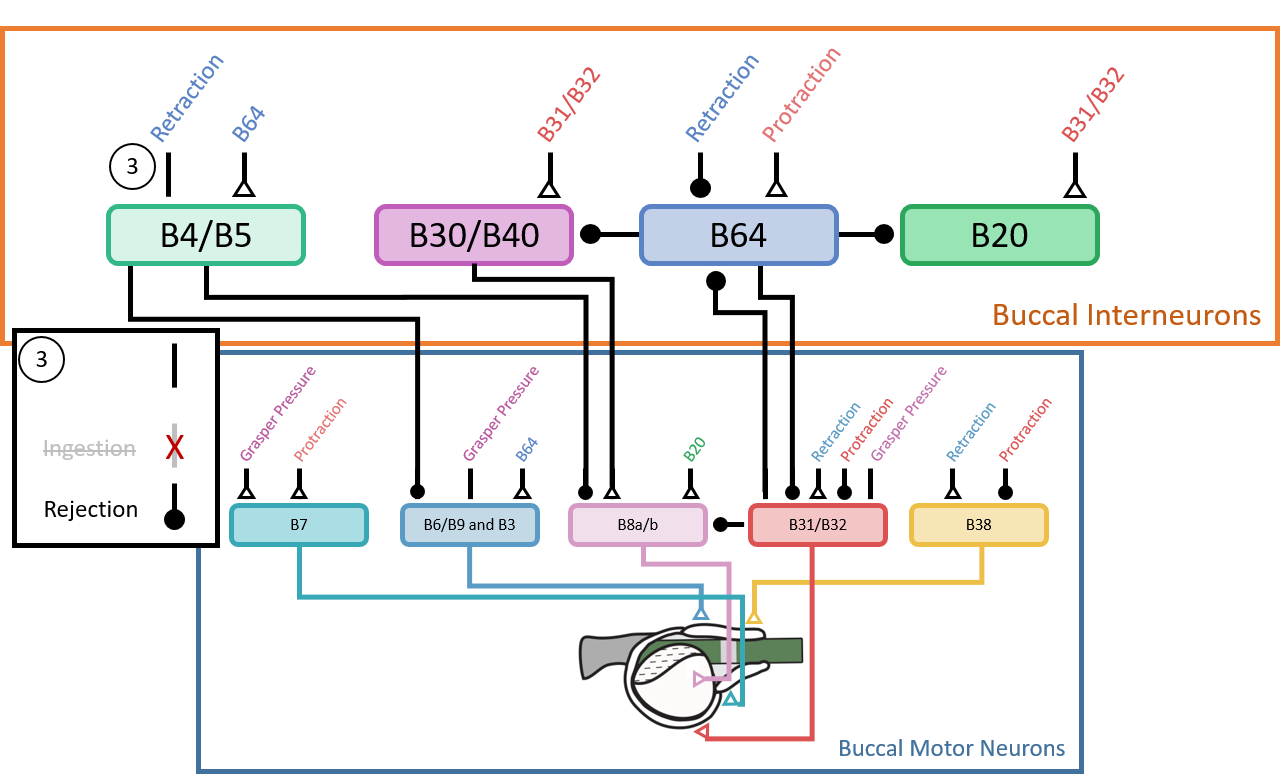}

\caption{Schematic of the local control layer added to the motor control layer shown previously. See Appendix \ref{logic} for logic formulations. The local control layer consists of known interneurons in the buccal ganglia based on the previous literature (see Tables \ref{tab:motor} and \ref{tab:interneurons}). In this diagram, the retraction-triggered proprioceptive feedback to B4/B5 varies with behavioral state as shown by the inset (3). This feedback is only present during rejection and is inhibitory during this behavior. See Appendix \ref{logic} for detailed circuit specifications.}
\label{fig:2layer}       
\end{figure*}

Building on the first layer of the model, we can add local coordination through the inclusion of known interneurons (Figure \ref{fig:2layer}). This layer coordinates the functional timing of the activity in the motor layer such that effective behaviors are generated. For the \textit{Aplysia} case study presented here, this layer includes key interneurons identified in prior literature including B30, B40, B64, and B20. In addition, we have included B4/B5 based on the existing literature documenting its importance (e.g. \cite{Gardner1977,Warman1995,Ye2006bRejections}) and our own preliminary data that it may play a role in mediating grasper release needed to switch rapidly to rejection (unpublished observations). In this model, B30 and B40 are represented as a single model neuron as they both serve primarily to provide inhibition to B8a/b during ingestive behaviors \cite{Jing2004}. The differences in how they are activated (B30 receives excitatory stimuli from CBI-4, and B40 receives excitatory input from CBI-2 \cite{Jing2004}) are included in the model neuron's logic (see Appendix \ref{logic}). The connectivity of this layer with the motor layer was established based on the previous literature. As with the motor layer, some sensory feedback pathways are proposed such that proprioceptive inputs can excite or inhibit specific interneurons based on the grasper position relative to model thresholds. 

\subsubsection{Global Coordination and Behavioral Switching}
\label{sec:boolean-network-global}

\begin{figure*}
\centering

  \includegraphics[width=\textwidth]{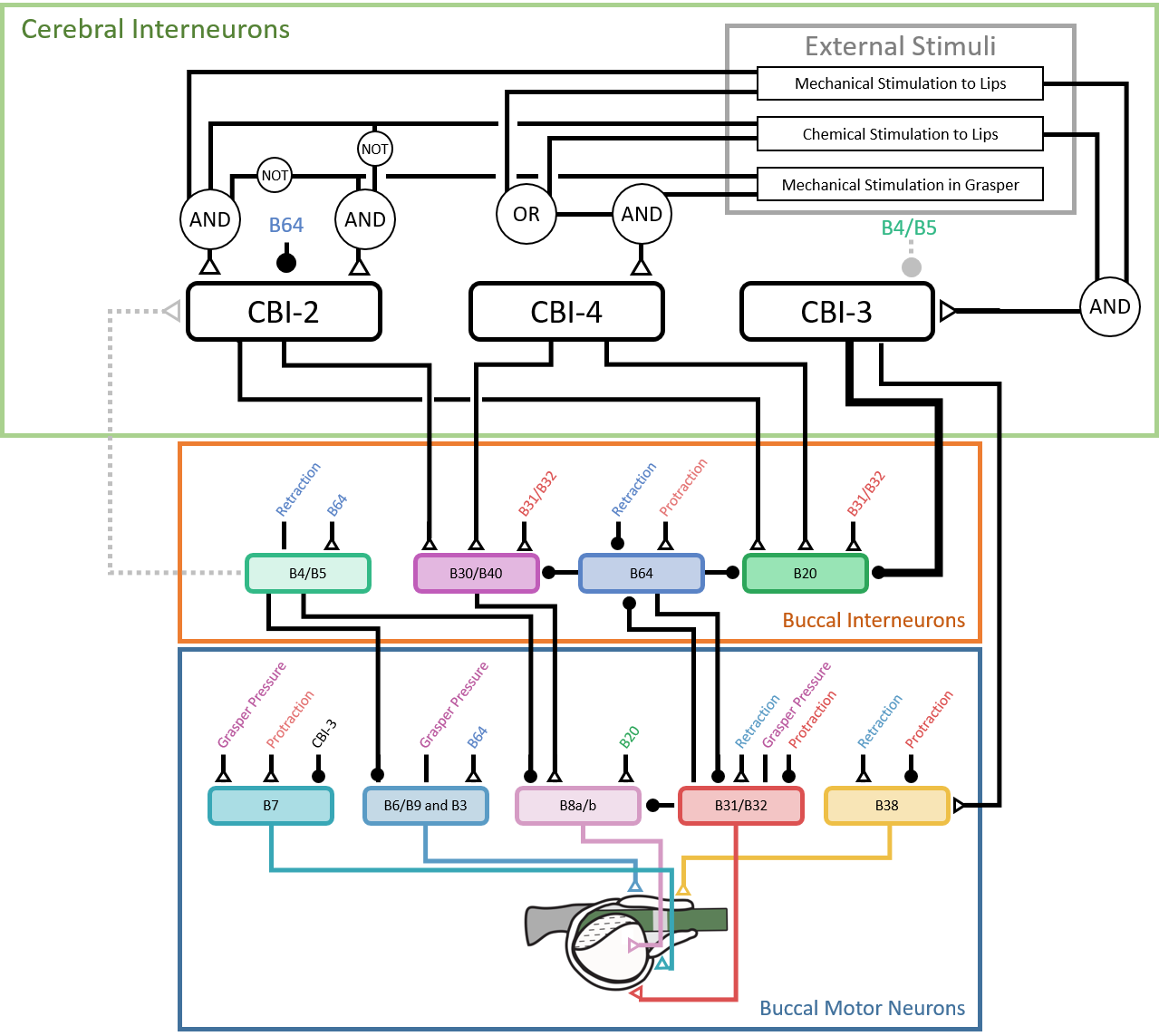}

\caption{Schematic of the full Boolean network controller including the cerebral interneurons for behavioral switching and global control. See Appendix \ref{logic} for logic formulations. External sensory cues are implemented as acting directly on relevant cerebral-buccal interneurons. However, such proprioceptive and exteroceptive feedback may be mediated through additional sensory neurons or interneurons in the actual neural circuitry of the animal. CBIs interact primarily with the local control layer to control behavioral switching. Strong inhibition from CBI-3 that overrides other inputs to B20 is shown with a bold connection. Dashed connections from B4/B5 represent hypothetical inhibition and excitation that occurs only if the presynaptic node is strongly activated (represented as a 2 in the modeling framework, rather than the Boolean 0 or 1).}
\label{fig:3layer}       
\end{figure*}

This two-layer model, when properly stimulated, can independently produce the three behaviors of interest. However, it does not allow coordinated behavioral switching based on external sensory cues. To add this capability, we add a cerebral ganglion layer, again referring to the existing literature, which responds to three external stimuli: mechanical and chemical stimulation of the lips, and mechanical stimulation in the grasper (Figure \ref{fig:3layer}). 
Cerebral-buccal interneurons 2 and 4 (CBI-2 and CBI-4) play critical roles in rejection as well as in biting and swallowing, respectively \cite{Jing2004}. The transition from egestive to ingestive behaviors appears to be handled, at least in part, by the inhibition of key buccal interneurons by CBI-3 \cite{Jing2001,Morgan2002}. Although there are other CBIs that have been shown to play some role in feeding behaviors, such as CBI-12 modulating the timing of protraction and retraction in swallowing \cite{Jing2005}, CBIs 2, 3, and 4 were selected as a minimum set to generate behavioral switching among the behaviors of interest. Using a demand-driven complexity framework, additional CBIs or CBI effects could be included in future iterations if such variations in timing were deemed necessary.

\subsection{Availability of Model Code}

The model was implemented in MATLAB, and source code is available at \url{https://github.com/CMU-BORG/Aplysia-Feeding-Boolean-Model}. Archived code is available through Zenodo (doi:10.5281/zenodo.3978414).

\section{Results}
\label{sec:results}

\subsection{Multifunctional Behavior Control in the \textit{Aplysia} Feeding Boolean Model}
\label{sec:results-behaviors}

\begin{figure*}
\centering

  \includegraphics[width=\textwidth]{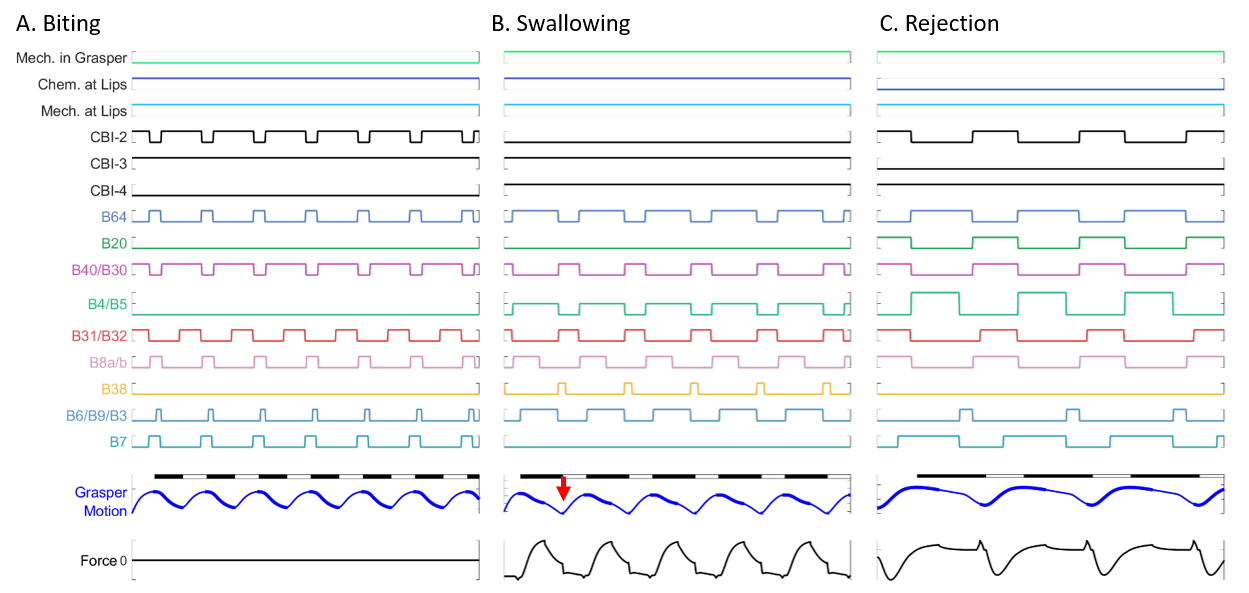}

\caption{The hybrid Boolean network model and simplified periphery is capable of producing the functional characteristics of the three targeted behaviors: biting (\textbf{A}), swallowing (\textbf{B}), and rejection (\textbf{C}). \textbf{A} In biting, chemical stimuli are present at the lips while mechanical stimuli are absent at the lips and at the grasper. This results in motion of the periphery which includes a strong protraction (open bars above grasper motion) followed by weaker retraction (closed bars), and grasper closure coincides with retraction. No force is applied to the seaweed as it is not yet grasped. Thickening of the grasper motion trace represents the position of the grasper when closing pressure would be great enough to hold an object firmly. In biting this has minimal effect as no material is present in the grasper. \textbf{B} In swallowing, both mechanical and chemical stimuli are present at the lips and mechanical stimuli are present in the grasper. Protraction of the grasper results in near zero force on the seaweed, whereas retraction of the grasper results in strong positive force on the seaweed. The arrow indicates the recoil of the grasper at the time it first releases the seaweed. Thickening of the grasper motion trace represents the position of the grasper when static friction between the grasper and object is present, indicating that the seaweed is being firmly grasped. \textbf{C} In rejection, a mechanical stimulus (inedible material) is present at both the lips and in the grasper, but chemical stimuli are absent. Grasper closure coincides with protraction. Protraction of the grasper results in increasingly negative forces (pushing the inedible material out) while retraction results in forces approaching zero. Thickening of the grasper motion trace represents the position of the grasper when static friction between the grasper and object is present, indicating that the tube is firmly grasped.}
\label{fig:ResultsAllBehaviors}       
\end{figure*}

Using the hybrid Boolean model approach, we developed a functional controller based on known neural circuitry while taking into consideration the effect of the peripheral biomechanics. Using only 20 of the possible thousands of neurons in the \textit{Aplysia} ganglia, the Boolean model is capable of producing multifunctional behaviors (Figure \ref{fig:ResultsAllBehaviors}). In the presence of mechanical and chemical stimulation at the lips, the controller generates rhythmic biting patterns characterized by a strong protraction followed by a relatively weaker retraction, with grasper closure in-phase with retraction. As no seaweed is in the grasper, no force is experienced by the seaweed (Figure \ref{fig:ResultsAllBehaviors}.A). If mechanical stimulation is applied to the grasper while mechanical and chemical stimulation is present at the lips, indicating the presence of edible material in the grasper, the model qualitatively reproduces swallowing with a weaker protraction phase followed by a strong retraction (Figure \ref{fig:ResultsAllBehaviors}.B). This results in high positive (ingestive) force being applied to the seaweed during the retraction phase. In contrast, the presence of mechanical stimulation at the lips and in the grasper without chemical stimulation at the lips indicates the presence of inedible material in the grasper. Under such conditions, the model successfully generates rejection-like behaviors (Figure \ref{fig:ResultsAllBehaviors}.C). The inedible material is grasped during the protraction phase, resulting in a negative (egestive) force being applied to it during protraction, pushing it out of the buccal cavity.

\begin{figure*}
\centering

  \includegraphics[width=0.6\textwidth]{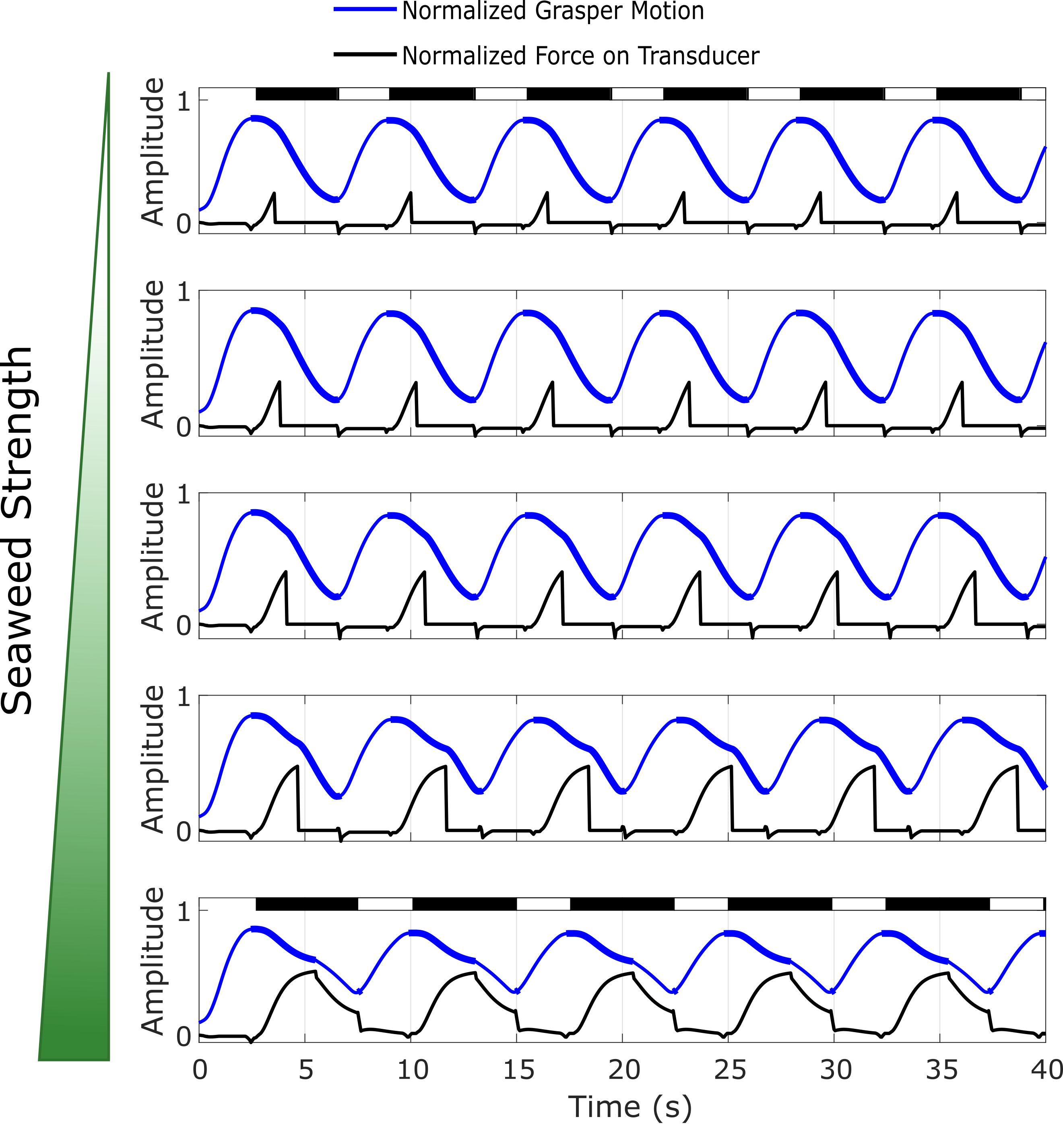}

\caption{Characteristic examples of grasper motion and measured force on the force transducer during swallowing with varying seaweed strength thresholds, $z_S$. Thickening of the grasper motion trace represents the position of the grasper when closing pressure would be great enough to hold an object firmly. In the top and bottom panels, protraction is indicated by the open bars above grasper motion and retraction by closed bars. Seaweed thresholds increase from $z_S = 0.1$ (top), where seaweed breaks early in retraction and force on the transducer quickly drops to zero, to $z_S = 0.5$ (bottom), at which point the seaweed does not break during the swallow attempt. The cycle period of swallows increases with increasing seaweed strength.}
\label{fig:SeaweedStrength}       
\end{figure*}

In addition to being multifunctional, the Boolean model framework exhibits robustness within a single behavior. During \textit{Aplysia} feeding, robustness is observed when animals attempt to feed on seaweeds of varying mechanical strength or that are attached to the substrate by a holdfast. Increasing mechanical load increases the duration of swallows overall and the retraction phase in particular \cite{hurwitz1992adaptation,Shaw2015,GillChiel2020}. The Boolean model presented here reproduces this phenomenon even though the behavior has not been explicitly programmed. The adjustment to seaweed strength is instead an emergent property of the control network and biomechanics. By including a force threshold in our biomechanical model, we can vary the force at which the seaweed ``breaks'', thereby allowing the grasper to move again as it is no longer anchored to the rigid force transducer (Figure \ref{fig:SchematicIntro}). Increasing the strength of the seaweed by increasing the value of this threshold results in a longer period for each swallow due to the increased retraction duration (Figure \ref{fig:SeaweedStrength}), as is observed in behaving animals \cite{GillChiel2020}.

\subsection{Behavioral Switching Based on Sensory Cues}
\label{sec:results-switching}

Truly multifunctional controllers need to be able to appropriately switch between behaviors. In addition to being able to reproduce distinct behavior through coordinated variation of motor neuron activation, the model can also switch between behaviors in response to changing sensory inputs. 

\begin{figure*}
\centering

  \includegraphics[width=\textwidth]{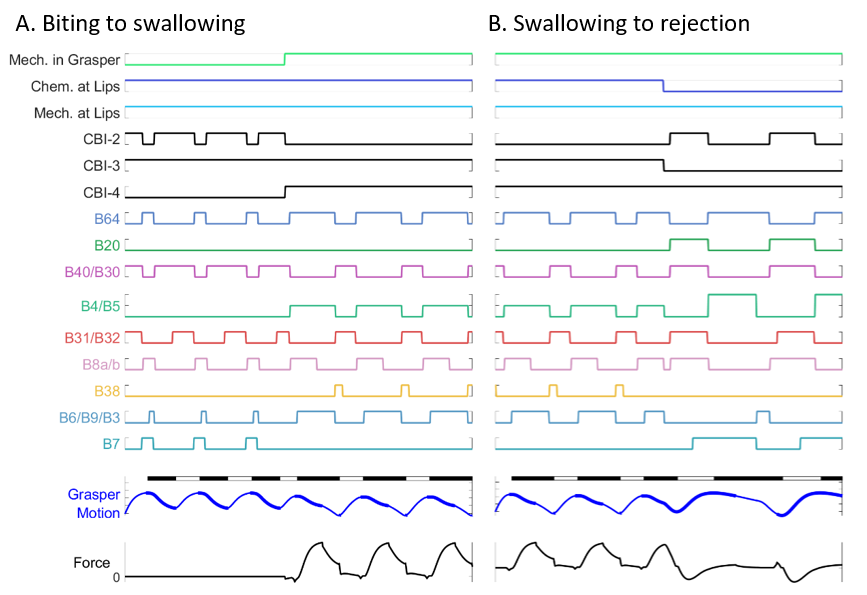}

\caption{By changing the combination of external stimuli, the hybrid Boolean network controller can appropriately switch between behaviors. \textbf{A} When mechanical and chemical stimuli are present at the lips, but no mechanical stimuli are in the grasper, the network produces biting-like behavior. A step change, halfway through the simulation, in the mechanical stimuli in the grasper representing a successful grasp attempt switches the network to producing swallow-like behavior. \textbf{B} When mechanical stimuli are present at the lips and in the grasper, and chemical stimuli are present at the lips, the model produces swallowing-like behavior. Loss of chemical stimuli at the lips halfway through the simulation triggers the model to initiate rejection-like behavior.}
\label{fig:ResultsBehaviorSwitching}       
\end{figure*}

In the animal, a change from biting to swallowing motor patterns is observed when seaweed is present at the lips and the grasper successfully grabs the seaweed, so the grasper now senses a mechanical stimulus. We assessed the controller's ability to reproduce this transition by applying a step change to the mechanical stimulus in the grasper near the peak of protraction during the biting cycle (Figure \ref{fig:ResultsBehaviorSwitching}.A). As a result, the model successfully transitioned from biting-like to swallowing-like neural and behavioral patterns.

Similarly, a transition from swallowing to rejection is observed when inedible food is detected in the grasper. This transition can be seen in the animal by inducing it to bite and swallow a polyethylene tube while simultaneously touching the lips with food, and removing the food stimulus after some length of tubing has been ingested. This behavioral transition is observed in the model when chemical stimuli are removed during swallowing. This results in a sensory state in which only mechanical stimuli are present both at the lips and in the grasper. As a consequence, the model successfully transitions from swallowing to rejection (Figure \ref{fig:ResultsBehaviorSwitching}.B). A sudden drop in force on the seaweed is observed as the grasper briefly releases the seaweed and transitions to grasping the seaweed during protraction in order to push the seaweed out of the feeding apparatus. \red{As implemented in the model, changes in activity due to sensory stimuli happen instantaneously. The impact of transition dynamics could be investigated in future model iterations through the inclusion of temporal dynamics in modeling the CBIs and sensory feedback.} 

\subsection{Using the Model to Propose Testable Hypotheses}
\label{sec:hypotheses}

The modeling framework provides a significant advantage over population-based neural control schemes such as machine learning because the network is explainable and grounded in an animal's neurobiology. As a consequence, the model is a tool not only for robotic control, but also for generating and testing potential neurobiological hypotheses. To mimic electrophysiology experiments, ``electrodes'' can be added to the Boolean logic statements for a given model neuron as excitatory or inhibitory inputs, and the Boolean architecture can be extended to include a strongly excited state wherein activity is set to 2 rather than 1, as was implemented for B4/B5. 

One such testable hypothesis is the role of the B4/B5 multi-action neurons in behavioral switching. Gardner has previously shown that these multi-action neurons have widespread outputs to many neurons within the buccal ganglia \cite{Gardner1977}. Furthermore, B4/B5 have been observed to be intensely activated during rejection and less so in biting and swallowing \cite{Warman1995}, an observation that is also seen in the animal data shown above (Figure \ref{fig:translation}). B4/B5 have also been observed to fire strongly in response to sudden increases in load on seaweed during swallowing \cite{GillSfN2018Poster}. The intense firing in B4/B5 may be critical for delaying the onset of activity in the jaw muscles during rejection. As the grasper protracts closed and retracts open during rejection, it passes through the lumen of the jaws as it rejects the inedible material. If the jaws closed prematurely, the grasper could be forced shut and food could be pulled back into the buccal cavity \cite{Ye2006bRejections}. These observations led us to hypothesize that strong activation of B4/B5 could be used to trigger transient rejection behavior. 

\begin{figure}

  \includegraphics[width=0.6\textwidth]{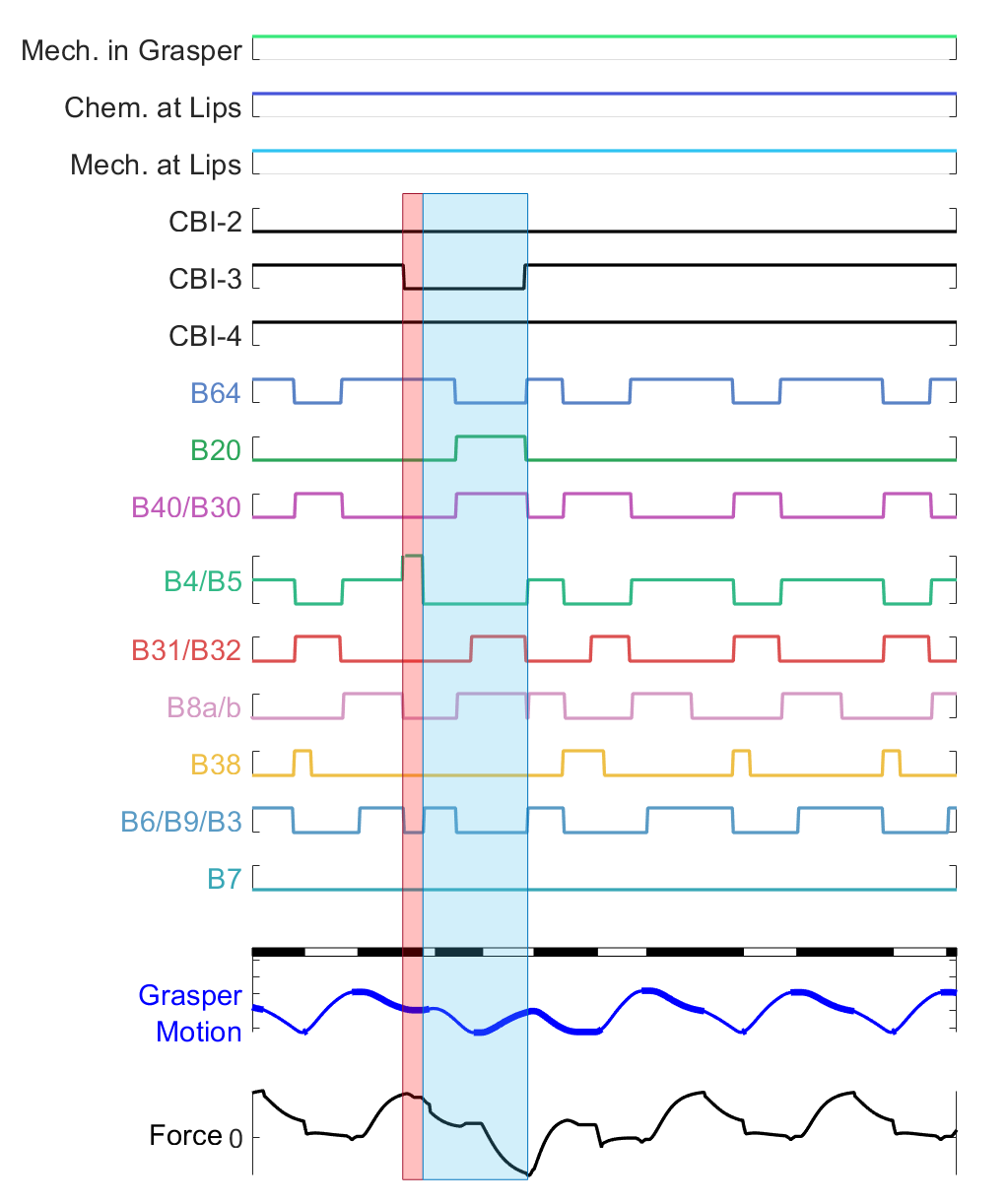}

\caption{In a hypothetical experiment in which postulated network connections are added, strong B4/B5 stimulation leads to transient egestive behavior. The modeling framework allows the network to be easily modified to accommodate the addition of ``electrodes'' to stimulate individual neurons, as well as individual timing properties, such as refractory periods. Here, strongly stimulating B4/B5 (red rectangular overlay) with the postulated connections for the B4/B5 neuron shown as dashed lines in Figure \ref{fig:3layer}, and with a refractory period affecting CBI-3, results in temporarily switching from swallowing-like to rejection-like behavior (blue rectangular overlay).}
\label{fig:ResultsB4B5}       
\end{figure}

To test this hypothesis, we have postulated connections from B4/B5 to CBI-2 (excitatory) and CBI-3 (inhibitory) and added them to the Boolean model, as well as an electrode to strongly excite B4/B5 transiently. In the actual animal, the postulated connections may be indirect. Additionally, the model makes it possible to easily include a refractory period associated with a connection. We have included one such refractory period for CBI-3, which again may be indirect, during which it remains inhibited after strong inhibition from B4/B5. To test the hypothesis that B4/B5 stimulation can temporarily switch behavior from ingestion to rejection in the model, we strongly excited B4/B5 as the grasper approached the peak of retraction. This strong excitation resulted in inhibition of B8a/b and therefore the pressure on the seaweed was released, causing an abrupt drop in force. The model then transitioned to rejection-like behavior for the duration of the CBI-3 refractory period, after which the model returned to swallowing-like behavior (Figure \ref{fig:ResultsB4B5}). In the absence of these postulated connections, the model does not transition to rejection-like behavior when B4/B5 is strongly excited (data not shown). The model thus makes specific testable predictions.

\section{Discussion}
\label{sec:discussion}

The hybrid Boolean model framework applied to \textit{Aplysia} feeding results in a multilayer controller based on the known neural circuitry and peripheral biomechanics of the animal. This model is capable of reproducing three key behaviors observed in feeding (Figure \ref{fig:ResultsAllBehaviors}), reproduces robustness within a behavior by adjusting to varying mechanical load during swallowing (Figure \ref{fig:SeaweedStrength}), captures the ability to switch between behaviors in response to sensory cues (Figure \ref{fig:ResultsBehaviorSwitching}), and provides a straightforward means of using the model to suggest testable hypotheses about circuit function (Figure \ref{fig:ResultsB4B5}). The model is easily extensible as additional neural units or sensory feedback pathways are identified or if additional features of the biomechanics need to be captured.

\subsection{Limitations}

There is a long-standing debate in the neurobiology community on the relative roles of central pattern generator-like circuitry, where a rhythmic pattern can be generated in the absence of sensory feedback, versus chain reflexes, where each phase of the pattern initiates sensory feedback critical for generating the next part of the behavior \cite{kuo2002relative}.  It is likely that both modalities contribute to behavior. For example, central pattern generators are heavily influenced by sensory feedback \cite{pearson1993common} and chain reflexes have central components \cite{Buschges2008,BaSSLER125,Pearson1987}. Since our model focuses on behavioral forces and movements, we constructed it to depend heavily on sensory feedback and less on the intrinsic internal dynamics of the neural circuit. As a consequence, removing all sensory inputs will stop the model from oscillating. This is an application of demand-driven complexity to reduce computational cost by focusing on a reduced set of internal connections, cells, and dynamics sufficient to qualitatively reproduce multifunctionality. Such intrinsic mechanisms, however, could be added to the model in future iterations. A testable hypothesis from these observations would be that in the intact behaving animal, sensory feedback may play as significant a role as intrinsic mechanisms during feeding. \red{This hypothesis is supported by recent related work highlighting the importance of sensory feedback in shaping the functional motor outputs in central pattern generator circuits underlying insect locomotion \cite{MantziarisBockemuehlBueschges2020DevNeuro-review-CPG-insect}. In \textit{Aplysia},} this question could be investigated in intact animal models as well as in suspended buccal mass preparations \cite{mcmanus2012vitro} to determine the relative importance of these factors.

Though the model provides an accessible framework for capturing known networks for multifunctional control, it still has many parameters that must be set. The model architecture can easily be implemented based on known circuit connectivity. Thresholds, time constants, and maximum forces can be approximated through measuring the relevant strength of synaptic inputs and relevant force and movement outputs.  However, if it is important for a modeling application to capture the detailed time series of neural and muscle activity, such as individual spikes or bursting, more details may be required (see Section \ref{sec:prior-models}). On the other hand, if the focus is on overall behavior and the system has slow muscles, fast details may not be as important for obtaining appropriate behavioral outputs. In future work, parameters could be found using automated approaches such as optimization or machine learning.

Additionally, the model implementation does not capture the cycle-to-cycle variability observed in actual animal behavior \cite{Cullins2015b}. Although deterministic control is useful in many robotic applications, variability plays a critical role in behavioral flexibility. Variability is observed not only between individuals, but also within a given individual as it repeats a behavior. Indeed, variability in biological control contributes to the overall success of behaviors and species \cite{Cullins2015b,Marder2011}. For example, by using different techniques to pull on seaweed, the animal may be able to effectively fatigue the material and cause it to break \cite{stewart2004consequences,koehl2006wave}. As a consequence, animals may vary a behavior even if the mechanical load is identical, which is not yet captured by our current model. \red{The importance of variability in cyclic neural activity is certainly not limited to \textit{Aplysia}. For example, highly variable motor patterns and muscle outputs have been observed in the stick insect, \textit{Carausis morosus}, which may help the animal locomote in complex environments or avoid predation \cite{Hooper2016,Hooper2007}. Variation within the system may also stem from tuning of temporal dynamic time constants, which impact how the muscles respond to variable motor-neuronal inputs \cite{Hooper2007b}.} Although many robots can be programmed to handle a variety of situations, fewer robots are capable of autonomous multifunctional behaviors \cite{Royakkers2015}. A biologically-inspired approach may allow variability to be effectively harnessed to improve autonomous robot performance. Moreover, harnessing variability might allow closed loop controllers of the nervous system to be fit to individual animals.  On the other hand, some variability observed in animal behavior may not be true stochasticity, but rather results from changing internal states due to neuromodulation (e.g. \cite{Cullins2015}). As the neural locus of such internal states and effects is identified, these variables could be included in our modeling framework in the future to better capture these effects. The model could be extended to include variability through the inclusion of stochastic processes for activity switching in each node. Furthermore, the faster-than-real-time speed (2-3 orders of magnitude on standard CPU hardware) of the model allows many instantiations with small parameter variations to be run in parallel, thereby capturing the variability observed both within a given behavior and between individual animals, or for finding optimal solutions. 

\red{Although we have demonstrated faster-than-real-time simulations for the \textit{Aplysia} feeding test case presented here, as more complex networks are implemented in this framework, the computational cost of modeling the network will increase. The computational cost for additional Boolean neurons will scale roughly linearly with the number of neurons. While this poses little problem for the \text{Aplysia} circuitry, where we envision doubling or even tripling the number of neurons in future models, when modeling mammalian systems with orders of magnitude more neurons, faster-than-real-time simulations may not be achievable without simplifying the model neurons to groups, or implementing the model using hardware logic.}

A final limitation of the current model is that, due to the Boolean, \red{instantaneous} nature of our external sensory cues, intermediate behaviors \cite{Morton1993}, such as repeatedly moving the seaweed back and forth, are not captured. \red{Additionally, the model does not include intermediate transition dynamics when switching behaviors, but these dynamics could easily be added.}  Previous literature has demonstrated the importance of such intermediate behaviors. For example, before rejections, the animal may attempt to reposition food and retry swallowing \cite{Morton1993,Katzoff2006}. This ability to selectively reposition, reject, and swallow along a continuum of behaviors is important for feeding efficiency \cite{Katzoff2006}. Although the  current model does not capture these behaviors, modifications to the interneuronal circuitry motivated by experimental findings and by hypothesized connections may better capture these behaviors. 
\red{Furthermore, this modeling framework can generate testable hypotheses for the mechanisms underlying behavioral switching, which can be investigated in simulation. Such switching mechanisms are important not only in \textit{Aplysia}, but also in locomotion when switching from forward to backwards walking \cite{Feng2020}, switching from standing to walking \cite{Bidaye2020}, and transitioning through speed-dependent gaits \cite{Danner2016}.}

\subsection{Conclusions and Future Directions}

The hybrid Boolean network control framework presented here leads to a bioinspired, computationally efficient controller capable of producing key multifunctional behaviors observed in the animal. Its computational efficiency stems, in part, from using a demand-driven complexity approach which minimizes the number of neurons and connections used to reproduce the desired behavior. This framework and the use of the semi-implicit integration scheme in Appendix \ref{integration} allow new nodes and connections to be added with relative ease through modification of the Boolean logic statements. As a consequence, additional neurons, temporal effects, connections, and sensory pathways can easily be added based on the existing literature \cite{Cropper2019} and future experimental results. Although additional neurons and connections were not necessary to produce multifunctionality,  including them may improve controller robustness. Our modeling framework will make it possible to clarify how sensory feedback affects behavior as additional connections are added. Moreover, our simple biomechanical model could readily be incorporated into a much more realistic neural circuit model \cite{Costa2020,Cataldo2006} to assess the role of sensory feedback on more detailed neural mechanisms, even though this may reduce computational efficiency.

More complex biomechanical models can also be interfaced with the hybrid Boolean controller to capture \textit{morphological intelligence}, i.e., how the structural biomechanics itself contributes to the control of the system. In the \textit{Aplysia} feeding system, morphological intelligence is demonstrated by the effect of changes in the grasper shape on the mechanical advantage of the I2 and I3 muscles \cite{Novakovic2006}. The phase in which shape changes occur can either increase the mechanical advantage of a muscle, allowing it to generate higher forces, or diminish the muscles effectiveness, creating regions where the timing of the control signal is less critical. \red{More realistic muscle models could further improve accuracy, but this will have to be balanced against the potential computational cost.} The simplicity of the modeling framework allows morphology and more detailed biomechanical models \cite{Neustadter2002a,Novakovic2006,Sutton2004b} to be integrated in future iterations. 

Although we have applied the model framework to \textit{Aplysia} feeding, the framework can be extended to many other multifunctional systems. In \textit{Aplysia} feeding, differences between the key behaviors are largely the result of shifts in phasing between muscle activity in the grasper relative to protraction and retraction (Figure \ref{fig:translation}). Similarly, in locomotion, changes in relative timing of swing and stance are observed as animals transition from walking to running \cite{Mann1980,Cappellini2006,Li1999}. Another multifunctional behavior observed in legged systems, hopping, uses the same periphery as walking and running but synchronizes the phase of muscle activity between legs \cite{Verstappen2000,MORITANI199134}. The framework can be adapted to such multifunctional behaviors through application of the multilayered controller design. Similar approaches have been previously reported in mammalian neural circuit controllers using more physiological neuron models \cite{molkov2015mechanisms,markin2016neuromechanical,ivashko2003modeling}. 

The modeling framework has several advantages over other control and modeling approaches. Similar to the discrete event-based neural network model recently reported by Bazenkov \textit{et al.} \cite{Bazenkov2020}, this modeling approach allows rapid simulation of multifunctional behavior. However, unlike such prior discrete models, the Boolean model framework presented here includes the known neural circuitry and simplified biomechanics of the periphery. The direct relationship to the underlying circuitry makes it possible to both generate and test specific neurobiological hypotheses; at the same time, the relative simplicity of the network makes it attractive as a basis for robot control. Furthermore, unlike current artificial neural network architectures, synthetic nervous systems including the hybrid Boolean model are explainable: the structure of the networks directly informs the functional outputs of the systems. Although the connections and trained weights of artificial neural networks may provide similar control capabilities, these networks must be trained on large datasets. Part of the strength of synthetic nervous systems is that they use a basis set of dynamics derived from biological neurons and thus can generate robust control even without additional training \cite{Szczecinski2015,Szczecinski2017,Szczecinski2017a,Hunt2015,Hunt2017}.

\begin{acknowledgements}
HJC and JPG were supported by NSF grant IOS1754869. VWW was supported by startup funding from the Carnegie Mellon University Department of Mechanical Engineering. VWW, HJC, and JPG were supported by NSF grant DBI2015317. HJC and PJT were supported by NIH grant R01 NS118606. We thank the anonymous reviewers for helpful comments on an earlier version of this manuscript.
\end{acknowledgements}

%
\section*{Conflict of interest}
The authors declare that they have no conflict of interest.

\bibliographystyle{spmpsci}      

\begin{thebibliography}{100}
\providecommand{\url}[1]{{#1}}
\providecommand{\urlprefix}{URL }
\expandafter\ifx\csname urlstyle\endcsname\relax
  \providecommand{\doi}[1]{DOI~\discretionary{}{}{}#1}\else
  \providecommand{\doi}{DOI~\discretionary{}{}{}\begingroup
  \urlstyle{rm}\Url}\fi

\bibitem{Aggarwal2019}
Aggarwal, S., Chugh, N.: Signal processing techniques for motor imagery brain
  computer interface: A review.
\newblock Array \textbf{1-2}(January), 100003 (2019).
\newblock \doi{10.1016/j.array.2019.100003}.
\newblock \urlprefix\url{https://doi.org/10.1016/j.array.2019.100003}

\bibitem{Ayers1995}
Ayers, J.: A reactive ambulatory robot architecture for operation in current
  and surge.
\newblock In: Autonomous Vehicles in Mine Countermeasures Symposium, April
  1995, pp. 1--14 (1995).
\newblock \urlprefix\url{http://www.neurotechnology.neu.edu/nps95mcm
  manuscript.html}

\bibitem{Ayers2020}
Ayers, J.: A conservative biomimetic control architecture for autonomous
  underwater robots.
\newblock Neurotechnology for Biomimetic Robots pp. 1--14 (2002).
\newblock \doi{10.7551/mitpress/4962.003.0019}

\bibitem{Ayers1977}
Ayers, J.L., Davis, W.J.: Neuronal control of locomotion in the lobster
  \textit{Homarus americanus}.
\newblock Journal of Comparative Physiology A \textbf{115}(1), 29--46 (1977).
\newblock \doi{10.1007/bf00667783}

\bibitem{BaSSLER125}
B{\"a}ssler, U.: Functional principles of pattern generation for walking
  movements of stick insect forelegs: The role of the femoral chordotonal organ
  afferences.
\newblock Journal of Experimental Biology \textbf{136}(1), 125--147 (1988).
\newblock \urlprefix\url{https://jeb.biologists.org/content/136/1/125}

\bibitem{Bazenkov2020}
Bazenkov, N.I., Boldyshev, B.A., Dyakonova, V., Kuznetsov, O.P.: Simulating
  small neural circuits with a discrete computational model.
\newblock Biological Cybernetics  (2020).
\newblock \doi{10.1007/s00422-020-00826-w}.
\newblock \urlprefix\url{https://doi.org/10.1007/s00422-020-00826-w}

\bibitem{Beck2007}
Beck, J.M., Pouget, A.: Exact inferences in a neural implementation of a hidden
  {Markov} model.
\newblock Neural Computation \textbf{19}(5), 1344--1361 (2007).
\newblock \doi{10.1162/neco.2007.19.5.1344}

\bibitem{Beer1999}
Beer, R.D., Chiel, H.J., Gallagher, J.C.: Evolution and analysis of model
  {CPG}s for walking: {II.} {General} principles and individual variability.
\newblock Journal of Computational Neuroscience \textbf{7}(2), 119--147 (1999).
\newblock \doi{10.1023/A:1008920021246}.
\newblock \urlprefix\url{https://doi.org/10.1023/A:1008920021246}

\bibitem{Beer1992b}
Beer, R.D., Chiel, H.J., Quinn, R.D., Espenschied, K.S.: A distributed neural
  network architecture for hexapod robot locomotion.
\newblock Neural Computation \textbf{4}(3), 356--365 (1992)

\bibitem{BEER1990169}
Beer, R.D., Chiel, H.J., Sterling, L.S.: A biological perspective on autonomous
  agent design.
\newblock Robotics and Autonomous Systems \textbf{6}(1), 169 -- 186 (1990).
\newblock \doi{https://doi.org/10.1016/S0921-8890(05)80034-X}.
\newblock
  \urlprefix\url{http://www.sciencedirect.com/science/article/pii/S092188900580034X}.
\newblock Designing Autonomous Agents

\bibitem{Bicanski2013}
Bicanski, A., Ryczko, D., Knuesel, J., Harischandra, N., Charrier, V., Ekeberg,
  {\"{O}}., Cabelguen, J.M., Ijspeert, A.J.: Decoding the mechanisms of gait
  generation in salamanders by combining neurobiology, modeling and robotics.
\newblock Biological Cybernetics \textbf{107}(5), 545--564 (2013).
\newblock \doi{10.1007/s00422-012-0543-1}

\bibitem{Bidaye2020}
Bidaye, S.S., Laturney, M., Chang, A.K., Liu, Y., Bockemühl, T., Büschges,
  A., Scott, K.: Two brain pathways initiate distinct forward walking programs
  in drosophila.
\newblock Neuron  (2020).
\newblock \doi{10.1016/j.neuron.2020.07.032}.
\newblock \urlprefix\url{https://doi.org/10.1016/j.neuron.2020.07.032}

\bibitem{BluemelEtAlBueschges2012a-Hill}
Bl{\"u}mel, M., Guschlbauer, C., Daun-Gruhn, S., Hooper, S.L., B{\"u}schges,
  A.: {Hill}-type muscle model parameters determined from experiments on single
  muscles show large animal-to-animal variation.
\newblock Biological cybernetics \textbf{106}(10), 559--571 (2012)

\bibitem{BluemelEtAlBueschges2012b-Using}
Bl{\"u}mel, M., Guschlbauer, C., Hooper, S.L., B{\"u}schges, A.: Using
  individual-muscle specific instead of across-muscle mean data halves muscle
  simulation error.
\newblock Biological Cybernetics \textbf{106}(10), 573--585 (2012)

\bibitem{BluemelEtAlBueschges2012c-Determining}
Bl{\"u}mel, M., Hooper, S.L., Guschlbauerc, C., White, W.E., B{\"u}schges, A.:
  Determining all parameters necessary to build {Hill}-type muscle models from
  experiments on single muscles.
\newblock Biological cybernetics \textbf{106}(10), 543--558 (2012)

\bibitem{Brosch2014}
Brosch, T., Neumann, H.: Interaction of feedforward and feedback streams in
  visual cortex in a firing-rate model of columnar computations.
\newblock Neural Networks \textbf{54}, 11--16 (2014).
\newblock \doi{10.1016/j.neunet.2014.02.005}.
\newblock \urlprefix\url{http://dx.doi.org/10.1016/j.neunet.2014.02.005}

\bibitem{Brown2018}
Brown, J.W., Caetano-Anoll{\'{e}}s, D., Catanho, M., Gribkova, E., Ryckman, N.,
  Tian, K., Voloshin, M., Gillette, R.: Implementing goal-directed foraging
  decisions of a simpler nervous system in simulation.
\newblock eNeuro \textbf{5}(1), 1--10 (2018).
\newblock \doi{10.1523/ENEURO.0400-17.2018}

\bibitem{Buschges2005}
B{\"{u}}schges, A.: Sensory control and organization of neural networks
  mediating coordination of multisegmental organs for locomotion.
\newblock Journal of Neurophysiology \textbf{93}(3), 1127--1135 (2005).
\newblock \doi{10.1152/jn.00615.2004}

\bibitem{Buschges2008}
B{\"{u}}schges, A., Akay, T., Gabriel, J.P., Schmidt, J.: Organizing network
  action for locomotion: insights from studying insect walking.
\newblock Brain Research Reviews \textbf{57}(1), 162--171 (2008).
\newblock \urlprefix\url{http://www.ncbi.nlm.nih.gov/pubmed/17888515}

\bibitem{Cappellini2006}
Cappellini, G., Ivanenko, Y.P., Poppele, R.E., Lacquaniti, F.: Motor patterns
  in human walking and running.
\newblock Journal of Neurophysiology \textbf{95}(6), 3426--3437 (2006).
\newblock \doi{10.1152/jn.00081.2006}

\bibitem{Cash1989}
Cash, D., Carew, T.J.: A quantitative analysis of the development of the
  central nervous system in juvenile aplysia californica.
\newblock Journal of Neurobiology \textbf{20}(1), 25--47 (1989).
\newblock \doi{10.1002/neu.480200104}.
\newblock
  \urlprefix\url{https://onlinelibrary.wiley.com/doi/abs/10.1002/neu.480200104}

\bibitem{Cataldo2006}
Cataldo, E., Byrne, J.H., Baxter, D.A.: Computational model of a central
  pattern generator.
\newblock {Lecture Notes in Computer Science (including subseries Lecture Notes
  in Artificial Intelligence and Lecture Notes in Bioinformatics)} \textbf{4210
  LNBI}, 242--256 (2006).
\newblock \doi{10.1007/11885191_17}

\bibitem{Chiel1997}
Chiel, H.J., Beer, R.D.: The brain has a body: Adaptive behavior emerges from
  interactions of nervous system, body and environment.
\newblock Trends in Neurosciences \textbf{20}(12), 553--557 (1997).
\newblock \doi{10.1016/S0166-2236(97)01149-1}

\bibitem{Beer1999a}
Chiel, H.J., Beer, R.D., Gallagher, J.C.: {Evolution and analysis of model
  {CPG}s for walking: {I.} {Dynamical} Modules}.
\newblock Journal of Computational Neuroscience \textbf{7}(2), 99--118 (1999).
\newblock \doi{10.1023/A:1008920021246}

\bibitem{Chiel1992}
Chiel, H.J., Crago, P., Mansour, J.M., Hathi, K.: Biomechanics of a muscular
  hydrostat: a model of lapping by a reptilian tongue.
\newblock Biological Cybernetics \textbf{67}(5), 403--415 (1992).
\newblock \doi{10.1007/BF00200984}

\bibitem{Chiel1988}
Chiel, H.J., Kupfermann, I., Weiss, K.R.: An identified histaminergic neuron
  can modulate the outputs of buccal-cerebral interneurons in \textit{Aplysia}
  via presynaptic inhibition.
\newblock The Journal of Neuroscience \textbf{8}(January), 49--63 (1988)

\bibitem{Chiel2009}
Chiel, H.J., Ting, L.H., Ekeberg, {\"{O}}., Hartmann, M.J.: The brain in its
  body: Motor control and sensing in a biomechanical context.
\newblock Journal of Neuroscience \textbf{29}(41), 12807--12814 (2009).
\newblock \doi{10.1523/JNEUROSCI.3338-09.2009}

\bibitem{Chiel1986}
Chiel, H.J., Weiss, K.R., Kupfermann, I.: An identified histaminergic neuron
  modulates feeding motor circuitry in \textit{Aplysia}.
\newblock Journal of Neuroscience \textbf{6}(8), 2427--2450 (1986).
\newblock \doi{10.1523/jneurosci.06-08-02427.1986}.
\newblock \urlprefix\url{http://www.ncbi.nlm.nih.gov/pubmed/3746416}

\bibitem{Church1994}
Church, P.J., Lloyd, P.E.: Activity of multiple identified motor neurons
  recorded intracellularly during evoked feeding-like motor programs in
  \textit{Aplysia}.
\newblock {Journal of Neurophysiology} \textbf{72}(4), 1794--1809 (1994).
\newblock \doi{10.1152/jn.1994.72.4.1794}

\bibitem{Church1993}
Church, P.J., Whim, M.D., Lloyd, P.E.: Modulation of neuromuscular transmission
  by conventional and peptide transmitters released from excitatory and
  inhibitory motor neurons in \textit{Aplysia}.
\newblock Journal of Neuroscience \textbf{13}(7), 2790--2800 (1993).
\newblock \doi{10.1523/jneurosci.13-07-02790.1993}

\bibitem{Connor1986}
Connor, J.A., Kretz, R., Shapiro, E.: Calcium levels measured in a presynaptic
  neurone of \textit{Aplysia} under conditions that modulate transmitter
  release.
\newblock The Journal of Physiology \textbf{375}(1), 625--642 (1986).
\newblock \doi{10.1113/jphysiol.1986.sp016137}.
\newblock
  \urlprefix\url{https://physoc.onlinelibrary.wiley.com/doi/abs/10.1113/jphysiol.1986.sp016137}

\bibitem{Costa2020}
Costa, R.M., Baxter, D.A., Byrne, J.H.: Computational model of the distributed
  representation of operant reward memory : combinatoric engagement of
  intrinsic and synaptic plasticity mechanisms.
\newblock Learning {\&} Memory \textbf{27}, 236--249 (2020).
\newblock \doi{10.1101/lm.051367.120}

\bibitem{Cropper2019}
Cropper, E.C., Jing, J., Weiss, K.R.: The feeding network of \textit{Aplysia}.
\newblock In: The Oxford Handbook of Invertebrate Neurobiology, December, pp.
  400--422. Oxford University Press (2019).
\newblock \doi{10.1093/oxfordhb/9780190456757.013.19}

\bibitem{CullinsChielJoVE2010}
Cullins, M.J., Chiel, H.J.: Electrode fabrication and implantation in
  \textit{{Aplysia} californica} for multi-channel neural and muscular
  recordings in intact, freely behaving animals.
\newblock Journal of Visualized Experiments (40), e1791 (2010).
\newblock \doi{10.3791/1791}.
\newblock
  \urlprefix\url{http://www.jove.com/video/1791/electrode-fabrication-implantation-aplysia-californica-for-multi}

\bibitem{Cullins2015b}
Cullins, M.J., Gill, J.P., McManus, J.M., Lu, H., Shaw, K.M., Chiel, H.J.:
  Sensory feedback reduces individuality by increasing variability within
  subjects.
\newblock Current Biology \textbf{25}(20), 2672--2676 (2015).
\newblock \doi{10.1016/j.cub.2015.08.044}.
\newblock \urlprefix\url{http://dx.doi.org/10.1016/j.cub.2015.08.044}

\bibitem{Cullins2015}
Cullins, M.J., Shaw, K.M., Gill, J.P., Chiel, H.J.: Motor neuronal activity
  varies least among individuals when it matters most for behavior.
\newblock Journal of Neurophysiology \textbf{113}(3), 981--1000 (2015).
\newblock \doi{10.1152/jn.00729.2014}.
\newblock
  \urlprefix\url{http://jn.physiology.org/lookup/doi/10.1152/jn.00729.2014}

\bibitem{Dallidis2014}
Dallidis, S.E., Karafyllidis, I.G.: Boolean network model of the
  \textit{Pseudomonas aeruginosa} quorum sensing circuits.
\newblock IEEE Transactions on Nanobioscience \textbf{13}(3), 343--349 (2014).
\newblock \doi{10.1109/TNB.2014.2345439}

\bibitem{Danner2016}
Danner, S.M., Wilshin, S.D., Shevtsova, N.A., Rybak, I.A.: Central control of
  interlimb coordination and speed-dependent gait expression in quadrupeds.
\newblock The Journal of Physiology \textbf{594}(23), 6947--6967 (2016).
\newblock \doi{10.1113/JP272787}.
\newblock
  \urlprefix\url{https://physoc.onlinelibrary.wiley.com/doi/abs/10.1113/JP272787}

\bibitem{DeJong2002}
{De Jong}, H.: Modeling and simulation of genetic regulatory systems: A
  literature review.
\newblock Journal of Computational Biology \textbf{9}(1), 67--103 (2002).
\newblock \doi{10.1089/10665270252833208}

\bibitem{Destexhe2009}
Destexhe, A., Sejnowski, T.J.: The {Wilson-Cowan} model, 36 years later.
\newblock Biological Cybernetics \textbf{101}(1), 1--2 (2009).
\newblock \doi{10.1007/s00422-009-0328-3}

\bibitem{Drushel1998}
Drushel, R.F., Neustadter, D.M., Hurwitz, I., Crago, P.E., Chiel, H.J.:
  Kinematic models of the buccal mass of \textit{Aplysia californica}.
\newblock {The Journal of Experimental Biology} \textbf{201}(Pt 10), 1563--83
  (1998).
\newblock \urlprefix\url{http://www.ncbi.nlm.nih.gov/pubmed/9556539}

\bibitem{Edwards2001}
Edwards, R., Siegelmann, H.T., Aziza, K., Glass, L.: {Symbolic dynamics and
  computation in model gene networks}.
\newblock Chaos \textbf{11}(1), 160--169 (2001).
\newblock \doi{10.1063/1.1336498}

\bibitem{Eisenberg1980}
Eisenberg, E., Hill, T.L., Chen, Y.: Cross-bridge model of muscle contraction.
  quantitative analysis.
\newblock Biophysical Journal \textbf{29}(2), 195--227 (1980).
\newblock \doi{10.1016/S0006-3495(80)85126-5}.
\newblock \urlprefix\url{http://dx.doi.org/10.1016/S0006-3495(80)85126-5}

\bibitem{Ekeberg1993}
Ekeberg, {\"{O}}.: A combined neuronal and mechanical model of fish swimming.
\newblock Biological Cybernetics \textbf{69}(5-6), 363--374 (1993).
\newblock \doi{10.1007/bf00199436}

\bibitem{Ekeberg1991}
Ekeberg, {\"{O}}., Wall{\'{e}}n, P., Lansner, A., Tr{\aa}v{\'{e}}n, H., Brodin,
  L., Grillner, S.: A computer based model for realistic simulations of neural
  networks.
\newblock Biological Cybernetics \textbf{65}(2), 81--90 (1991).
\newblock \doi{10.1007/bf00202382}

\bibitem{Ermentrout2010}
Ermentrout, B.: Neural networks as spatio-temporal pattern-forming systems.
\newblock Reports on Progress in Physics \textbf{61}(1998), 353--430 (2010)

\bibitem{Evans1998}
Evans, C.G., Cropper, E.C.: {Proprioceptive input to feeding motor programs in
  Aplysia}.
\newblock {Journal of Neuroscience} \textbf{18}(19), 8016--8031 (1998).
\newblock \doi{10.1523/jneurosci.18-19-08016.1998}

\bibitem{Feng2020}
Feng, K., Sen, R., Minegishi, R., D{\"u}bbert, M., Bockem{\"u}hl, T.,
  B{\"u}schges, A., Dickson, B.J.: Distributed control of motor circuits for
  backward walking in drosophila.
\newblock bioRxiv  (2020).
\newblock \doi{10.1101/2020.07.11.198663}.
\newblock
  \urlprefix\url{https://www.biorxiv.org/content/early/2020/07/12/2020.07.11.198663}

\bibitem{Gardner1977}
Gardner, D.: Interconnections of identified multiaction interneurons in buccal
  ganglia of \textit{Aplysia}.
\newblock Journal of Neurophysiology \textbf{40}(2), 349--361 (1977).
\newblock \doi{10.1152/jn.1977.40.2.349}

\bibitem{Georgopoulos1692}
Georgopoulos, A.P., Ashe, J., Smyrnis, N., Taira, M.: The motor cortex and the
  coding of force.
\newblock Science \textbf{256}(5064), 1692--1695 (1992).
\newblock \doi{10.1126/science.256.5064.1692}.
\newblock \urlprefix\url{https://science.sciencemag.org/content/256/5064/1692}

\bibitem{Georgopoulos1988}
Georgopoulos, A.P., Kettner, R.E., Schwartz, A.B.: Primate motor cortex and
  free arm movements to visual targets in three-dimensional space. {II.}
  {Coding} of the direction of movement by a neuronal population.
\newblock Journal of Neuroscience \textbf{8}(8), 2928--2937 (1988).
\newblock \doi{10.1523/jneurosci.08-08-02928.1988}

\bibitem{Giacomantonio2010}
Giacomantonio, C.E., Goodhill, G.J.: A {Boolean} model of the gene regulatory
  network underlying mammalian cortical area development.
\newblock PLoS Computational Biology \textbf{6}(9) (2010).
\newblock \doi{10.1371/journal.pcbi.1000936}

\bibitem{GillChiel2020}
Gill, J.P., Chiel, H.J.: Rapid adaptation to changing mechanical load by
  ordered recruitment of identified motor neurons.
\newblock eNeuro \textbf{7}(3) (2020).
\newblock \doi{10.1523/ENEURO.0016-20.2020}.
\newblock
  \urlprefix\url{https://www.eneuro.org/content/7/3/ENEURO.0016-20.2020}

\bibitem{gill_neurotic_2020}
Gill, J.P., Garcia, S., Ting, L.H., Wu, M., Chiel, H.J.: \textit{neurotic}:
  {Neuroscience} tool for interactive characterization.
\newblock eNeuro \textbf{7}(3) (2020).
\newblock \doi{10.1523/ENEURO.0085-20.2020}.
\newblock
  \urlprefix\url{https://www.eneuro.org/content/7/3/ENEURO.0085-20.2020}

\bibitem{GillSfN2018Poster}
Gill, J.P., Vorster, A.P.A., Lyttle, D.N., Keller, T.A., Stork, S.C., Chiel,
  H.J.: Neural correlates of adaptive responses to changing load in feeding
  \textit{{Aplysia}} (2018).
\newblock
  \urlprefix\url{https://www.abstractsonline.com/pp8/#!/4649/presentation/17445}.
\newblock Poster presented at Society for Neuroscience 48th Annual Meeting, San
  Diego, CA

\bibitem{Glaser2017}
Glaser, J.I., Chowdhury, R.H., Perich, M.G., Miller, L.E., Kording, K.P.:
  Machine learning for neural decoding (2017).
\newblock Http://arxiv.org/abs/1708.00909

\bibitem{Golowasch2002}
Golowasch, J., Goldman, M.S., Abbott, L.F., Marder, E.: Failure of averaging in
  the construction of a conductance-based neuron model.
\newblock Journal of Neurophysiology \textbf{87}(2), 1129--1131 (2002).
\newblock \doi{10.1152/jn.00412.2001}

\bibitem{Harischandra2010}
Harischandra, N., Cabelguen, J.M., Ekeberg, {\"{O}}.: A {3D} musculo-mechanical
  model of the salamander for the study of different gaits and modes of
  locomotion.
\newblock Frontiers in Neurorobotics \textbf{4}(DEC), 1--10 (2010).
\newblock \doi{10.3389/fnbot.2010.00112}

\bibitem{Harris2002}
Harris, S.E., Sawhill, B.K., Wuensche, A., Kauffman, S.: A model of
  transcriptional regulatory networks based on biases in the observed
  regulation rules.
\newblock Complexity \textbf{7}(4), 23--40 (2002).
\newblock \doi{10.1002/cplx.10022}

\bibitem{Haselgrove1973}
Haselgrove, J.C., Huxley, H.E.: X-ray evidence for radial cross-bridge movement
  and for the sliding filament model in actively contracting skeletal muscle.
\newblock Journal of Molecular Biology \textbf{77}(4) (1973).
\newblock \doi{10.1016/0022-2836(73)90222-2}

\bibitem{Hausknecht2015}
Hausknecht, M., Stone, P.: Deep recurrent q-learning for partially observable
  mdps.
\newblock AAAI Fall Symposium - Technical Report \textbf{FS-15-06}, 29--37
  (2015)

\bibitem{Heuer1995}
Heuer, H., Schmidt, R.A., Ghodsian, D.: Generalized motor programs for rapid
  bimanual tasks: a two-level multiplicative-rate model.
\newblock Biological Cybernetics \textbf{73}(4), 343--356 (1995).
\newblock \doi{10.1007/BF00199470}

\bibitem{A.V.Hill1938}
Hill, A.: The heat of shortening and the dynamic constants of muscle.
\newblock Proceedings of the Royal Society of London. Series B - Biological
  Sciences \textbf{126}(843), 136--195 (1938).
\newblock \doi{10.1098/rspb.1938.0050}

\bibitem{Hodgkin1952}
Hodgkin, A., Huxley, A.: A quantitative description of membrane current and its
  application to conduction and excitation in nerve.
\newblock The Journal of Physiology \textbf{117}(4) (1952)

\bibitem{Hooper2016}
Hooper, S.L., Guschlbauer, C., von Uckermann, G., Büschges, A.: Natural neural
  output that produces highly variable locomotory movements.
\newblock Journal of Neurophysiology \textbf{96}(4), 2072--2088 (2006).
\newblock \doi{10.1152/jn.00366.2006}.
\newblock \urlprefix\url{https://doi.org/10.1152/jn.00366.2006}.
\newblock PMID: 16775206

\bibitem{Hooper2007}
Hooper, S.L., Guschlbauer, C., von Uckermann, G., Büschges, A.: Different
  motor neuron spike patterns produce contractions with very similar rises in
  graded slow muscles.
\newblock Journal of Neurophysiology \textbf{97}(2), 1428--1444 (2007).
\newblock \doi{10.1152/jn.01014.2006}.
\newblock \urlprefix\url{https://doi.org/10.1152/jn.01014.2006}.
\newblock PMID: 17167058

\bibitem{Hooper2007b}
Hooper, S.L., Guschlbauer, C., von Uckermann, G., Büschges, A.: Different
  motor neuron spike patterns produce contractions with very similar rises in
  graded slow muscles.
\newblock Journal of Neurophysiology \textbf{97}(2), 1428--1444 (2007).
\newblock \doi{10.1152/jn.01014.2006}.
\newblock \urlprefix\url{https://doi.org/10.1152/jn.01014.2006}.
\newblock PMID: 17167058

\bibitem{horchler2015designing}
Horchler, A.D., Daltorio, K.A., Chiel, H.J., Quinn, R.D.: Designing responsive
  pattern generators: stable heteroclinic channel cycles for modeling and
  control.
\newblock Bioinspiration \& biomimetics \textbf{10}(2), 026001 (2015)

\bibitem{Hosman2019}
Hosman, T., Vilela, M., Milstein, D., Kelemen, J.N., Brandman, D.M., Hochberg,
  L.R., Simeral, J.D.: {BCI} decoder performance comparison of an {LSTM}
  recurrent neural network and a {Kalman} filter in retrospective simulation.
\newblock International IEEE/EMBS Conference on Neural Engineering, NER
  \textbf{2019-March}, 1066--1071 (2019).
\newblock \doi{10.1109/NER.2019.8717140}

\bibitem{Huang1990}
Huang, Z., Satterlie, R.A.: Neuronal mechanisms underlying behavioral switching
  in a pteropod mollusc.
\newblock Journal of Comparative Physiology A \textbf{166}(6), 875--887 (1990).
\newblock \doi{10.1007/BF00187335}

\bibitem{Hunt2015}
Hunt, A., Schmidt, M., Fischer, M., Quinn, R.: A biologically based neural
  system coordinates the joints and legs of a tetrapod.
\newblock Bioinspiration and Biomimetics \textbf{10}(5) (2015).
\newblock \doi{10.1088/1748-3190/10/5/055004}

\bibitem{Hunt2017}
Hunt, A., Szczecinski, N., Quinn, R.: Development and training of a neural
  controller for hind leg walking in a dog robot.
\newblock Frontiers in Neurorobotics \textbf{11}(APR), 1--16 (2017).
\newblock \doi{10.3389/fnbot.2017.00018}

\bibitem{Hurwitz1994}
Hurwitz, I., Goldstein, R.S., Susswein, A.J.: Compartmentalization of
  pattern-initiation and motor functions in the b31 and b32 neurons of the
  buccal ganglia of \textit{Aplysia californica}.
\newblock {Journal of Neurophysiology} \textbf{71}(4), 1514--27 (1994).
\newblock \doi{10.1152/jn.1994.71.4.1514}.
\newblock \urlprefix\url{http://www.ncbi.nlm.nih.gov/pubmed/8035232}

\bibitem{hurwitz1992adaptation}
Hurwitz, I., Susswein, A.J.: Adaptation of feeding sequences in \textit{Aplysia
  oculifera} to changes in the load and width of food.
\newblock Journal of experimental biology \textbf{166}(1), 215--235 (1992)

\bibitem{Hurwitz1996}
Hurwitz, I., Susswein, A.J.: B64, a newly identified central pattern generator
  element producing a phase switch from protraction to retraction in buccal
  motor programs of \textit{Aplysia californica}.
\newblock Journal of Neurophysiology \textbf{75}(4), 1327--1344 (1996).
\newblock \doi{10.1152/jn.1996.75.4.1327}

\bibitem{ivashko2003modeling}
Ivashko, D.G., Prilutsky, B.I., Markin, S.N., Chapin, J.K., Rybak, I.A.:
  Modeling the spinal cord neural circuitry controlling cat hindlimb movement
  during locomotion.
\newblock Neurocomputing \textbf{52}, 621--629 (2003)

\bibitem{Izhikevich2003}
Izhikevich, E.: Simple model of spiking neurons.
\newblock IEEE Transactions on Neural Networks \textbf{14}(6), 1569--1572
  (2003).
\newblock \doi{10.1109/TNN.2003.820440}.
\newblock \urlprefix\url{http://www.ncbi.nlm.nih.gov/pubmed/18244602}

\bibitem{IZHIKEVICH2000}
Izhikevich, E.M.: Neural excitability, spiking and bursting.
\newblock International Journal of Bifurcation and Chaos \textbf{10}(06),
  1171--1266 (2000).
\newblock \doi{10.1142/S0218127400000840}

\bibitem{Jaques2017}
Jaques, N., Gu, S., Turner, R.E., Eck, D.: Workshop track -iclr 2017 tuning
  recurrent neural networks with re- inforcement learning.
\newblock In: ICLR 2017, pp. 1--13 (2017)

\bibitem{Jing2004}
Jing, J., Cropper, E.C., Hurwitz, I., Weiss, K.R.: The construction of movement
  with behavior-specific and behavior-independent modules.
\newblock {Journal of Neuroscience} \textbf{24}(28), 6315--6325 (2004).
\newblock \doi{10.1523/JNEUROSCI.0965-04.2004}

\bibitem{Jing2017}
Jing, J., Cropper, E.C., Weiss, K.R.: Network functions of electrical coupling
  present in multiple and specific sites in behavior-generating circuits.
\newblock Elsevier Inc. (2017).
\newblock \doi{10.1016/B978-0-12-803471-2.00005-9}.
\newblock \urlprefix\url{http://dx.doi.org/10.1016/B978-0-12-803471-2.00005-9}

\bibitem{Jing2001}
Jing, J., Weiss, K.R.: Neural mechanisms of motor program switching in
  \textit{Aplysia}.
\newblock {Journal of Neuroscience} \textbf{21}(18), 7349--7362 (2001).
\newblock \doi{10.1523/jneurosci.21-18-07349.2001}

\bibitem{Jing2002}
Jing, J., Weiss, K.R.: Interneuronal basis of the generation of related but
  distinct motor programs in \textit{Aplysia}: Implications for current
  neuronal models of vertebrate intralimb coordination.
\newblock {Journal of Neuroscience} \textbf{22}(14), 6228--6238 (2002).
\newblock \doi{10.1523/jneurosci.22-14-06228.2002}

\bibitem{Jing2005}
Jing, J., Weiss, K.R.: Generation of variants of a motor act in a modular and
  hierarchical motor network.
\newblock {Current Biology} \textbf{15}(19), 1712--1721 (2005).
\newblock \doi{10.1016/j.cub.2005.08.051}

\bibitem{Kabotyanski1998}
Kabotyanski, E.A., Baxter, D.A., Byrne, J.H.: Identification and
  characterization of catecholaminergic neuron {B65}, which initiates and
  modifies patterned activity in the buccal ganglia of \textit{Aplysia}.
\newblock Journal of Neurophysiology \textbf{79}(2), 605--21 (1998).
\newblock \doi{10.1152/jn.1998.79.2.605}.
\newblock \urlprefix\url{http://www.ncbi.nlm.nih.gov/pubmed/9463425}

\bibitem{KamaliSarvestani2013}
{Kamali Sarvestani}, I., Kozlov, A., Harischandra, N., Grillner, S., Ekeberg,
  {\"{O}}.: A computational model of visually guided locomotion in lamprey.
\newblock Biological Cybernetics \textbf{107}(5), 497--512 (2013).
\newblock \doi{10.1007/s00422-012-0524-4}

\bibitem{kandel1976cellular}
Kandel, E.: Cellular Basis of Behavior: An Introduction to Behavioral
  Neurobiology.
\newblock Books in psychology. W. H. Freeman (1976).
\newblock \urlprefix\url{https://books.google.com/books?id=gHZ1QgAACAAJ}

\bibitem{Katzoff2006}
Katzoff, A., Ben-Gedalya, T., Hurwitz, I., Miller, N., Susswein, Y.Z.,
  Susswein, A.J.: Nitric oxide signals that \textit{Aplysia} have attempted to
  eat, a necessary component of memory formation after learning that food is
  inedible.
\newblock Journal of Neurophysiology \textbf{96}(3), 1247--1257 (2006).
\newblock \doi{10.1152/jn.00056.2006}.
\newblock \urlprefix\url{https://doi.org/10.1152/jn.00056.2006}.
\newblock PMID: 16738221

\bibitem{kauffman1993origins}
Kauffman, S.A.: The origins of order: Self-organization and selection in
  evolution.
\newblock Oxford University Press, USA (1993)

\bibitem{koch1998methods}
Koch, C., Segev, I., et~al.: Methods in neuronal modeling: {From} ions to
  networks.
\newblock MIT press (1998)

\bibitem{koehl2006wave}
Koehl, M.A.: Wave-swept shore: the rigors of life on a rocky coast.
\newblock Univ of California Press (2006)

\bibitem{kuo2002relative}
Kuo, A.D.: The relative roles of feedforward and feedback in the control of
  rhythmic movements.
\newblock Motor control \textbf{6}(2), 129--145 (2002)

\bibitem{Kupfermann1974}
Kupfermann, I.: Feeding behavior in \textit{{Aplysia}}: a simple system for the
  study of motivation.
\newblock Behavioral Biology \textbf{10}(1), 1--26 (1974).
\newblock \doi{10.1016/S0091-6773(74)91644-7}.
\newblock
  \urlprefix\url{http://www.sciencedirect.com/science/article/pii/S0091677374916447}

\bibitem{Latash1999}
Latash, M.: Progress in motor control: Bernstein's traditions in movement
  studies.
\newblock Journal of Athletic Training \textbf{34}(3), 1999 (1999)

\bibitem{Li1999}
Li, L., {Van Den Bogert}, E.C., Caldwell, G.E., {Van Emmerik}, R.E., Hamill,
  J.: Coordination patterns of walking and running at similar speed and stride
  frequency.
\newblock Human Movement Science \textbf{18}(1), 67--85 (1999).
\newblock \doi{10.1016/S0167-9457(98)00034-7}

\bibitem{LU2012137}
Lu, C.W., Patil, P.G., Chestek, C.A.: Chapter seven - current challenges to the
  clinical translation of brain machine interface technology.
\newblock In: C.~Hamani, E.~Moro (eds.) Emerging Horizons in Neuromodulation,
  \emph{International Review of Neurobiology}, vol. 107, pp. 137--160. Academic
  Press (2012).
\newblock \doi{https://doi.org/10.1016/B978-0-12-404706-8.00008-5}.
\newblock
  \urlprefix\url{http://www.sciencedirect.com/science/article/pii/B9780124047068000085}

\bibitem{LuJoVE2013}
Lu, H., McManus, J.M., Chiel, H.J.: Extracellularly identifying motor neurons
  for a muscle motor pool in \textit{{Aplysia} californica}.
\newblock Journal of Visualized Experiments (73), e50189 (2013).
\newblock \doi{10.3791/50189}.
\newblock
  \urlprefix\url{http://www.jove.com/video/50189/extracellularly-identifying-motor-neurons-for-muscle-motor-pool}

\bibitem{Lyttle2017}
Lyttle, D.N., Gill, J.P., Shaw, K.M., Thomas, P.J., Chiel, H.J.: Robustness,
  flexibility, and sensitivity in a multifunctional motor control model.
\newblock Biological Cybernetics \textbf{111}(1), 25--47 (2017).
\newblock \doi{10.1007/s00422-016-0704-8}

\bibitem{Mann1980}
Mann, R.A., Hagy, J.: Biomechanics of walking, running, and sprinting.
\newblock The American Journal of Sports Medicine \textbf{8}(5), 345--350
  (1980).
\newblock \doi{10.1177/036354658000800510}

\bibitem{MantziarisBockemuehlBueschges2020DevNeuro-review-CPG-insect}
Mantziaris, C., Bockem{\"u}hl, T., B{\"u}schges, A.: Central pattern generating
  networks in insect locomotion.
\newblock Developmental Neurobiology  (2020)

\bibitem{Marder2011}
Marder, E., Taylor, A.L.: Multiple models to capture the variability in
  biological neurons and networks.
\newblock Nature Neuroscience \textbf{14}(2), 133--138 (2011)

\bibitem{markin2016neuromechanical}
Markin, S.N., Klishko, A.N., Shevtsova, N.A., Lemay, M.A., Prilutsky, B.I.,
  Rybak, I.A.: A neuromechanical model of spinal control of locomotion.
\newblock In: Neuromechanical modeling of posture and locomotion, pp. 21--65.
  Springer (2016)

\bibitem{McCulloch1943}
McCulloch, W.S., Pitts, W.: A logical calculus of the ideas immanent in nervous
  activity.
\newblock The Bulletin of Mathematical Biophysics \textbf{5}(4), 115--133
  (1943).
\newblock \doi{10.1007/BF02478259}

\bibitem{mcmanus2012vitro}
McManus, J.M., Lu, H., Chiel, H.J.: An \textit{in vitro} preparation for
  eliciting and recording feeding motor programs with physiological movements
  in \textit{Aplysia californica}.
\newblock JoVE (Journal of Visualized Experiments) (70), e4320 (2012)

\bibitem{McManus2014}
McManus, J.M., Lu, H., Cullins, M.J., Chiel, H.J.: Differential activation of
  an identified motor neuron and neuromodulation provide \textit{Aplysia}'s
  retractor muscle an additional function.
\newblock Journal Neurophysiology \textbf{112}(4), 778--791 (2014)

\bibitem{Mihalas2009a}
Mihalaş, S., Niebur, E.: A generalized linear integrate-and-fire neural model
  produces diverse spiking behaviors.
\newblock Neural Computation \textbf{21}(3), 704--718 (2009).
\newblock \doi{10.1162/neco.2008.12-07-680}.
\newblock \urlprefix\url{https://www.ncbi.nlm.nih.gov/pubmed/18928368
  https://www.ncbi.nlm.nih.gov/pmc/articles/PMC2954058/}

\bibitem{molkov2015mechanisms}
Molkov, Y.I., Bacak, B.J., Talpalar, A.E., Rybak, I.A.: Mechanisms of
  left-right coordination in mammalian locomotor pattern generation circuits:
  {A} mathematical modeling view.
\newblock PLoS Comput Biol \textbf{11}(5), e1004270 (2015)

\bibitem{Morgan2002}
Morgan, P.T., Jing, J., Vilim, F.S., Weiss, K.R.: Interneuronal and peptidergic
  control of motor pattern switching in \textit{Aplysia}.
\newblock {Journal of Neurophysiology} \textbf{87}(1), 49--61 (2002).
\newblock \doi{10.1152/jn.00438.2001}

\bibitem{MORITANI199134}
Moritani, T., Oddsson, L., Thorstensson, A.: Phase-dependent preferential
  activation of the soleus and gastrocnemius muscles during hopping in humans.
\newblock Journal of Electromyography and Kinesiology \textbf{1}(1), 34--40
  (1991).
\newblock \doi{https://doi.org/10.1016/1050-6411(91)90024-Y}.
\newblock
  \urlprefix\url{http://www.sciencedirect.com/science/article/pii/105064119190024Y}

\bibitem{MORTON1994413}
Morton, D., Chiel, H.: Neural architectures for adaptive behavior.
\newblock Trends in Neurosciences \textbf{17}(10), 413 -- 420 (1994).
\newblock \doi{https://doi.org/10.1016/0166-2236(94)90015-9}.
\newblock
  \urlprefix\url{http://www.sciencedirect.com/science/article/pii/0166223694900159}

\bibitem{Morton1993}
Morton, D.W., Chiel, H.J.: \textit{In vivo} buccal nerve activity that
  distinguishes ingestion from rejection can be used to predict behavioral
  transitions in \textit{Aplysia}.
\newblock {Journal of Comparative Physiology. A, Sensory, Neural, and
  Behavioral Physiology} \textbf{172}(1), 17--32 (1993).
\newblock \doi{10.1007/bf00214712}.
\newblock \urlprefix\url{https://doi.org/10.1007/bf00214712}

\bibitem{Morton1993b}
Morton, D.W., Chiel, H.J.: The timing of activity in motor neurons that produce
  radula movements distinguishes ingestion from rejection in \textit{Aplysia}.
\newblock Journal of Comparative Physiology A \textbf{173}(5), 519--536 (1993).
\newblock \doi{10.1007/BF00197761}

\bibitem{Mulgaonkar2016}
Mulgaonkar, Y., Araki, B., Koh, J.S., Guerrero-Bonilla, L., Aukes, D.M.,
  Makineni, A., Tolley, M.T., Rus, D., Wood, R.J., Kumar, V.: The flying
  monkey: A mesoscale robot that can run, fly, and grasp.
\newblock Proceedings - IEEE International Conference on Robotics and
  Automation \textbf{2016-June}, 4672--4679 (2016).
\newblock \doi{10.1109/ICRA.2016.7487667}

\bibitem{Neustadter2002a}
Neustadter, D.M., Drushel, R.F., Crago, P.E., Adams, B.W., Chiel, H.J.: A
  kinematic model of swallowing in \textit{Aplysia californica} based on
  radula/odontophore kinematics and \textit{in vivo} magnetic resonance images.
\newblock Journal of Experimental Biology \textbf{205}(20), 3177--3206 (2002)

\bibitem{neustadter2007kinematics}
Neustadter, D.M., Herman, R.L., Drushel, R.F., Chestek, D.W., Chiel, H.J.: The
  kinematics of multifunctionality: comparisons of biting and swallowing in
  \textit{Aplysia californica}.
\newblock Journal of Experimental Biology \textbf{210}(2), 238--260 (2007)

\bibitem{Nicolelis2009}
Nicolelis, M.A., Lebedev, M.A.: Principles of neural ensemble physiology
  underlying the operation of brain-machine interfaces.
\newblock Nature Reviews Neuroscience \textbf{10}(7), 530--540 (2009).
\newblock \doi{10.1038/nrn2653}

\bibitem{Novakovic2006}
Novakovic, V.A., Sutton, G.P., Neustadter, D.M., Beer, R.D., Chiel, H.J.:
  Mechanical reconfiguration mediates swallowing and rejection in
  \textit{Aplysia californica}.
\newblock {Journal of Comparative Physiology A: Neuroethology, Sensory, Neural,
  and Behavioral Physiology} \textbf{192}(8), 857--870 (2006).
\newblock \doi{10.1007/s00359-006-0124-7}

\bibitem{Oishi2014}
Oishi, K., Klavins, E.: Framework for engineering finite state machines in gene
  regulatory networks.
\newblock ACS Synthetic Biology \textbf{3}(9), 652--665 (2014).
\newblock \doi{10.1021/sb4001799}

\bibitem{Packard1985}
Packard, N., Wolfram, S.: Two-dimensional cellular automata.
\newblock Journal of Statistical Physics \textbf{38}(March), 901--946 (1985).
\newblock \doi{10.1201/9780429494093-6}

\bibitem{Payne2013}
Payne, J.L., Wagner, A.: Constraint and contingency in multifunctional gene
  regulatory circuits.
\newblock PLoS Computational Biology \textbf{9}(6) (2013).
\newblock \doi{10.1371/journal.pcbi.1003071}

\bibitem{pearson1993common}
Pearson, K.: Common principles of motor control in vertebrates and
  invertebrates.
\newblock Annual review of neuroscience \textbf{16}(1), 265--297 (1993)

\bibitem{Pearson1987}
Pearson, K.G.: Central Pattern Generation: {A} Concept Under Scrutiny, pp.
  167--185.
\newblock Springer US, Boston, MA (1987).
\newblock \doi{10.1007/978-1-4615-9492-5_10}.
\newblock \urlprefix\url{https://doi.org/10.1007/978-1-4615-9492-5_10}

\bibitem{Piazzesi1995}
Piazzesi, G., Lombardi, V.: A cross-bridge model that is able to explain
  mechanical and energetic properties of shortening muscle.
\newblock Biophysical Journal \textbf{68}(5), 1966--1979 (1995).
\newblock \doi{10.1016/S0006-3495(95)80374-7}

\bibitem{Prescott2016}
Prescott, T.J., Ayers, J.L., Grasso, F., Verschure, P.F.M.J.: Chapter 17.
  {Embodied} models and neurorobotics.
\newblock In: From Neuron to Cognition via Computational Neuroscience, chap.
  Embodied M, pp. 483--512. MIT Press (2016)

\bibitem{prinz2004similar}
Prinz, A.A., Bucher, D., Marder, E.: Similar network activity from disparate
  circuit parameters.
\newblock Nature neuroscience \textbf{7}(12), 1345--1352 (2004)

\bibitem{Ravn1995}
Ravn, A.P., Rischel, H., Holdgaard, M., Eriksen, T.J., Conrad, F., Andersen,
  T.O.: Hybrid control of a robot — a case study.
\newblock Hybrid Systems {II} pp. 391--404 (1995).
\newblock \doi{10.1007/3-540-60472-3_20}

\bibitem{RiveraTorres2018}
{Rivera Torres}, P.J., {Serrano Mercado}, E.I., {Anido Rif{\'{o}}n}, L.:
  Probabilistic {Boolean} network modeling of an industrial machine.
\newblock Journal of Intelligent Manufacturing \textbf{29}(4), 875--890 (2018).
\newblock \doi{10.1007/s10845-015-1143-4}

\bibitem{Roschard2003}
R{\"{o}}schard, J., Roces, F.: Cutters, carriers and transport chains:
  Distance-dependent foraging strategies in the grass-cutting ant \textit{Atta
  vollenweideri}.
\newblock Insectes Sociaux \textbf{50}(3), 237--244 (2003).
\newblock \doi{10.1007/s00040-003-0663-7}

\bibitem{Rosin2013}
Rosin, D.P., Rontani, D., Gauthier, D.J., Sch{\"{o}}ll, E.: Experiments on
  autonomous boolean networks.
\newblock Chaos \textbf{23}(2) (2013).
\newblock \doi{10.1063/1.4807481}

\bibitem{Royakkers2015}
Royakkers, L., van Est, R.: A literature review on new robotics: Automation
  from love to war.
\newblock International Journal of Social Robotics \textbf{7}(5), 549--570
  (2015).
\newblock \doi{10.1007/s12369-015-0295-x}.
\newblock \urlprefix\url{http://dx.doi.org/10.1007/s12369-015-0295-x}

\bibitem{Saadatpour2010}
Saadatpour, A., Albert, I., Albert, R.: Attractor analysis of asynchronous
  {Boolean} models of signal transduction networks.
\newblock Journal of Theoretical Biology \textbf{266}(4), 641--656 (2010).
\newblock \doi{10.1016/j.jtbi.2010.07.022}

\bibitem{S.C.Kleene1951}
{S.C. Kleene}: {Representation of events in nerve nets and finite automata}.
\newblock Tech. rep., U.S. Air Force Project RAND (1951)

\bibitem{Schwartz1988}
Schwartz, A.B., Kettner, R.E., Georgopoulos, A.P.: Primate motor cortex and
  free arm movements to visual targets in three-dimensional space. {I.}
  {Relations} between single cell discharge and direction of movement.
\newblock Journal of Neuroscience \textbf{8}(8), 2913--2927 (1988).
\newblock \doi{10.1523/jneurosci.08-08-02913.1988}

\bibitem{selverston1992dynamic}
Selverston, A.I.: Dynamic biological networks: the stomatogastric nervous
  system.
\newblock MIT press (1992)

\bibitem{SELVERSTON1976215}
Selverston, A.I., Russell, D.F., Miller, J.P., King, D.G.: The stomatogastric
  nervous system: Structure and function of a small neural network.
\newblock Progress in Neurobiology \textbf{7}, 215 -- 289 (1976).
\newblock \doi{https://doi.org/10.1016/0301-0082(76)90008-3}.
\newblock
  \urlprefix\url{http://www.sciencedirect.com/science/article/pii/0301008276900083}

\bibitem{Sewak2019}
Sewak, M.: Deep reinforcement learning.
\newblock Deep Reinforcement Learning pp. 1--9 (2019).
\newblock \doi{10.1007/978-981-13-8285-7}

\bibitem{Shadmehr1970}
Shadmehr, R.: A mathematical muscle model.
\newblock ReCALL  (1970)

\bibitem{Shaw2015}
Shaw, K.M., Lyttle, D.N., Gill, J.P., Cullins, M.J., Mcmanus, J.M., Lu, H.,
  Thomas, P.J., Chiel, H.J.: The significance of dynamical architecture for
  adaptive responses to mechanical loads during rhythmic behavior.
\newblock J Computational Neuroscience \textbf{38}, 25--51 (2015).
\newblock \doi{10.1007/s10827-014-0519-3}

\bibitem{Shea-Brown2006}
Shea-Brown, E., Rinzel, J., Rakitin, B.C., Malapani, C.: A firing rate model of
  {Parkinsonian} deficits in interval timing.
\newblock Brain Research \textbf{1070}(1), 189--201 (2006).
\newblock \doi{10.1016/j.brainres.2005.10.070}

\bibitem{Shoham2005}
Shoham, S., Paninski, L.M., Fellows, M.R., Hatsopoulos, N.G., Donoghue, J.P.,
  Normann, R.A.: Statistical encoding model for a primary motor cortical
  brain-machine interface.
\newblock IEEE Transactions on Biomedical Engineering \textbf{52}(7),
  1312--1322 (2005).
\newblock \doi{10.1109/TBME.2005.847542}

\bibitem{Siegle2018}
Siegle, L., Schwab, J.D., K{\"{u}}hlwein, S.D., Lausser, L., T{\"{u}}mpel, S.,
  Pfister, A.S., K{\"{u}}hl, M., Kestler, H.A.: A boolean network of the
  crosstalk between {IGF} and wnt signaling in aging satellite cells.
\newblock PLoS ONE \textbf{13}(3), 1--24 (2018).
\newblock \doi{10.1371/journal.pone.0195126}

\bibitem{Stamhuis2005}
Stamhuis, E., Aerts, P., Nauwelaerts, S.: {Swimming and jumping in a
  semi-aquatic frog}.
\newblock Animal Biology \textbf{55}(1) (2005)

\bibitem{Stehouwer1992}
Stehouwer, D.J.: Development of anuran locomotion: Ethological and
  neurophysiological considerations.
\newblock Journal of Neurobiology \textbf{23}(10), 1467--1485 (1992).
\newblock \doi{10.1002/neu.480231008}

\bibitem{stewart2004consequences}
Stewart, H.L.: Consequences of flexural stiffness and buoyancy for hydrodynamic
  forces, light interception and dispersal of a tropical alga.
\newblock University of California, Berkeley (2004)

\bibitem{Sussillo2012}
Sussillo, D., Nuyujukian, P., Fan, J.M., Kao, J.C., Stavisky, S.D., Ryu, S.,
  Shenoy, K.: A recurrent neural network for closed-loop intracortical
  brain-machine interface decoders.
\newblock Journal of Neural Engineering \textbf{9}(2) (2012).
\newblock \doi{10.1088/1741-2560/9/2/026027}

\bibitem{Susswein1988}
Susswein, A.J., Byrne, J.H.: Identification and characterization of neurons
  initiating patterned neural activity in the buccal ganglia of
  \textit{Aplysia}.
\newblock The Journal of Neuroscience \textbf{8}(6), 2049--61 (1988).
\newblock \urlprefix\url{http://www.ncbi.nlm.nih.gov/pubmed/3385489}

\bibitem{SUSSWEIN2012304}
Susswein, A.J., Chiel, H.J.: Nitric oxide as a regulator of behavior: New ideas
  from \textit{Aplysia} feeding.
\newblock Progress in Neurobiology \textbf{97}(3), 304 -- 317 (2012).
\newblock \doi{https://doi.org/10.1016/j.pneurobio.2012.03.004}.
\newblock
  \urlprefix\url{http://www.sciencedirect.com/science/article/pii/S0301008212000366}

\bibitem{Sutton2004a}
Sutton, G.P., Macknin, J.B., Gartman, S.S., Sunny, G.P., Beer, R.D., Crago,
  P.E., Neustadter, D.M., Chiel, H.J.: Passive hinge forces in the feeding
  apparatus of \textit{Aplysia} aid retraction during biting but not during
  swallowing.
\newblock {Journal of Comparative Physiology A: Neuroethology, Sensory, Neural,
  and Behavioral Physiology} \textbf{190}(6), 501--514 (2004).
\newblock \doi{10.1007/s00359-004-0517-4}

\bibitem{Sutton2004b}
Sutton, G.P., Mangan, E.V., Neustadter, D.M., Beer, R.D., Crago, P.E., Chiel,
  H.J.: Neural control exploits changing mechanical advantage and context
  dependence to generate different feeding responses in \textit{Aplysia}.
\newblock {Biological Cybernetics} \textbf{91}(5), 333--345 (2004).
\newblock \doi{10.1007/s00422-004-0517-z}

\bibitem{Szczecinski2015}
Szczecinski, N.S., Chrzanowski, D.M., Cofer, D.W., Terrasi, A.S., Moore, D.R.,
  Martin, J.P., Ritzmann, R.E., Quinn, R.D.: Introducing mantisbot: Hexapod
  robot controlled by a high-fidelity, real-time neural simulation.
\newblock IEEE International Conference on Intelligent Robots and Systems
  \textbf{2015-December}(September), 3875--3881 (2015).
\newblock \doi{10.1109/IROS.2015.7353922}

\bibitem{Szczecinski2017a}
Szczecinski, N.S., Hunt, A.J., Quinn, R.D.: A functional subnetwork approach to
  designing synthetic nervous systems that control legged robot locomotion.
\newblock Frontiers in Neurorobotics \textbf{11}(August) (2017).
\newblock \doi{10.3389/fnbot.2017.00037}.
\newblock
  \urlprefix\url{http://journal.frontiersin.org/article/10.3389/fnbot.2017.00037/full}

\bibitem{Szczecinski2017}
Szczecinski, N.S., Quinn, R.D.: Leg-local neural mechanisms for searching and
  learning enhance robotic locomotion.
\newblock Biological Cybernetics \textbf{112}(1-2), 99--112 (2018).
\newblock \doi{10.1007/s00422-017-0726-x}.
\newblock \urlprefix\url{https://doi.org/10.1007/s00422-017-0726-x}

\bibitem{Tal1997a}
Tal, D., Schwartz, E.L.: Computing with the leaky integrate-and-fire neuron:
  logarithmic computation and multiplication.
\newblock Neural Computation \textbf{9}(2), 305--18 (1997).
\newblock \urlprefix\url{http://www.ncbi.nlm.nih.gov/pubmed/9117905}

\bibitem{TEYKE1991307}
Teyke, T., Weiss, K.R., Kupfermann, I.: Activity of identified cerebral neuron
  correlates with food-induced arousal in \textit{Aplysia}.
\newblock Neuroscience Letters \textbf{133}(2), 307--310 (1991).
\newblock \doi{https://doi.org/10.1016/0304-3940(91)90595-K}.
\newblock
  \urlprefix\url{http://www.sciencedirect.com/science/article/pii/030439409190595K}

\bibitem{Verstappen2000}
Verstappen, M., Aerts, P., {Van Damme}, R.: Terrestrial locomotion in the
  black-billed magpie: Kinematic analysis of walking, running and out-of-phase
  hopping.
\newblock Journal of Experimental Biology \textbf{203}(14), 2159--2170 (2000)

\bibitem{Wang2018}
Wang, Y., Truccolo, W., Borton, D.A.: Decoding hindlimb kinematics from primate
  motor cortex using long short-term memory recurrent neural networks.
\newblock Proceedings of the Annual International Conference of the IEEE
  Engineering in Medicine and Biology Society, EMBS \textbf{2018-July},
  1944--1947 (2018).
\newblock \doi{10.1109/EMBC.2018.8512609}

\bibitem{Warman1995}
Warman, E.N., Chiel, H.J.: A new technique for chronic single-unit
  extracellular recording in freely behaving animals using pipette electrodes.
\newblock Journal of Neuroscience Methods \textbf{57}(2), 161--169 (1995).
\newblock \doi{10.1016/0165-0270(94)00144-6}

\bibitem{Webster2013}
Webster, V.A., Lonsberry, A.J., Horchler, A.D., Shaw, K.M., Chiel, H.J., Quinn,
  R.D.: A segmental mobile robot with active tensegrity bending and
  noise-driven oscillators.
\newblock In: 2013 IEEE/ASME International Conference on Advanced Intelligent
  Mechatronics: Mechatronics for Human Wellbeing, AIM 2013, pp. 1373--1380.
  Wollongong, Australia (2013).
\newblock \doi{10.1109/AIM.2013.6584286}

\bibitem{Weiss1986}
Weiss, K.R., Chiel, H.J., Koch, U., Kupfermann, I.: Activity of an identified
  histaminergic neuron, and its possible role in arousal of feeding behavior in
  semi-intact \textit{Aplysia}.
\newblock The Journal of Neuroscience \textbf{6}(August), 2403--2415 (1986).
\newblock \doi{20026578}

\bibitem{Wilson1972}
Wilson, H.R., Cowan, J.D.: Excitatory and inhibitory interactions in localized
  populations of model neurons.
\newblock Biophysical Journal \textbf{12}(1), 1--24 (1972).
\newblock \doi{10.1016/S0006-3495(72)86068-5}

\bibitem{Wilson1973}
Wilson, H.R., Cowan, J.D.: A mathematical theory of the functional dynamics of
  cortical and thalamic nervous tissue.
\newblock Kybernetik \textbf{13}(2), 55--80 (1973).
\newblock \doi{10.1007/BF00288786}

\bibitem{Wood2017}
Wood, K.C., Blackwell, J.M., Geffen, M.N.: {Cortical inhibitory interneurons
  control sensory processing}.
\newblock Current Opinion in Neurobiology \textbf{46}, 200--207 (2017).
\newblock \doi{10.1016/j.conb.2017.08.018}

\bibitem{Xie2018}
Xie, Z., Schwartz, O., Prasad, A.: Decoding of finger trajectory from {ECoG}
  using deep learning.
\newblock Journal of Neural Engineering \textbf{15}(3) (2018).
\newblock \doi{10.1088/1741-2552/aa9dbe}

\bibitem{Ye2006aSwallows}
Ye, H., Morton, D.W., Chiel, H.J.: Neuromechanics of coordination during
  swallowing in \textit{Aplysia californica}.
\newblock Journal of Neuroscience \textbf{26}(5), 1470--1485 (2006).
\newblock \doi{10.1523/JNEUROSCI.3691-05.2006}.
\newblock
  \urlprefix\url{http://www.jneurosci.org/cgi/doi/10.1523/JNEUROSCI.3691-05.2006}

\bibitem{Ye2006bRejections}
Ye, H., Morton, D.W., Chiel, H.J.: Neuromechanics of multifunctionality during
  rejection in \textit{Aplysia californica}.
\newblock Journal of Neuroscience \textbf{26}(42), 10743--10755 (2006)

\bibitem{yu1997nonisometric}
Yu, S.N., Crago, P., Chiel, H.: A nonisometric kinetic model for smooth muscle.
\newblock American Journal of Physiology-Cell Physiology \textbf{272}(3),
  C1025--C1039 (1997)

\bibitem{Yu1999}
Yu, S.N., Crago, P.E., Chiel, H.J.: Biomechanical properties and a kinetic
  simulation model of the smooth muscle {I2} in the buccal mass of
  \textit{Aplysia}.
\newblock Biological Cybernetics \textbf{81}, 505--513 (1999).
\newblock \doi{10.1007/s004220050579}

\bibitem{Zahalak1990}
Zahalak, G.I., Ma, S.P.: Muscle activation and contraction: Constitutive
  relations based directly on cross-bridge kinetics.
\newblock Journal of Biomechanical Engineering \textbf{112}(1), 52--62 (1990).
\newblock \doi{10.1115/1.2891126}.
\newblock \urlprefix\url{https://doi.org/10.1115/1.2891126}

\bibitem{Zajac1989}
Zajac, F.E.: Muscle and tendon: properties, models, scaling, and application to
  biomechanics and motor control.
\newblock Critical Reviews in Biomedical Engineering \textbf{17}(4), 359--411
  (1989).
\newblock \urlprefix\url{http://europepmc.org/abstract/MED/2676342}

\bibitem{ziv_simulator_1994}
Ziv, I., Baxter, D.A., Byrne, J.H.: Simulator for neural networks and action
  potentials: description and application.
\newblock Journal of Neurophysiology \textbf{71}(1), 294--308 (1994).
\newblock \urlprefix\url{http://jn.physiology.org/content/71/1/294}

\end{thebibliography}

%
%

\appendix 

\newpage

\section{Appendices}

\subsection{Semi-Implicit Integration Scheme} \label{integration}
Suppose a continuously varying quantity $x$  satisfies the initial value problem
\begin{equation}
    \frac{dx(t)}{dt}=\frac{-(x(t)-x_\infty(y_2(t),\ldots,y_n(t)))}{\tau},\quad x(t_0)=x_0
\end{equation}
where $x_\infty(t)$ is set by the other variables in our system, say $\{y_i\}_{i=2}^n$, generally following some nonlinear dependencies, and $\tau$ is a fixed time constant.   
We would like to implement a numerical approximation to the exact solution for $x$, namely
\begin{equation}
\label{eq:exact_ODE_solution}
    x(t)=e^{-(t-t_0)/\tau}x_0+\frac1\tau\int_{t_0}^t e^{-(t-s)/\tau}x_\infty(y_2(s),\ldots,y_n(s))\,ds,
\end{equation}
along with the remaining variables that satisfy their own differential equations. 
Euler's forward method is convenient to implement but prone to numerical instability.  
Euler's backward or implicit method is numerically stable but computationally expensive, as it requires solving an implicit equation at each step.  Both methods proceed from a discrete approximation of the derivative, namely
\begin{equation}
    \label{eq:discrete_approx_dxdt}
    \frac{x(t+h)-x(t)}{h}\approx \frac{dx}{dt}.
\end{equation}
In both cases we create an update rule $x(t)\to x(t+h)$, by evaluating the right hand side of \eqref{eq:discrete_approx_dxdt} at either time $t$ or time $t+h$, and solving for $x(t+h)$.
\paragraph{Forward:}
\begin{align}
 \frac{x(t+h)-x(t)}{h}&=\frac{
 -(x(t)-x_\infty(y_2(t),\ldots,y_n(t))}{\tau}\\
 x(t+h)&=x(t)-\frac{x(t)-x_\infty(y_2(t),\ldots,y_n(t)}{\tau}h
 \label{eq:forward_Euler}
\end{align}
\paragraph{Backward:}
\begin{align}
 \frac{x(t+h)-x(t)}{h}&=
 \frac{-(x(t+h)-x_\infty(y_2(t+h),\ldots,y_n(t+h))}{\tau}\\
 x(t+h)&=\frac{\tau x(t)+h x_\infty(y_2(t+h),\ldots,y_n(t+h))}{\tau+h}.
\label{eq:backward_Euler} 
\end{align}
Since the variables $y_2,\ldots,y_n$ appear on the right hand side of \eqref{eq:backward_Euler} evaluated at the later time point, $t+h$, \eqref{eq:backward_Euler} is part of a system of $n$ nonlinear equations that must be solved simultaneously to determine the system state at $t+h$.  Both numerical schemes \eqref{eq:forward_Euler} and \eqref{eq:backward_Euler} are first-order accurate, meaning that the truncation error between the true solution \eqref{eq:exact_ODE_solution} and the numerical approximation scales as $\mathcal{O}(h^2)$ on each time step, with a global error (after $T/h$ time steps for a simulation of total runtime $T$) that is $\mathcal{O}(h)$.  

\paragraph{Semi-implicit:}
In our model implementation, we use a semi-implicit method based
on the approximation
\begin{equation}
    \frac{x(t+h)-x(t)}{h}\approx \frac{-(x(t+h)-x_\infty(y_2(t),\ldots,y_n(t))}{\tau},
\end{equation}
namely
\begin{framed}
\begin{equation}
\label{eq:semi_implicit_method}
    x(t+h)=\frac{\tau x(t)+hx_\infty(y_2(t),\ldots,y_n(t))}{\tau+h}.
\end{equation}
\end{framed}
At each time step we update $x$ using a weighted average of its past value $x(t)$ and its target value $x_\infty(t)$, with the (short) timestep $h$ and the intrinsic time constant $\tau$ providing the relative weight of past and future.  We expect an accurate approximation to \eqref{eq:exact_ODE_solution} provided $h\ll\tau$.  As we show below, the method is first-order accurate, and numerically stable, but it does not require solving an implicit equation at each time step.  Thus this method combines the advantages of both the forward and backward methods.  The method may be seen as an example of operator splitting.\footnote{Citation: MacNamara, Shev, and Gilbert Strang. ``Operator splitting." Splitting Methods in Communication, Imaging, Science, and Engineering. Springer, 2016. 95-114.}

To see that \eqref{eq:semi_implicit_method} is first-order accurate, we assume that $x(t)$ is smooth enough to have Taylor expansions through the 2nd order.  Thus, for $h\ll 1$ we may write
\begin{align}
    x(t+h)&=x(t)+h\frac{dx}{dt}(t)+\mathcal{O}(h^2),\quad\text{ as }h\to 0\\
    &=x(t)+h\frac{x_\infty(\mby(t))-x(t)}{\tau}+\mathcal{O}(h^2)\\
    &=x(t)\frac{\tau+h}{\tau+h}+h\frac{x_\infty(\mby(t))-x(t)}{\tau+h}\left(\frac{\tau+h}{\tau}\right)+\mathcal{O}(h^2)\\
    &=\frac{\tau x(t)}{\tau+h}+\frac{hx(t)}{\tau+h}+\left(\frac{\tau+h}{\tau}\right)\frac{hx_\infty(\mby(t))}{\tau+h}-\left(\frac{\tau+h}{\tau}\right)\frac{hx(t)}{\tau+h}+\mathcal{O}(h^2)\\
    &=\frac{\tau x(t)}{\tau+h}+\frac{hx_\infty(\mby(t))}{\tau+h}+\frac{hx(t)}{\tau+h}-\frac{hx(t)}{\tau+h}+\mathcal{O}(h^2)\\
    &=\frac{\tau x(t)}{\tau+h}+\frac{hx_\infty(\mby(t))}{\tau+h}+\mathcal{O}(h^2),\quad\text{ as }h\to 0.
\end{align}
Thus, the semi-implicit scheme \eqref{eq:semi_implicit_method} is first-order accurate in the time step $h$.  

To see that \eqref{eq:semi_implicit_method} is numerically stable, suppose that we fix $\mby$ so that $x_\infty(\mby)=c,$ a constant.  Clearly if $x(t)=c$ then $x(t+h)=c$ as well, so $x=c$ is a fixed point of the iteration \eqref{eq:semi_implicit_method}, under this assumption.  Numerical stability follows if we can show that $x=c$ is a stable fixed point for all $h>0$, as we now establish.
Let $x(t_0+nh)=c+a_n,$ with $a_0$ arbitrary.  Then 
\begin{align}
    a_{n+1}&=x(t_0+nh+h)-c\\
    &=\frac{\tau x(t_0+nh)+hc}{\tau+h}-c\\
    &=\frac{\tau (c+a_n)+hc}{\tau+h}-c\\
    &=\frac{\tau}{\tau+h}a_n\to 0,\quad \text{ as }n\to\infty
\end{align}
no matter the size of the timestep $h>0$.  Thus the scheme \eqref{eq:semi_implicit_method} is both (first-order) accurate and numerically stable.

\newcommand{\mbb}{\mathbf{b}}
\newcommand{\mbx}{\mathbf{x}}

The head and grasper position variables $\xh$, $\xg$ form a linearly coupled pair, for which we can extend the semi-implicit algorithm given in one-dimensional form above.  
In general, consider a nonhomogeneous linear system expressed in terms of a vector $\mbx$, a matrix $A$, and a forcing vector $\mbb$:
\begin{equation}
\label{eq:nonhomogeneous linear ODE}
    \frac{d\mbx}{dt}=A(t)\mbx(t)+\mbb(t).
\end{equation}
To set up a semi-implicit first-order iteration scheme, observe that
\begin{align}
    \frac{\mbx(t+h)-\mbx(t)}{h}&=A(t)\mbx(t+h)+\mbb(t)+\mathcal{O}(h^2),\quad\text{ so }\\
    \mbx(t+h)-hA(t)\mbx(t+h)&=\mbx(t)+h\mbb(t)+\mathcal{O}(h^2),\quad\text{ therefore }\\
    \mbx(t+h)&=\left(I-hA(t)\right)^{-1}(\mbx(t)+h\mbb(t))+\mathcal{O}(h^2).
\end{align}
Dropping the $\mathcal{O}(h^2)$ term and writing the update  scheme in MATLAB style notation gives
\begin{framed}
\begin{equation}
\label{eq:semi-implicit-vector}
\mbx(t+h)=(I-h*A(t))\backslash(\mbx(t)+h*\mbb(t)).
\end{equation}
\end{framed}
(Here the backslash notation $M\backslash\mbu$ stands for $M^{-1}\mbu$, i.e., the least-squares solution $\mby$ to the linear system $M\mby=\mbu$.) 
Comparing this update scheme with \eqref{eq:semi_implicit_method}
it is easy to check that they are consistent for a single variable by setting $A=-1/\tau$ and $b=x_\infty/\tau$.

In our case the head and grasper positions $\xh$ and $\xg$ comprise a linearly coupled system, $\mbx$, and the  coupling matrix $A$ is $2\times 2$, so $I-hA$ can be inverted explicitly, provided $h$ is smaller than the reciprocal of the largest positive eigenvalue of $A$.  
(If no eigenvalues of $A$ are positive real numbers, then $I-hA$ can always be inverted.)
For a general $2\times 2$ system the update rule reads
\begin{equation}
\label{eq:semi-implicit update 2x2}
    \left(\begin{array}{cc}x_1(t+h) \\ x_2(t+h)\end{array}\right) = 
    \frac1{1-h\text{Tr}A +h^2\det A}
    \left(
    \begin{array}{c}
    (1-hA_{22})(x_1+h b_1)+hA_{12}(x_2+hb_2)\\
    hA_{21}(x_1+hb_1)+(1-hA_{11})(x_2+hb_2)
    \end{array}
    \right),
\end{equation}
with all time-varying elements of the right hand side evaluated at time $t$.
In \eqref{eq:semi-implicit update 2x2} $\text{Tr}A$ and $\det A$ denote the trace and determinant of $A$, respectively.  
Thus, truncating terms of order $\mathcal{O}(h^2)$ and higher gives a first-order semi-implicit update scheme for two-component state and forcing vectors $\mbx$ and $\mbb$:
\begin{framed}
\begin{equation}
\label{eq:semi-implicit update 2x2 first order}
    \mbx(t+h) = 
    \frac1{1-h\text{Tr}A(t)}\left[
    \left(I+h\left(
    \begin{array}{rr}
    -A_{22} & A_{12}\\
    A_{21} & -A_{11}    
    \end{array}
    \right)\right)\mbx(t)+h\mbb(t)\right].
\end{equation}
\end{framed}
In \eqref{eq:semi-implicit update 2x2 first order} $I$ denotes the $2\times 2$ identity matrix.

\newpage

\renewcommand{\arraystretch}{1.25}

\subsection{Table of Symbols} \label{symbols}

\begin{center}
\tiny
\begin{tabular}{@{}ll@{}}
\toprule

\textbf{Symbol} & \textbf{Meaning} \\

\midrule

$\xhref$            & reference position \\
$\xh$               & head position relative to reference \eqref{eq:dxdt} \\
$\xg$               & grasper position relative to reference \eqref{eq:dxdt} \\
$\xgh = \xg-\xh$    & grasper position relative to head \\

\midrule

$N_i$       & Boolean state of neuron $i$ \\
$A_i$       & muscle activation associated with muscle $i$ \\
$T_i$       & muscle tension associated with muscle $i$ \\
$\tau_m$    & muscle activation time constant \\
$c_h$       & damping coefficient for head movements \\
$c_g$       & damping coefficient for grasper movements \\

\midrule

$T_\text{I2}$    & I2 protractor muscle tension \eqref{eq:T_I2} \\
$T_\text{I3} $   & I3 retractor muscle tension \eqref{eq:T_I3} \\
$T_\text{hinge}$      & hinge muscle tension \eqref{eq:T_hinge} \\
$P_\text{I4}$                 & grasper pressure \eqref{eq:P_I4} \\
$P_\text{I3,ant.}$     & anterior I3 pinch pressure \eqref{eq:P_I3,ant} \\

\midrule

$F_\text{I2}$    & I2 protractor muscle force \eqref{eq:F_I2} \\
$F_\text{I3} $   & I3 retractor muscle force \eqref{eq:F_I3} \\
$F_\text{hinge}$      & hinge muscle force \eqref{eq:F_hinge} \\
$F_\text{I4}$    & grasper closing force \eqref{eq:F_I4} \\
$F_\text{I3,ant.}$     & anterior I3 pinch force \eqref{eq:F_I3,ant} \\

\midrule

$F_\text{I2,max}$    & scaling parameter for I2 protractor muscle force \\
$F_\text{I3,max} $   & scaling parameter for I3 retractor muscle force  \\
$F_\text{hinge,max}$      & scaling parameter for hinge muscle force  \\
$F_\text{I4,max}$                 & scaling parameter for grasper closing force \\
$F_\text{I3,ant.,max}$     & scaling parameter for anterior I3 pinch force \\

\midrule

$F_h$    & net force on head \eqref{eq:F_h} \\
$F_g$    & net force on grasper \eqref{eq:F_g} \\
$F_o$    & net force on object, force on transducer during swallowing \eqref{eq:F_o} \\

\midrule

$[\text{hinge stretched}] = [\xgh>0.5]$                 & Boolean state of hinge stretch \\
$\unbroken = [F_o \le z_s]$                             & Boolean state of seaweed unbroken \\

\midrule

$\ChemLips$       & Boolean state of chemical stimulus at lips \\
$\MechLips$       & Boolean state of mechanical stimulus at lips \\
$\MechGrasper$    & Boolean state of mechanical stimulus in grasper \\

\midrule

$K_g$               & grasper spring constant \\
$K_h$               & head spring constant \\
$x_{g/h}^0$         & rest length of grasper spring \\
$x_{h}^0$           & rest length of the head spring \\

\midrule

$\mu_{s,g}$         & coefficient of static friction between grasper and seaweed \\
$\mu_{k,g}$         & coefficient of kinetic friction between grasper and seaweed \\
$\mu_{s,h}$         & coefficient of static friction between head (jaws) and seaweed \\
$\mu_{k,h}$         & coefficient of kinetic friction between head (jaws) and seaweed \\

\bottomrule

\end{tabular}
\end{center}

\newpage

\subsection{Boolean Logic of \textit{Aplysia} Feeding Control} \label{logic}

The following sections detail the logic implementations for the activity of each neuron in the controller network. \red{All neurons, except for B4/B5, are implemented as standard Boolean elements and are either OFF (0), or ON (1). B4/B5 is a ternary unit which is either OFF (0), weakly ON (1), or strongly ON (2). The activity state of B4/B5 must be checked using conditional logic prior to negation, as $\lneg N_{B4/B5}$ is undefined (see Section~\ref{sec:boolean-network}).}
These interactions include known direct connections between neurons based on previous literature as well as hypothesized connections that may be direct or indirect, and indirect sensory feedback pathways. 
Sensory feedback pathways involving proprioception of the grasper position and pressure exerted by the closed grasper are gated by logic tests of position and pressure relative to user specified thresholds. Allowing these thresholds to vary in response to activity levels of interneurons and external sensory cues provides an approximation of neuromodulation. 
All time varying elements on the right side of the logic equations are at time $(j)$.
\\

\noindent
\textbf{Cerebral Interneurons}
\begin{enumerate}

\item{Metacerebral Cell}
\begin{equation}
    \MCC (j+1) = [\text{arousal}]
\end{equation}

\item{CBI-2}\\
CBI-2 is activated by sensory inputs present in biting and rejection, but not in swallowing.
\begin{equation}
    \begin{split}
        \CBItwo (j+1) = & \MCC \land (\lneg \Bsixfour) \land \big(\left(\MechLips \land \ChemLips \land \lneg \MechGrasper \right) \lor \\
        &\left( \MechGrasper \land \lneg \ChemLips \right) \big)
    \end{split}
\end{equation}

With the hypothesized connections in Section \ref{sec:hypotheses}, these equation changes to:

\begin{equation}
    \begin{split}
        \CBItwo (j+1) = & \MCC \land (\lneg \Bsixfour) \land \big(\left(\MechLips \land \ChemLips \land \lneg \MechGrasper \right) \lor \\
        &\left( \MechGrasper \land \lneg \ChemLips \right) \lor \left(\Bfour \geq 2 \right)\big)
    \end{split}
\end{equation}

\item{CBI-3}\\
CBI-3 is activated by sensory inputs present in biting and swallowing, but not in rejection. 

\begin{equation}
    \begin{split}
        \CBIthree (j+1) = & \MCC \land \MechLips \land \ChemLips 
    \end{split}
\end{equation}

With the equations and refractory period proposed in Section \ref{sec:hypotheses}, the logic implementation for CBI-3 changes to include a gating state variable based on whether or not the neuron is in a refractory state following strong inhibition. Similar logic could be added to other nodes in the network as needed based on animal experiments. This period was included here as part of the hypothesis that strong activation of B4/B5 triggers rejection in animals that are swallowing. This hypothesis and an assessment of whether this refractory period occurs in CBI-3 in animal preparations or whether this effect is due to another mechanism could be tested experimentally. The equation becomes:

\begin{equation}
    \begin{split}
        \CBIthree (j+1) = & \MCC \land \MechLips \land \ChemLips \\
        & \land  \left(\Bfour < 2 \right) \land \left(\lneg [\text{refractory}_\text{CBI-3}]\right) 
    \end{split}
\end{equation}

\item{CBI-4}\\
CBI-4 is activated by sensory inputs present in swallowing and rejection, but not in biting.
\begin{equation}
    \CBIfour (j+1) = \MCC \land \left(\MechLips \lor \ChemLips\right) \land \MechGrasper 
\end{equation}

\end{enumerate}

\par

\noindent
\textbf{Buccal Interneurons}

\begin{enumerate}

\item{B64} \\
Activity in $\Bsixfour$ is influenced by the activity of the $\MCC$ and $\Bthreeone$. 
It is also excited by protraction and inhibited by retraction. 
The proprioceptive feedback is implemented as:

\begin{equation}
    \begin{split}
        \text{B64}_\text{proprioception} = & (\CBIthree \land ((\MechGrasper \land [\text{protracted}_{\Bsixfour,\text{swallow}}]) \lor \\
        &\quad ((\lneg \MechGrasper)\land [\text{protracted}_{\Bsixfour,\text{bite}}]))\quad ) \lor \\
        &((\lneg \CBIthree)\land [\text{protracted}_{\Bsixfour,\text{reject}}])
    \end{split}
\end{equation}

\noindent
where, 
\begin{equation}
    [\text{protracted}_{\Bsixfour,\text{swallow}}] = [\xgh > z_{\text{B64},\text{swallow}}]
\end{equation}
\begin{equation}
    [\text{protracted}_{\Bsixfour,\text{bite}}] = [\xgh > z_{\text{B64},\text{bite}}]
\end{equation}
\begin{equation}
    [\text{protracted}_{\Bsixfour,\text{reject}}] = [\xgh > z_{\text{B64},\text{reject}}]
\end{equation}

This amounts to the threshold being depend on the behavior with different threshold values for bites, swallows, and rejections.

\begin{equation}
    \begin{split}
        \Bsixfour (j+1) = & \MCC \land (\lneg \Bthreeone) \land \text{B64}_\text{proprioception}
    \end{split}
\end{equation}

\item{B4/B5}\\

$\Bfour$ has been shown to have varying effects when firing strongly vs. weakly. To represent this in the modeling framework, quiescence is represented as 0, weak firing as 1 and strong firing as 2. The neurons are quiescent during biting, and they fire weakly during the retraction phase of swallowing. The neurons fire strongly when stimulated with the external electrode and during the retraction phase of rejection. During rejection, B4/B5 is observed to cease firing, allowing B3/B6/B9 to fire briefly at the end of the behavior. To implement this, we have used a proprioceptive feedback pathway which inhibits the activity of B4/B5 once the grasper has reached a user-specified level of retraction.

\begin{equation}
    \begin{split}
        \Bfour (j+1) = \MCC \land \bigg( &(\lneg [\text{electrode}_\text{B4/B5}]) \land \big( 2 (\lneg \CBIthree) \land \Bsixfour\\ 
        &\land [\text{protracted}_{\Bfour}] + \\
        &\CBIthree \land \MechGrasper \land \Bsixfour \big ) + \\
        &2 \, [\text{electrode}_\text{B4/B5}] \bigg)
    \end{split}
\end{equation}

where,
\begin{equation}
    [\text{protracted}_{\Bfour}] = [\xgh > z_{\text{B4/B5}}]
\end{equation}

\item{B20}
\begin{equation}
    \begin{split}
        \Btwenty (j+1) = \MCC \big(\CBItwo \lor \CBIfour \lor \Bthreeone \big) \land \lneg \CBIthree \land \lneg \Bsixfour
    \end{split}
\end{equation}

\item{B40/B30}

$\Bforty$ has fast inhibitory and slow excitatory connections to $\Beight$. To capture this, we record the time (j) at which $\Bforty$ transitions between states for later use in the $\Beight$ activity calculations (see below). First, the activity of $\Bforty$ in the next time step is determined:

\begin{equation}
    \begin{split}
        \Bforty (j+1) = \MCC \land \big(\CBItwo \lor \CBIfour \lor \Bthreeone \big) \land \lneg \Bsixfour
    \end{split}
\end{equation}

After calculating the new activity, we assess transitions as defined by the following pseudocode:

if $(\Bforty (j) == 0 \, \text{AND} \, \Bforty (j+1) == 1)$, then set $t_{\Bforty\text{,on}}$ = j;

if $(\Bforty (j) == 1 \, \text{AND} \, \Bforty (j+1) == 0)$, then set $t_{\Bforty\text{,off}}$ = j;

\end{enumerate}

\noindent
\textbf{Buccal Motor Neurons}

\begin{enumerate}

\item{B31/B32}

$\Bthreeone$ receives input from interneurons and proprioceptive feedback. To capture possible modulation of $\Bthreeone$ and generate multifunctional behavior under different sensory cues, behavior-dependent proprioceptive inputs are implemented. Though the resulting full equation for $\Bthreeone$ activity is large, it can be broken down to three sections: (1) if $\CBIthree$ is active and there is sensory stimuli in the grasper (swallowing), (2) if $\CBIthree$ is active and there is NOT sensory stimuli in the grasper (biting), and (3) if $\CBIthree$ is NOT active (rejection).

\begin{equation}
    \begin{split}
        \Bthreeone (j+1) = & \MCC \land \bigg( \CBIthree \land \\ 
        &\MechGrasper 
        ( (\lneg \Bsixfour) \land ((![\text{pressure}_{\Bthreeone,\text{ingestion}}]) \lor \CBItwo) \land \\
        &\quad ((\lneg \Bthreeone)\land [\text{retracted}_{\Bthreeone,\text{swallow},\text{off}}] +\\
        & \quad \Bthreeone \land [\text{retracted}_{\Bthreeone,\text{swallow},\text{on}}]))\\
        &(\lneg \MechGrasper) 
        ( (\lneg \Bsixfour) \land ((![\text{pressure}_{\Bthreeone,\text{ingestion}}]) \lor \CBItwo) \land \\
        &\quad ((\lneg \Bthreeone)\land 
        [\text{retracted}_{\Bthreeone,\text{bite},\text{off}}] +\\
        &\quad \Bthreeone \land [\text{retracted}_{\Bthreeone,\text{bite},\text{on}}])) + \\
        &(\lneg \CBIthree) 
        \land \big((\lneg \Bsixfour) \land [\text{pressure}_{\Bthreeone,\text{rejection}}] \land \\
        &  \quad (\CBItwo \lor \CBIfour) \land \\
        &\quad ((\lneg \Bthreeone)\land [\text{retracted}_{\Bthreeone,\text{reject},\text{off}}] + \\
        &  \quad \Bthreeone \land [\text{retracted}_{\Bthreeone,\text{reject},\text{on}}] \big) \bigg)
    \end{split}
\end{equation}

\noindent
where,
\begin{equation}
    [\text{pressure}_{\Bthreeone,\text{ingestion}}] = [P_g > 0.5 p_\text{max}]
\end{equation}

\begin{equation}
    [\text{pressure}_{\Bthreeone,\text{rejection}}] = [P_g > 0.25 p_\text{max}]
\end{equation}

\begin{equation}
    [\text{retracted}_{\Bthreeone,\text{swallow},\text{off}}] = [\xgh < z_{\Bthreeone,\text{swallow},\text{off}}]
\end{equation}
\begin{equation}
    [\text{retracted}_{\Bthreeone,\text{swallow},\text{on}}] = [\xgh < z_{\Bthreeone,\text{swallow},\text{on}}]
\end{equation}

\begin{equation}
    [\text{retracted}_{\Bthreeone,\text{bite},\text{off}}] = [\xgh < z_{\Bthreeone,\text{bite},\text{off}}]
\end{equation}
\begin{equation}
    [\text{retracted}_{\Bthreeone,\text{bite},\text{on}}] = [\xgh < z_{\Bthreeone,\text{bite},\text{on}}]
\end{equation}

\begin{equation}
    [\text{retracted}_{\Bthreeone,\text{reject},\text{off}}] = [\xgh < z_{\Bthreeone,\text{reject},\text{off}}]
\end{equation}
\begin{equation}
    [\text{retracted}_{\Bthreeone,\text{reject},\text{on}}] = [\xgh < z_{\Bthreeone,\text{reject},\text{on}}]
\end{equation}

\item{B6/B9/B3}

\begin{equation}
    \begin{split}
        \Bsix (j+1) =  \MCC &\land \Bsixfour\land (\lneg (\Bfour \geq 2))  \land \bigg(\\
        &\big((\CBIthree \land (\lneg \MechGrasper)) \land [\text{pressure}_{\Bsix,\text{bite}}] \big) + \\
        &\big((\CBIthree\land \MechGrasper)  \land [\text{pressure}_{\Bsix,\text{swallow}}] \big) + \\
        & (\lneg \CBIthree) \land (\lneg [\text{pressure}_{\Bsix,\text{reject}}]) \big) \bigg)
    \end{split}
\end{equation}

\noindent
where,
\begin{equation}
    [\text{pressure}_{\Bsix,\text{bite}}] = [P_g > z_{\Bsix,\text{bite},\text{pressure}})]\end{equation}

\begin{equation}
    [\text{pressure}_{\Bsix,\text{swallow}}] = [P_g > z_{\Bsix,\text{swallow},\text{pressure}})]\end{equation}

\begin{equation}
    [\text{pressure}_{\Bsix,\text{reject}}] = [P_g > z_{\Bsix,\text{reject},\text{pressure}}]
\end{equation}

\item{B8a/b}

$\Beight$ receives fast inhibitory and slow excitatory input from $\Bforty$ \cite{Jing2004,Cropper2019}. In the Boolean framework here we implement this as an excitatory input immediately following cessation of $\Bforty$ activity for a user specified duration ($\text{duration}_{\Bforty,\text{excite}}$). Prior to calculating a new value for $\Beight$ we first check whether the synaptic connection from $\Bforty$ is excitatory with the following statements:\\

\par

if $(\Bforty (j) == 0 \, \text{AND} \, j < (t_{\Bforty\text{,off}} + \text{duration}_{\Bforty,\text{excite}}))$, then set $\Bforty,\text{excite} = 1$\\
else set $\Bforty,\text{excite} = 0$

\begin{equation}
    \begin{split}
        \Beight (j+1) = &\MCC \land (\lneg (\Bfour \geq 2)) \land \\
        &((\CBIthree \land (\Btwenty \lor (\Bforty,\text{excite})) \land \\
        & \quad \quad (\lneg \Bthreeone)) + \\
        & \quad ((\lneg \CBIthree) \land \Btwenty))
    \end{split}
\end{equation}

\item{B7}
\label{appendix item B7}

\begin{equation}
    \begin{split}
        \Bseven (j+1) = &\MCC \\
        &\land\big(\big((\lneg \CBIthree \lor \MechGrasper ) \land ([\text{protracted}_{\Bseven,\text{reject}}] \lor [\text{pressure}_{\Bseven}])\big) + \\
        &\big((\CBIthree \land \lneg \MechGrasper) \land ([\text{protracted}_{\Bseven,\text{bite}}] \lor [\text{pressure}_{\Bseven}])\big)\big)
    \end{split}
\end{equation}

\noindent
where,
\begin{equation}
    [\text{protracted}_{\Bseven,\text{reject}}] = [\xgh > z_{\text{B7,\text{reject}}}]
\end{equation}
\begin{equation}
    [\text{protracted}_{\Bseven,\text{bite}}] = [\xgh > z_{\text{B7,\text{bite}}}]
\end{equation}
\begin{equation}
    [\text{pressure}_{\Bseven}] = [P_g > z_{\Bseven,\text{pressure}}]
\end{equation}

\item{B38}

\begin{equation}
    \begin{split}
        \Bthreeeight (j+1) = \MCC \land \MechGrasper \land \bigg( \CBIthree \land  [\text{retracted}_{\Bthreeeight}] \bigg)
    \end{split}
\end{equation}

\noindent
where,
\begin{equation}
    [\text{retracted}_{\Bthreeeight}] = [\xgh < z_{\text{B38}}]
\end{equation}

\end{enumerate}

\newpage

\subsection{Muscle Forces} \label{MuscleForces}

Contact forces, such as the pressure resulting from grasper closure and force due to the anterior pinch, are implemented as second-order responses to neural activation using the semi-implicit integration scheme, Eq. \eqref{eq:semi_implicit_method}, as shown in the following equations.
\begin{enumerate}
    \item{Grasper Pressure}
    
    \begin{equation}\label{eq:P_I4}
        P_\text{I4}(t+h) = \frac{\tau_\text{I4} P_\text{I4}(t) + h A_\text{I4}(t)}{\tau_\text{I4} + h}
    \end{equation}

    \begin{equation}
        A_\text{I4}(t+h) = \frac{\tau_\text{I4} A_\text{I4}(t) + h \Beight(t)}{\tau_\text{I4} + h}
    \end{equation}
    
    \item{Pinch Pressure}
    
    \begin{equation}\label{eq:P_I3,ant}
        P_\text{I3,ant.}(t+h) = \frac{\tau_\text{I3,ant.} P_\text{I3,ant.} (t) + h A_\text{I3,ant.}(t) }{\tau_\text{I3,ant.} +h}
    \end{equation}

    \begin{equation}
        A_\text{I3,ant.}(t+h) = \frac{\tau_\text{I3,ant.} A_\text{I3,ant.}(t) + h (\Bthreeeight(t) + \Bsix (t))}{\tau_\text{I3,ant.} + h}
    \end{equation}
\end{enumerate}

\noindent
Muscle tensions for the remaining musculature were calculated using a second-order response to the neural activity as outlined in the following equations.
\begin{enumerate}
    \item{I3 Tension}
    
    \begin{equation}\label{eq:T_I3}
        T_\text{I3}(t+h) = \frac{\tau_\text{I3} T_\text{I3}(t) + h A_\text{I3}(t)}{\tau_\text{I3} + h}
    \end{equation}
    
    \begin{equation}
        A_\text{I3}(t+h) = \frac{\tau_\text{I3} A_\text{I3}(t) + h \Bsix(t)}{\tau_\text{I3} + h}
    \end{equation}
    
    \item{I2 Tension}
    
    Time constants for I2 were tuned independently for ingestion and egestion to account for the experimental observations that egestions have a longer period than ingestions.  Such variation in responsiveness of the animal may exist due to differences in neuromodulation between the behaviors. Therefore the time constant for I2 is calculated as:
    
    \begin{equation}
        \tau_\text{I2} =\CBIthree \;\tau_\text{I2,ingestion} + (\lneg \CBIthree) \;\tau_\text{I2,egestion}
    \end{equation}
    
    \begin{equation}\label{eq:T_I2}
        T_\text{I2}(t+h) = \frac{\tau_\text{I2} T_\text{I2} + h A_\text{I2}}{\tau_\text{I2} + h}
    \end{equation}
    
    \begin{equation}
        A_\text{I2}(t+h) = \frac{\tau_\text{I2} A_\text{I2}(t) + h \Bthreeone(t)}{\tau_\text{I2} + h}
    \end{equation}

    \item{Hinge Tension}
    
    \begin{equation}\label{eq:T_hinge}
        T_\text{hinge}(t+h) = \frac{\tau_\text{hinge} T_\text{hinge}(t) + h A_\text{hinge}(t)}{\tau_\text{hinge} + h}
    \end{equation}
    
    \begin{equation}
        A_\text{hinge}(t+h) = \frac{\tau_\text{hinge} A_\text{hinge}(t) + h \Bseven(t)}{\tau_\text{hinge} + h}
    \end{equation}
    
\end{enumerate}

\subsection{Biomechanical Model} \label{biomechanics}

The motions of the head and grasper are calculated based on the quasi-static equations of motion:

\renewcommand{\arraystretch}{2}
\begin{equation}\label{eq:dxdt}
    \frac{d}{dt} \xvec = {\begin{bmatrix}
\frac{F_g}{c_g}\\
\frac{F_h}{c_h}
\end{bmatrix}}
\end{equation}

\noindent
where $\xh$ is the position of the head relative to the ground frame, $\xg$ is the position of the grasper relative to the ground frame, and $c_h$ and $c_g$ are the damping coefficients for the motion of the head and grasper, respectively. The forces on the grasper and head can be calculated as outlined in the following sections.

\subsubsection{Forces on the grasper}

The positive direction for $\xg$ corresponds to protraction (Figure \ref{fig:SchematicIntro}). The sum of forces on the grasper is

\begin{equation}\label{eq:F_g}
    F_g = {F}_\text{I2} + F_{sp,g} - {F}_\text{I3} - {F}_\text{hinge} + {F}_{f,g}
\end{equation}

\noindent
where the component forces are defined and calculated as follows:

\vspace{10pt}
\noindent
${F}_\text{I2}$: The force due to the I2 muscle. This value is dependent on the tension of the muscle as well as the mechanical advantage. It is scaled by a tunable maximum parameter, $F_\text{I2,max}$, and is calculated as follows:

\begin{equation} \label{eq:F_I2}
    {F}_\text{I2} = F_\text{I2,max} T_\text{I2}(t)(1-\xgh)
\end{equation}

\noindent
where $\xgh = \xg - \xh$ is the position of the grasper relative to the head.

\vspace{10pt}
\noindent
$F_{sp,g}$: The force in the spring connecting the grasper to the head. This spring represents the surrounding musculature of the esophagus, buccal mass, and extrinsic muscles which are not explicitly modeled. This is calculated as:

\begin{equation}
    F_{sp,g} = K_g (x_{g/h}^0-\xgh)
\end{equation}

\noindent
where $K_g$ is the spring constant and $x_{g/h}^0$ is the rest length of the spring.

\vspace{10pt}
\noindent
${F}_\text{I3}$: The force due to the I3 muscle which pushes the grasper backwards during retraction. This force is due to tension in I3 closing the muscular toroids. This value is dependent on the tension of the muscle as well as the mechanical advantage. It is scaled by a tunable maximum parameter, $F_\text{I3,max}$, and is calculated as follows:

\begin{equation} \label{eq:F_I3}
    {F}_\text{I3} = F_\text{I3,max} T_\text{I3}(t)(\xgh-0)
\end{equation}

\vspace{10pt}
\noindent
${F}_\text{hinge}$: The force due to the hinge. This value is dependent on the tension of the muscle as well as the mechanical advantage. It is scaled by a tunable maximum parameter, $F_\text{hinge,max}$, and is calculated as follows:

\begin{equation} \label{eq:F_hinge}
   {F}_\text{hinge} = [\text{hinge stretched}]F_\text{hinge,max} T_\text{hinge}(t)(\xgh-0.5)
\end{equation}

\noindent
where $[\text{hinge stretched}] = [\xgh>0.5]$ determines whether the hinge is sufficiently stretched to produce any force \cite{Sutton2004a}.

\vspace{10pt}
\noindent
$F_{f,g}$: Friction resulting from the grasper closing on an object. To determine $F_{f,g}$ it is necessary to check if the grasper is slipping against the object by checking the inequality:

\begin{equation}\label{eq:grasper-static-friction-condition}
    |{F}_\text{I2} + F_{sp,g} - {F}_\text{I3} - {F}_\text{hinge}| \leq |\mu_{s,g} {F}_\text{I4}|
\end{equation}

\noindent
where $\mu_{s,g}$ is the coefficient for static friction between the grasper and the object. ${F}_\text{I4}$ is the normal force due to the grasper muscle I4 closing on the object. This is calculated directly as the grasper pressure defined in the previous appendix applied to a unit area scaled by a parameter.

\begin{equation}\label{eq:F_I4}
    {F}_\text{I4} = F_{I4,\text{max}} P_\text{I4}(t)
\end{equation}

If the condition in Eq. \eqref{eq:grasper-static-friction-condition} is true, then the contact is in a state of static friction and $F_{f,g}$ is calculated as:

\begin{equation}
    |F_{f,g}| = {F}_\text{I2} + F_{sp,g} - {F}_\text{I3} - {F}_\text{hinge}
\end{equation}

If the condition in Eq. \eqref{eq:grasper-static-friction-condition} is not true, the contact is sliding and is in a state of kinetic friction, and $F_{f,g}$ is calculated as:

\begin{equation}
    |F_{f,g}| = \mu_{k,g} {F}_\text{I4}
\end{equation}

\noindent
where $\mu_{k,g}$ is the coefficient for kinetic friction between the grasper and the seaweed.

The sign of the friction force is dependent on which direction the grasper would be moving without the friction present, and ${F}_{f,g}$ can be calculated as:

\begin{equation}
 {F}_{f,g} = -\text{sgn}({F}_\text{I2} + F_{sp,g} - {F}_\text{I3} - {F}_\text{hinge}) |F_{f,g}|
\end{equation}

\subsubsection{Forces on head}

The forces on the head are calculated as:
\begin{equation}\label{eq:F_hFull}
    F_h = F_{sp,h} - F_{sp,g} - F_\text{I2} + F_\text{I3} + F_\text{hinge} + F_{f,h}
\end{equation}

\noindent
The muscles and grasper spring exert forces on the head equal and opposite to those on the grasper. As the muscles contract and apply forces to move the grasper forward this also stretches the spring between the grasper and head proportionally to the muscle force. For the quasi-static model, acceleration is assumed to be negligible and therefore the forces on the grasper must equal zero. 

\begin{equation}
    0 = F_{sp,g} + F_\text{I2} - F_\text{I3} - F_\text{hinge} + F_{f,g}
\end{equation}

Solving for the spring forces, $F_{sp,g}$, and substituting into Eq. \eqref{eq:F_hFull} yields:

\begin{equation}
    F_h = F_{sp,h} + \cancel{F_\text{I2}} - \cancel{F_\text{I3}} - \cancel{F_\text{hinge}} + F_{f,g} - \cancel{F_\text{I2}} + \cancel{F_\text{I3}} + \cancel{F_\text{hinge}} + F_{f,h}
\end{equation}

\noindent
which simplifies to:

\begin{equation}\label{eq:F_h}
    F_h = F_{sp,h} + F_{f,g} + F_{f,h}
\end{equation}

\noindent
where $F_{sp,h}$ is the spring force between the head and neck of the animal, $F_{f,g}$ is the previously calculated friction force between the grasper and the object, and $F_{f,h}$ is the friction force resulting from the jaws pinching on the object. These components are calculated as follows.

\begin{equation}
    F_{sp,h} = K_h (x_{h}^0-x_{h})
\end{equation}

\noindent
where $K_h$ is the spring constant and $x_{h}^0$ is the rest length of the spring.

To determine the value of $F_{f,h}$ it is necessary to check if the jaws are slipping relative to the seaweed by checking the following inequality:

\begin{equation}\label{eq:jaws-static-friction-condition}
    |F_{sp,h} + F_{f,g}| \leq |\mu_{s,h} {F}_\text{I3,ant.}|
\end{equation}

\noindent
where $\mu_{s,h}$ is the coefficient for static friction between the jaws and the seaweed. ${F}_\text{I3,ant.}$ is the normal force due to the anterior portion of the I3 jaw muscle closing on the seaweed. This is calculated directly as the pinch pressure defined in the previous appendix applied to a unit area, scaled by a parameter, and multiplied by a mechanical advantage term:

\begin{equation}\label{eq:F_I3,ant}
    {F}_\text{I3,ant.} = F_{\text{I3,ant.,max}} P_\text{I3,ant.}(t)(1-\xgh).
\end{equation}

If the condition in Eq. \eqref{eq:jaws-static-friction-condition} is true, the jaws are in static friction and $F_{f,h}$ is calculated as:

\begin{equation}
    |F_{f,h}| = F_{sp,h} + F_{f,g}
\end{equation}

If the condition in Eq. \eqref{eq:jaws-static-friction-condition} is not true, the jaws are slipping and $F_{f,h}$ is calculated as:

\begin{equation}
    |F_{f,h}| = \mu_{k,h} {F}_\text{I3,ant.}
\end{equation}

\noindent
where $\mu_{k,h}$ is the coefficient for kinetic friction between the jaws and the seaweed.

The sign of the friction force is dependent on which direction the head would be moving without the friction present and ${F}_{f,h}$ can be calculated as:

\begin{equation}
 {F}_{f,h} = -\text{sgn}(F_{sp,h} + F_{f,g}) |F_{f,h}|
\end{equation}

\subsubsection{Force on objects}

The force on the object if unbroken is equal to the sum of the friction forces where we use the conventions that positive force indicates tension on the force transducer:

\begin{equation} \label{eq:F_o}
    F_o = F_{f,g} + F_{f,h}
\end{equation}

If $F_o \leq z_s$, where $z_s$ is the user defined seaweed strength, the seaweed is not broken and the motion of the bodies is calculated based on the forces calculated in the previous sections. If $F_o > z_s$, the seaweed is broken and can no longer transmit forces to the head or grasper. Therefore the forces on the head and grasper are recalculated as:

\begin{equation}
    F_h = F_{sp,h}
\end{equation}

\begin{equation}
    F_g = {F}_\text{I2} + F_{sp,g} - {F}_\text{I3} - {F}_\text{hinge}
\end{equation}

A Boolean tracking variable [unbroken] is used to track whether the seaweed is intact (1) or broken (0). Once the seaweed breaks, it is not restored until the grasper has completed a new protraction and grasp motion. For this model, we have implemented this by resetting $[\text{unbroken}] = 1$ if at the current timestep $[\text{unbroken}] ==0$  AND $\xgh < 0.3$ AND $\xgh(j+1)>\xgh(j)$. These thresholds were tuned manually for this implementation.

\subsubsection{Updating Grasper and Head Positions}

All of the forces in this biomechanical model are linearly dependent on the position of the head, $\xh$, and grasper, $\xg$. Therefore they can each be rewritten in the form:

\begin{equation}
    F = \textbf{A}_F \xvec + \textbf{b}_F
\end{equation}

As a consequence, the equations of motion can be rewritten in the form 
\begin{equation}
    \frac{d}{dt} \xvec = {\begin{bmatrix}
\frac{A_{11}}{c_h} & \frac{A_{12}}{c_h}\\
\frac{A_{21}}{c_g} & \frac{A_{22}}{c_g}
\end{bmatrix}} \xvec + \begin{bmatrix}
b_{1} \\
b_{2}
\end{bmatrix}
\end{equation}

This can then be integrated with the semi-implicit integration scheme in Appendix \ref{integration} as:

\begin{equation}
    \mbx(t+h) = 
    \frac1{1-h\text{Tr}A(t)}\left[
    \left(I+h\left(
    \begin{array}{rr}
    -A_{22} & A_{12}\\
    A_{21} & -A_{11}    
    \end{array}
    \right)\right)\mbx(t)+h\mbb(t)\right].
\end{equation}

\end{document}